\documentclass{article}

\usepackage{arxiv}

\usepackage{amsmath}
\usepackage[table]{xcolor}
\usepackage{graphicx}
\usepackage[utf8]{inputenc} 
\usepackage[T1]{fontenc}    
\usepackage{hyperref}       
\usepackage{url}            
\usepackage{booktabs}       
\usepackage{amsfonts}       
\usepackage{nicefrac}       
\usepackage{microtype}      

\usepackage{color}
\definecolor{red}{rgb}{1,0.2,0.2}
\definecolor{green}{rgb}{0.2,1,0.5}
\definecolor{blue}{rgb}{0,0,1}
\definecolor{lightblue}{rgb}{0.3,0.5,1}
\definecolor{lightgray}{gray}{0.9}

\makeatletter

\def\hlinewd#1{%
\noalign{\ifnum0=`}\fi\hrule \@height #1 \futurelet
\reserved@a\@xhline}
\makeatother

\title{Entanglement Routing in Quantum Networks:\\
A Comprehensive Survey}

\author{
Amar Abane \\
National Institute of Standards and Technology \\
Gaithersburg, MD 20899, USA \\
\And
Michael Cubeddu \\
~~~~~~~~~~~~~~~~~~Aliro Technologies, Inc.~~~~~~~~~~~~~~~~~~ \\
Brighton, MA 02135, USA \\
\And
Van Sy Mai \\
National Institute of Standards and Technology \\
Gaithersburg, MD 20899, USA \\
\And
Abdella Battou \\
National Institute of Standards and Technology \\
Gaithersburg, MD 20899, USA \\
}

\begin{document}
\maketitle

\begin{abstract}
Entanglement routing in near-term quantum networks consists of choosing the optimal sequence of short-range entanglements to combine through swapping operations to establish end-to-end entanglement between two distant nodes. Similar to traditional routing technologies, a quantum routing protocol uses network information to choose the best paths to satisfy a set of end-to-end entanglement requests. However, in addition to network state information, a quantum routing protocol must also take into account the requested entanglement fidelity, the probabilistic nature of swapping operations, and the short lifetime of entangled states. 
In this work, we formulate a practical entanglement routing problem and analyze and categorize the main approaches to address it, drawing comparisons to, and inspiration from, classical network routing strategies where applicable. We classify and discuss the studied quantum routing schemes into reactive, proactive, opportunistic, and virtual routing. 
\end{abstract}
\newpage
\tableofcontents

\section{Introduction}
In the emerging field of quantum networking, establishing efficient routing mechanisms emerges as a fundamental challenge to enable reliable quantum information transfer across distant quantum devices \cite{IBM50qubit, Google72qubit, castelvecchi2017china}. 
Quantum communication is made possible through entanglement \cite{bennett1993first}, a phenomenon where distant particles (e.g., photons) exhibit strongly correlated behaviors in a way that defies classical physics. Two independent pairs of entangled particles can be combined to produce a single long-distance entangled pair in which the local particle of the first entanglement is entangled with the remote particle of the second one, via a process known as entanglement swapping \cite{zukowski1993swapping, riedmatten2005swapping}. 
Swapping is implemented by quantum repeaters, which are responsible for extending the range of quantum communication to longer distances. 
Entanglement routing, analogous to traffic management in classical networking, includes mechanisms for coordinating quantum data flow and operations across a quantum repeater network.

\subsection{Background}
In recent years, several protocol stack abstractions have been proposed for quantum networks \cite{illiano2022quantum}, all identifying entanglement as the main resource for quantum communication. Thus, quantum routing must deal with certain properties unique to quantum physics -- properties with no counterpart in classical routing. For example, quantum signals are fragile and cannot be amplified, copied, or indefinitely stored as with classical communication signals. This distinction has major implications for the architecture of quantum networks, as these properties prohibit near-term quantum networks from being direct-transmission networks, or networks where quantum information seamlessly flows from source to destination. 

Given the maturity of current quantum hardware technologies, the most viable way to use entanglement for transmitting quantum data between two distant nodes is to establish end-to-end (E2E) entangled pairs between the two end nodes. 
Consequently, routing in near-term quantum networks consists of choosing the best sequence of adjacent entanglements to stitch together using swapping to establish E2E entanglements between two end nodes. Similarly to classical networks, routing in quantum networks is supported by a routing protocol to collect network information, and a path computation algorithm to select the best paths to satisfy E2E entanglement requests.

The generated E2E entangled pairs can be consumed in a variety of ways in support of a multitude of applications. 
In addition to quantum data transfer via teleportation, other well-known quantum networking applications include quantum cryptography, quantum computing, and quantum sensing. Each of these application instances may impose different requirements and, as such, may place different constraints on the E2E entanglement requests for the application to be successful. For example, many applications typically consume a stream of individual E2E entangled pairs and may specify a desired E2E entanglement generation rate to achieve performance. Other applications may place requirements on the quality of the E2E entanglement, known as fidelity. More concretely, applications like quantum key distribution (QKD) may desire a high-rate stream of E2E entangled pairs, where all the produced pairs fall above a minimum fidelity threshold, whereas a distributed quantum computing application may prioritize a high-fidelity stream of E2E entanglement for more reliable teleportation of quantum data amongst processors. Such requirements may be specified as quality-of-service (QoS) parameters in the E2E entanglement requests --- it is the responsibility of the routing protocol to determine ways to optimally provision paths and resources to service these requests.

Due to the inherent quantum physical properties, routing in quantum networks is more challenging than in classical routing. For example, path computation for an E2E entanglement request does not only depend on the number of hops, the link costs, or throughput. It must also consider the requested E2E entanglement fidelity, where higher-fidelity entanglements may take more time to be generated. Path computation must also take into account the probabilistic nature of entanglement generation and swapping, as well as the short lifetime of entangled particles. If a swapping operation fails, the two entanglements involved are destroyed and must be regenerated while the other entanglements along the path decay. Furthermore, establishing entanglements and performing swapping relies on Local Operations and Classical Communication (LOCC) for carrying and processing measurements and control messages, which in turn adds latency, complicates signaling, and may impact the overall success probability of the routing process.

The complex challenge of entanglement routing in quantum networks has gained substantial interest recently. 
Researchers have considered network modeling, path computation algorithms, protocol design, and theoretical and experimental studies to explore various aspects of entanglement routing through diverse approaches.

In this paper, we review research efforts on entanglement routing in near-term quantum networks, including algorithms, protocol designs, and studies. The survey aims to provide a comprehensive description of the quantum routing problem, categorize and discuss the main approaches proposed to address it, and draw comparisons to analogous classical networking technologies.
Our objective is to bridge the gap between theoretical advancements and practical implementations, shedding light on the most promising and realistic solutions that can be adapted for emerging intermediate-scale quantum networks.

Intermediate-scale quantum networks are currently being deployed at a metropolitan scale around the world \cite{craddock2024automated}. They may consist of a few tens of nodes arranged in a non-trivial topology (relative to the trivial topology of a linear quantum repeater chain) with point-to-point distances of a few 100s of kilometers. These networks are intended to serve multiple users and applications simultaneously. 
The repeaters expected to be deployed in these networks are referred to as first-generation quantum repeaters (see Section \ref{ssec:qbit_to_qnet}), with future versions including quantum error correction (QEC) capabilities. 
This aligns with the recent research efforts, where most entanglement routing approaches proposed assume first-generation quantum repeaters or first-generation quantum routers\footnote{A quantum router is a quantum repeater that includes routing decisions capabilities with more than two quantum network interfaces.}.

With quantum technologies rapidly evolving, the absence of a reference architecture for intermediate-scale quantum networks means that practical deployments can vary significantly, depending on the network's purpose and the target applications. Given this variability, a one-size-fits-all quantum network architecture comparable to classical networks has yet to emerge.
In light of this context, we adopt a modular approach to categorize and discuss the diverse entanglement routing strategies documented in the literature, to accommodate the varying architectural needs and future developments in quantum networks.
Therefore, many concepts covered in this survey such as route computation and fidelity support can also be independently applied to repeater-less quantum local-area networks (QLAN) \cite{bathaee2023multiplexing, vanmilligen2024entanglement, sichen2022firstRequest}. In such QLANs, optical switches and wavelength division multiplexing (WDM) may be deployed to support re-configurable topologies and optical paths, but, for example, quantum routing will not need to take into account entanglement swapping operations when servicing E2E entanglement requests.
Similarly, as long-distance connections between QLANs are achieved through a linear chain of repeaters that perform swapping operations without involving path computation, aspects of these architectures (e.g., swapping, fidelity support) are also covered by this survey.

\subsection{Scope}
This survey reviews literature presenting or studying routing schemes to compute paths and establish E2E entanglements. Before 2017, only a few studies had been published on routing entanglements and considered adaptations of the Dijkstra algorithm for entanglement path selection \cite{van2013path, di2012optimal}. Several innovative approaches have since been proposed, moving beyond Dijkstra's algorithm.

While quantum channels can be implemented over free-space optical links using satellites and optical ground stations, most of the works reviewed consider quantum channels implemented on optical fiber. Hence, this survey focuses on entanglement distribution over optical fiber networks. 

Moreover, we consider only routing for bipartite entanglement. 
This choice is motivated by several reasons. First, two of the three protocol stack models proposed for quantum networks are designed for bipartite entanglements \cite{van2008system, kozlowski2019towards}. Second, the problem is well-defined mathematically and is the most studied to date. Third, because of its technological feasibility, bipartite entanglement networks are more likely to be implemented sooner than multipartite entanglement networks.

\subsection{Key Contributions}
We present several key contributions that collectively enhance the understanding and future development of routing in quantum networks. These include:
\begin{enumerate}
\item A discussion of quantum communications concepts through a lens of realistic quantum network design, with an alignment of these concepts to relevant classical terminology, providing a concrete reference for those familiar with traditional networking concepts.
\item The definition of a taxonomy for entanglement routing concepts, based on a distinction between routing and forwarding phases and their respective functions to provide a modular approach to quantum routing and network design.
\item A formal definition of the entanglement routing problem that comprehensively covers the major aspects necessary for effective routing.
\item A detailed discussion that encompasses entanglement routing schemes, swapping strategies, and path computation algorithms and metrics, to offer a holistic perspective on the current state of entanglement routing.
\item An exploration of practical aspects in protocol design and network operation, with an identification of the main challenges and open questions for future research and development.
\end{enumerate}

\subsection{Related Work}\label{sec:rw}
Recent literature has contributed significantly to understanding and addressing the unique challenges of entanglement routing in quantum networks. 
The research in \cite{dupuy2023survey} considers the transition from point-to-point quantum communications to wide-area quantum networks, or the quantum internet. The authors identify the main challenges in this transition including the handling of longer distances, entanglement routing, and multi-commodity support. Their proposed framework categorizes the tasks of a quantum network into four distinct phases, each defined by its timescale (ranging from months to nanoseconds). These phases encompass network design, management, path selection, and swapping. 
The study predominantly focuses on two types of routing: on-demand entanglement generation and proactive advance entanglement generation, providing a simple categorization of routing strategies. Our review extends beyond this, by providing a more complete and detailed organization of entanglement routing functions. Moreover, various discussions are provided to bridge quantum networking concepts with general networking terminology, offering a concrete and realistic perspective.
In \cite{kar2023routing}, the authors explore the design challenges and opportunities in routing for quantum networks. They classify existing routing techniques into two main categories: simple routing and routing with link purification. Simple routing encompasses strategies like redundant routing (alternative paths with redundant links), concurrent routing (multiple paths provisioned simultaneously), multi-user routing (non-overlapping paths using intermediate nodes), and opportunistic routing. On the other hand, routing with fidelity involves limiting hops or utilizing purification techniques \cite{yan2023purification} to maintain entanglement quality. Our survey offers a more extensive review of entanglement routing works, presenting a broader classification and a more comprehensive examination of the various challenges and solutions involved in quantum routing.

\subsection{Structure}
The structure of this paper is designed as follows. 
In the remainder of this section, we briefly introduce the key concepts in quantum communication along with the main characteristics of a first-generation quantum repeater network. For readers unfamiliar with these fundamental operations of quantum networking, Appendix \ref{appendix:q_comm} offers a more detailed description.

Section~\ref{sec:routing_problem} presents a detailed mathematical formulation of a quantum routing model that highlights key challenges and differences with classical routing. 
In Section~\ref{sec:quantum_routing}, we explore proposed approaches for addressing the entanglement routing problem, and propose a taxonomy for classifying and examining them. The section also covers path computation algorithms, along with strategies for entanglement swapping, fidelity support, and communication reliability. We end Section~\ref{sec:quantum_routing} with a discussion of the various approaches introduced, focusing on their interplay and their practical limitations.

In Section \ref{sec:protocols_entanglement_routing}, we discuss practical approaches for implementing entanglement routing protocols by drawing inspiration and analogies from classical networking architectures and protocols.
Section \ref{sec:classical_control} extends the discussion with an exploration of key aspects in operating a deployed quantum entanglement network. We include considerations for the classical control plane, the interactions between quantum and classical networking, and the key aspects of operating and managing the network.

Section \ref{sec:open_questions} highlights the main challenges to address in the current landscape of entanglement routing and introduces key open questions.
Finally, Section \ref{sec:conclusion} synthesizes the findings, discusses the implications of our review, and outlines potential directions for future research.

\subsection{Quantum Preliminaries}

\begin{figure}[!t]
\centering
\includegraphics[width=.8\linewidth]{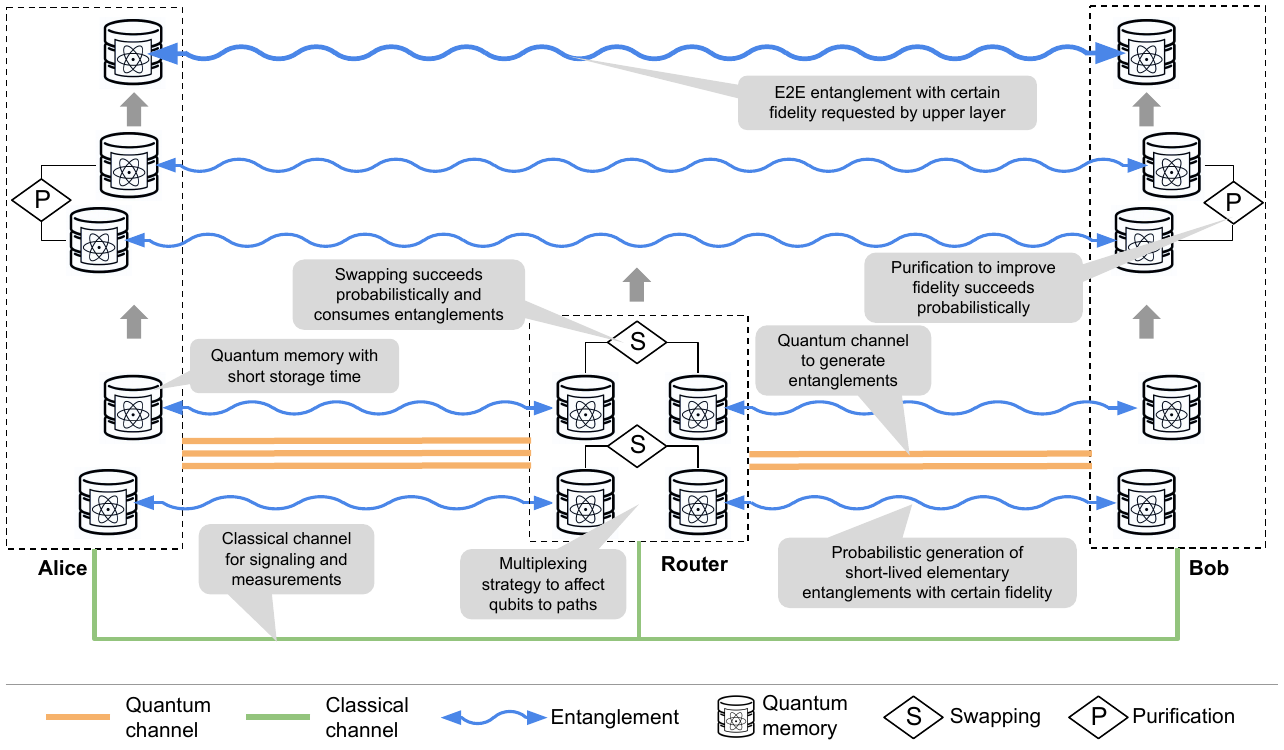}
\caption{\label{fig:network} Key quantum communication protocols for servicing end-to-end entanglement requests. Alice and Bob use quantum and classical channels to produce multiple entangled pairs with an intermediate router node. The router conducts swapping to produce the end-to-end entangled pairs, which may be further purified by Alice and Bob to meet the fidelity requirements of an application. The probabilistic nature of these protocols must be taken into account in entanglement routing strategies.}
\end{figure}

Figure~\ref{fig:network} summarizes the key components and characteristics of quantum communication through an illustration of an E2E entanglement distribution between two quantum nodes connected via one quantum repeater (router).
A brief glossary of these quantum communication concepts is presented below, while Appendix~\ref{appendix:q_comm} provides a detailed background of the primary hardware components and protocols that comprise an operational quantum network.

\paragraph{Qubits and Quantum States} 
Quantum information is represented by quantum bits (qubits) and usually encoded in the quantum state of particles such as photons, electrons, and atoms.

\paragraph{Elementary Entanglement} 
Bipartite entanglement is a special connection between two quantum states, where their properties are linked together regardless of the distance between them. These pairs of entangled qubits are known as \textit{Bell pairs} or Einstein-Podolsky-Rosen (EPR) pairs (terms that we will use interchangeably throughout this paper). Elementary entanglement is defined as bipartite entanglement shared between two neighboring nodes (i.e., directly connected through an optical channel).

\paragraph{Entanglement Generation and Heralding} 
As the generation of entangled pairs is nondeterministic, near-term quantum communications will rely on a process called heralding in which two nodes mutually acknowledge the confirmed presence or absence of an entangled pair between them when attempting an entanglement creation. As such, a Heralded Entanglement Generation (HEG) protocol requires classical signaling messages between the two nodes, which incur additional communication overhead and thus have critical implications for quantum routing.

\paragraph{Quantum Memory}
Qubits may be stored in quantum memory using a variety of technologies. Memories are  characterized by key parameters such as their storage time representing the time interval beyond which the stored quantum state is irreversibly degraded and can no longer be used. This results from entanglement decoherence where the entangled pair of particles degrades over time due to interactions with their surrounding environment.
Other memory parameters may include storage efficiency, retrieval efficiency, and acceptable wavelength range.

\paragraph{Entanglement Swapping} 
Entanglement swapping is to perform a Bell state measurement on two independent pairs of entangled qubits. Given the inherently lossy nature of quantum channels, entanglement swapping via repeaters becomes key to distributing long-distance entanglements. Specifically, a repeater is placed between Alice and Bob to split their distance into two smaller distances. Two elementary entanglements are generated; one between the Alice and the repeater and one between the repeater and the Bob. 
The repeater then interferes and measures its two local (and ideally indistinguishable) qubits, thereby destroying the two elementary entanglements in the process, and sends the measurement outcome to Bob. If successful, this procedure establishes entanglement between the two remote qubits on Alice and Bob.
The success of the swapping operation is also probabilistic. 

\paragraph{Fidelity and Purification} 
The probability that a pair of entangled qubits is in a specific, desired state is quantified by a value known as the fidelity of an entanglement. 
Decoherence, fiber loss, or environmental disturbances can prevent a pair of qubits from achieving or maintaining a maximally-entangled state. Purification techniques such as Heralded Entanglement Purification (HEP) have been developed to improve fidelity \cite{yan2023purification} by converting two or more low-fidelity Bell pairs into a single pair with higher fidelity. This process can sometimes fail and requires classical communications to notify both end nodes about the outcome of the purification effort \cite{singh2021quantum}.

\paragraph{Quantum Repeaters} 
The main function of a repeater is to capture (flying) qubits, store them in quantum memory, and implement the purification and swapping processes described above. Entanglement distribution over quantum repeaters is subject to two main types of errors \cite{muralidharan2016optimal}: loss errors arising from the attenuation of fibers, and operational errors stemming from inaccuracies in manipulating and measuring quantum states. The evolution of quantum repeaters can be delineated into three distinct generations based on their error mitigation strategies \cite{muralidharan2016optimal}; see Appendix~\ref{appendix:q_comm} for further details. 

\paragraph{Teleportation} 
Quantum teleportation is a process where a qubit is recreated at a distant destination node by utilizing an entangled link, while simultaneously destroying the original qubit at the source node. As long as the source and destination nodes share a Bell pair, teleportation can occur over any distance without physically transferring the quantum particle that encodes the qubit.
As mentioned, quantum teleportation serves as an exemplary application of utilizing E2E entanglements, highlighting just one of the potential ways to leverage these entangled states in quantum communication and computation systems.

\section{A Quantum Routing Problem}\label{sec:routing_problem}
In this section, we describe a general entanglement routing model in a centralized, offline and synchronous setting adapted from previous work in \cite{pant2019routing, shi2020concurrent, chakraborty2020entanglement, chang2022order}. Time is loosely synchronized and slotted, and within each time slot, there are two main phases, namely the \textit{external phase} for generating elementary entanglements and the \textit{internal phase} for establishing long-distance entanglements via swapping. Here, we assume that the duration of a time slot is chosen appropriately depending on hardware components so that established entanglements do not decohere before being used within one time slot.\footnote{The duration of a time slot is actually an important factor that affects the quantity/quality of entanglements since it is related to the entanglement generation rates as well as coherence times of quantum memories. It is also crucial for practical implementations of routing algorithms because near-term quantum communication relies on heralded entanglement generation that requires classical signaling between quantum nodes. For simplicity, we assume in the this section that the duration of a time slot is sufficient for carrying out necessary operations of a routing algorithm before entanglements decohere, including processing time of routing protocols, entanglement generation/swapping and signaling, and their consumption by applications.} 
To convey the main idea and key elements of entanglement routing, we will focus on entanglement routing and swapping for simplicity and do not consider purification in the formulation described below. 

\paragraph{Topology}
Consider a quantum network described by an undirected graph $\mathcal{G}=(\mathcal{V},\mathcal{E})$, where $\mathcal V$ is the set of nodes and $\mathcal E$ is the set of edges. Here, each node $u \in \mathcal V$ is a quantum node, equipped with a limited number of qubits to create Bell pairs. All nodes are connected via a classical network. An edge $(u,v) \in \mathcal E$ existing between two nodes $u$ and $v$ means that they share one or more quantum channels allowing for qubit transmission. We refer to the graph $\mathcal G$ formed by the nodes and physical channels as the physical topology of the quantum network. 

\paragraph{Quantum Link}
Since quantum channels are inherently lossy, each attempt to create an entanglement via a channel only succeeds with a certain probability. For an edge $(u,v)$, this probability is proportional to $e^{-\alpha L_{uv}}$, where $L_{uv}$ is the physical length of the channel and $\alpha$ is a constant depending on the physical media (e.g., optical fiber, free space). If an attempt succeeds, the two nodes $u$ and $v$ share an entangled pair, i.e., there is a \textit{quantum link} between $u$ and $v$. Let us denote by $p_{uv}$ as the overall success probability of a quantum link, taking into account efficiencies of entanglement sources and photon detectors, and the number of attempts allowed in one phase within a time slot. Here, for simplicity, we can assume that the physical media is the same for all channels and the success probability $p_{uv}$ is the same for all the channels connecting $u$ and $v$. 

Each edge $(u,v) \in \mathcal E$ is then characterized by a capacity $C_{uv}$ representing the maximum number of quantum links that can be established before swapping within a time slot. For simplicity, one can consider $C_{uv}$ as the number of parallel channels between $u$ and $v$. In general, however, the capacity $C_{uv}$ can be different from the number of channels, taking into account different multiplexing modes (including time, space, and wavelength multiplexing) and possibly the limited numbers of qubits at $u$ and $v$.\footnote{The special case with $C_{uv}=1$ has been studied extensively in the literature.} We refer to the (multi-)graph of nodes and edges associated with quantum links as the \textit{logical topology} or \textit{virtual topology}, which could be time-varying. It is straightforward to consider the number of successful quantum links between $u$ and $v$ in each time slot as a random variable following a Binomial distribution with parameters $C_{uv}$ and $p_{uv}$. In this model, the probability of having exactly $k$ entanglements on edge $(u,v)$ in each time slot is given by
    \begin{equation}
        p_k(u,v) = {C_{uv} \choose k} p_{uv}^k (1-p_{uv})^{C_{uv}-k}, \quad k=0,1,\ldots, C_{uv}. \label{eq_link_ent_prob}
    \end{equation}

\paragraph{Swapping in a Path}
A (quantum) path between two nodes is simply a concatenation of contiguous edges with positive capacities, where an E2E entanglement can be established by creating quantum links on all the edges and performing quantum swapping at the intermediate nodes. In particular, to combine two adjacent entanglements, say $(u,v)$ and $(v,w)$, node $v$ attempts a swapping on its corresponding local qubits which may succeed with a probability $q_v$ resulting in an $(u,w)$ entanglement.\footnote{Using linear optics platforms, the swap success probability is bounded by $q_v \le 0.5$. However, other platforms have achieved higher success probabilities\cite{bayerbach2023bell}.} 
In a more general case, suppose that node $v$ has $l$ $(u,v)$-entanglements and $r$ $(v,w)$-entanglements. Then, assuming that swapping operations can be performed between any pair of qubits in memory of a node,\footnote{i.e., a BSM can be performed between any pair of locally held qubits so that a quantum node can also act as a quantum switch.} node $v$ can attempt at most $\min\{l,r\}$ swaps to establish multiple entanglements between $u$ and $w$. 
As a result, by swapping at all the intermediate nodes, elementary entanglements are consumed to generate end-to-end entanglements. 

In the following, we will refer to a set of rules for performing swapping in a path as a \textit{swapping policy}. There are different swapping policies resulting in different swapping schemes as we will detail in subsection~\ref{sssec_swapping} below. In this section, we present two groups of policies to accommodate the routing model described here, namely heralded and unheralded swapping. 
\begin{itemize}
    \item \textbf{Unheralded swapping}: In this scenario, routers locally perform swapping without awaiting the swapping outcomes of other routers. This allows all the nodes to carry out swapping operations independently and in parallel once all adjacent quantum links are available. 
    Consider a general path of length $n$ given without loss of generality by $\{(0,1),(1,2),(2,3),\ldots(n-1,n)\}$ with corresponding capacities $C_{i-1,i}$ for $i=1,\ldots,n$. For any $i<j \le n$, we can define the capacity of the subpath from $i$ to $j$ as
    \begin{align}
        C_{ij} = \min_{i<l\le j}C_{l-1,l}\label{eq_subpath_capacity}
    \end{align}
    Let $\tilde p_k(0,i)$ denote the probability that among the first $i$ hops, at least one hop has exactly $k$ quantum links while other hops have at least $k$ quantum links, where $1\le k \le C_{0i}$. This probability can be computed recursively as follows: 
    \begin{align}
        \tilde p_k(0,i) &= \tilde p_k(0,i-1)\sum_{l=k}^{C_{i-1,i}}p_l(i-1,i) + p_k(i-1,i)\sum_{l = k+1}^{C_{0,i-1}} \tilde p_l(0,i-1), \quad i=2,\ldots,n \label{eq_long_ent_prob_unheralded}
    \end{align}
    with $\tilde p_k(0,1) = p_k(0,1)$ and $p_l(i-1,i)$ is computed as in \eqref{eq_link_ent_prob}.\footnote{Note that \cite{shi2020concurrent} used a similar expression to \eqref{eq_long_ent_prob_unheralded} but with $C_{0n}$ as the upper limit for both summations on the right-hand side, thereby leading to a lower success probability.} 
    Since edges can have different numbers of successful quantum links in each time slot, all the nodes will need to agree on how to perform internal swapping effectively instead of randomly pairing internal qubits. An easy way to do this is to assign IDs to quantum links and pair them in a common order (e.g., ascending, descending). Under such qubit binding and swapping, the probability of having exactly $k$ end-to-end entanglements, denoted by $p_k(0,n)$, is computed as
    \begin{align}
        p_k(0,n) &= \sum_{l=k}^{C_{0n}}\tilde p_l(0,n) {l \choose k} \bar{q}^k (1-\bar{q})^{l-k}, \quad k=1,\ldots, C_{0n}, \quad \bar{q} := \prod_{i=1}^{n}q_i
    \end{align}
    with $p_0(0,n) = 1-\sum_{k=1}^{C_{0n}} p_k(0,n)$ and $p_k(0,n) = 0$ for all $k> C_{0n}$. Here, note that, because swapping operations are independent, the probability of successfully connecting all elementary entanglements is simply the product of individual swapping probabilities, which does not depend on any swapping order.

    \item  \textbf{Heralded swapping}: In this scenario, routers carry out swapping based on the swapping outcomes of other routers, giving rise to a swapping order for the path in each time slot. Compared to unheralded swapping, heralded swapping aims to make better swapping choices at each node at the expense of increasing the amount of heralding signals. Again, let us consider a path of length $n$ given by $\{(0,1),(1,2),(2,3),\ldots(n-1,n)\}$ where the capacity $C_{ij}$ of the subpath from $i$ to $j$ is defined as in \eqref{eq_subpath_capacity}. With abuse of notation, let $p_i(x,y)$ and $p_j(y,z)$ denote the probabilities of having exactly $i$ $(x,y)$-entanglements and $j$ $(y,z)$-entanglements at node $y$, where $i\le C_{xy}$, $j\le C_{yz}$, and $x$, $y$ and $z$ need not be adjacent nodes. 
    Thus, the probability of having exactly $k$ $(x,z)$-entanglements after at most $C_{xz}$ swapping attempts at node $y$ can be computed as follows \cite{chang2022order}:
    \begin{equation}  
    p_k(x,z) = \sum_{i=k}^{C_{xy}}\sum_{j=k}^{C_{yz}}  p_i(x,y) p_j(y,z) {\min\{i,j\} \choose k} q_y^k (1-q_y)^{\min\{i,j\}-k}, \quad k=1,\ldots {C_{xz}}, \label{eq_long_ent_prob}
    \end{equation}
    with $ p_0(x,z) = 1-\sum_{k=1}^{C_{xz}} p_k(x,z)$. This allows us to find probabilities of long-distance entanglements $p_k(0,n)$ for any swapping order. The probability distribution of end-to-end entanglements in this case is more complicated to compute than the unheralded case. 
\end{itemize}

\paragraph{Path Throughput} 
As shown above, given a path $\pi$ in the graph $\mathcal{G}$ with a certain edge capacity $\mathcal{C}_{\pi}:=\{C^{\pi}_{uv}: (u,v) \in \pi\}$, end-to-end entanglements can be established by swapping at all intermediate nodes, which results in different success probabilities depending on how swapping operations are carried out. Let $\mathcal{Q}$ be a swapping policy and $p_k^{\mathcal{Q}}(\pi)$ denote the probability of having exactly $k$ end-to-end entanglements on path $\pi$ under policy $\mathcal{Q}$. Then we can define the expected throughput as follows
    \begin{equation} 
    \mathrm{EXT}(\pi; \mathcal{Q}) := \sum_{k=1}^{C_{\pi}} k\times p_k^{\mathcal{Q}}(\pi), \quad \textrm{with } C_{\pi}=\min \mathcal{C}_{\pi}, \label{eq_path_thruput}
    \end{equation}
    where $C_{\pi}$ is also known as the (minimum) width of the path. Note that the throughput can be computed in $O(|\pi| \max\mathcal{C}_{\pi})$ time for unheralded swapping and in $O\big(|\pi| (\max \mathcal{C}_{\pi})^2\big)$ time for the case of heralded swapping. In both cases, the time complexity scales linearly with the path length, but unlike the former, the latter is quadratic in the maximum width of the path.

\paragraph{Demands/Requests}
A demand can be defined as a tuple $r = (s,d, \delta, \underline{F})$, representing a request to deliver $\delta$ E2E entanglements with a minimum fidelity $\underline{F}$ per time unit between two end-nodes $s$ and $d$ \cite{chakraborty2020entanglement}. A request may also include other requirements, such as a desired latency $\bar{l}$, but we do not consider them here for simplicity. 

Let $\mathcal{P}_{r}$ denote the set of feasible paths serving request $r$ and $\mathcal{Q}_{\mathcal{P}_{r}}$ the set of corresponding swapping policies, i.e.,  $\mathcal{Q}_{\mathcal{P}_{r}} = \{\mathcal{Q}_{\pi}: \pi \in \mathcal{P}_{r}\}$, where a swapping policy $\mathcal{Q}_{\pi}$ can depend on the path itself and thus can also be a design parameter. The expected end-to-end entanglements for this request in a time slot can then be defined as 
\begin{equation}
    R(\mathcal{P}_{r}, \mathcal{Q}_{\mathcal{P}_{r}}) := \sum_{\pi \in \mathcal{P}_{r}} \mathrm{EXT}(\pi; \mathcal{Q}_{\pi}) \label{eq_request_thruput}
\end{equation}
The expected number of entangled qubit pairs delivered per time unit for a request that satisfies its fidelity (and possibly latency) constraints is referred to as the end-to-end (expected) throughput of the request, which is denoted by $\tilde R(\mathcal{P}_{r}, \mathcal{Q}_{\mathcal{P}_{r}})$ and upper bounded by $R(\mathcal{P}_{r}, \mathcal{Q}_{\mathcal{P}_{r}})$.

\paragraph{Objective}
Consider a quantum network $\mathcal G=(\mathcal V, \mathcal E)$ that is required to serve a set of requests $\mathcal{R}=\{r_i\}_{i=1}^m$.\footnote{For simplicity, we assume here that $\mathcal{G}$ and $\mathcal{R}$ are fixed within a period of interest. In general, one can consider time-varying graph $\mathcal{G}$ due to failures as well as time-varying requests with possibly estimated or unknown arrival patterns.} A common objective of routing algorithms is to maximize the total number of end-to-end entanglements delivered for all requests. The goal of a routing algorithm is then to efficiently determine a set of paths $\{\mathcal{P}_r:\, r\in \mathcal{R}\}$ together with allocated capacity $\{\mathcal{C}_{\pi}:\, \pi \in  \mathcal{P}_r\}$ and corresponding swapping policies $\{\mathcal{Q}_{\pi}:\, \pi \in  \mathcal{P}_r\}$ to either satisfy all requests or to maximize the overall throughput in the network \cite{van2014quantum, caleffi2017optimal, chakraborty2020entanglement,shi2020concurrent}. 
To this end, we can then consider a utility function $U_r\big(\delta_r, R(\mathcal{P}_{r}, \mathcal{Q}_{\mathcal{P}_{r}})\big)$ for each request $r\in \mathcal{R}$ with a desired flow rate of $\delta_r$. 
The problem can be formulated as maximizing the following aggregated utility function\footnote{To provide certain levels of fairness, the overall objective function can be defined as a weighted sum of all utilities or throughputs.}
    \begin{equation} 
     \sum_{r\in \mathcal{R}} U_r\big(\delta_r, R(\mathcal{P}_{r}, \mathcal{Q}_{\mathcal{P}_{r}})\big) \label{eq_routing_objective}
    \end{equation}
    with decision variables: $\{\mathcal{P}_{r}, \mathcal{C}_{\pi}, \mathcal{Q}_{\pi}, \forall \pi \in \mathcal{P}_{r}, \forall r\in \mathcal{R} \}$ and 
    subject to the following:  
    \begin{itemize}
        \item Feasible paths: each  $\pi \in \mathcal{P}_r$ has no loops.
        \item Capacity: $C^{\pi}_{uv} \ge 0, \forall (u,v) \in \mathcal{E}$ and 
                \begin{equation}
                     \sum_{\pi \in \cup_{r\in \mathcal{R}} \mathcal{P}_r} C^{\pi}_{uv} \le C_{uv}, \quad \forall (u,v) \in \mathcal{E} \label{eq_constraint_capacity}
                 \end{equation}

        \item Fidelity: Since the fidelity of an entanglement drops with each swapping, a minimum fidelity requirement $\underline{F}_r$ can be replaced with a path length constraint for each path $\pi \in \mathcal{P}_{r}$ without considering entanglement distillations \cite{chakraborty2020entanglement}. In particular, assuming all the mixed entangled states are Werner states, the following hop constraint can be used instead 
    \begin{equation}
            |\pi| \le h_r := \frac{\log(\frac{4\underline{F}_r-1}{3})}{\log(\frac{4F_0-1}{3})}, \quad \forall \pi \in \mathcal{P}_{r}, \forall r\in \mathcal{R} \label{eq_constraint_fidelity}
        \end{equation}
where $F_0$ is the minimum fidelity of each elementary Bell pair. 
    \end{itemize}

\begin{figure}[!t]
\centering
\includegraphics[width=0.7\linewidth]{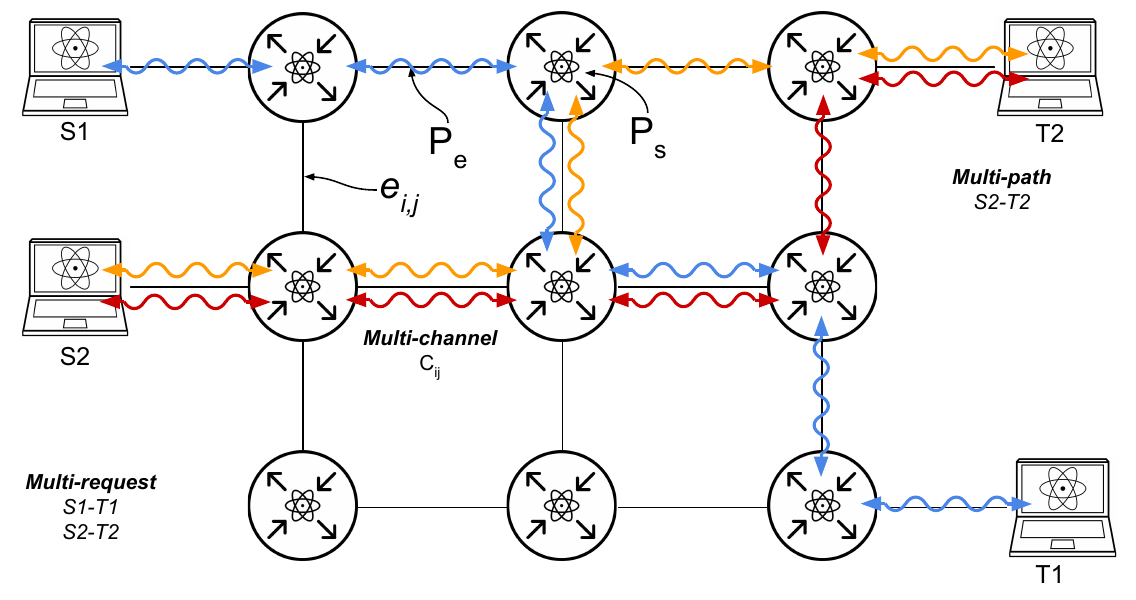}
\caption{\label{fig:entanglements}Entanglement routing on a grid network topology (adapted from \cite{li2021effective}). Edge $(i,j) \in \mathcal{E}$ supports multiple entangled pairs, whose capacity is denoted as $C_{ij}$. Entanglement generation between \textsc{S1} (\textsc{S2}) and \textsc{T1} (\textsc{T2}) is requested and one or more paths are provisioned for each request. The probabilities of successfully building an entangled pair between adjacent nodes and performing swapping are denoted as $p_{ab}$ for $(a,b) \in \mathcal{E}$ and $q_b$ for $b \in \mathcal{V}$, respectively.}
\end{figure}

Figure \ref{fig:entanglements} illustrates the routing concepts in a quantum network, where multiple quantum links can be created over an edge (multi-channel) and the routing algorithm can accept one or multiple requests at a time (multi-request). Here, entanglement generation between \textsc{S1} (resp. \textsc{S2}) and \textsc{T1} (resp. \textsc{T2}) is requested and one or multiple paths are provisioned for each request (multi-path). 
We emphasize that while not exhaustive, the problem formulation presented here captures key challenges in quantum routing. Previous research has often focused on simplified or approximate versions of this problem as we will explain below and in the next sections. 

\paragraph{Routing Decisions and Complexity} Given inputs to the routing problem, including network graph $G=(\mathcal{V}, \mathcal{E})$, edge capacity $C_{uv}$ with corresponding quantum link success probability $p_{uv}$ for all $(u,v) \in \mathcal{E}$, swapping probability $q_v$ for all $v\in \mathcal{V}$ and the set of requests $\mathcal{R}=\{r_i = (s_i,d_i, \delta_i, \underline{F}_i)\}_{i=1}^m$, the outputs of a routing algorithm after solving the above problem are the routing decisions, namely, 
\begin{itemize}
    \item A set of paths $\mathcal{P}_r$ for serving each request $r \in \mathcal{R}$, together with allocated capacity $C^{\pi}_{uv}$ for each edge $(u,v)$ along any path $\pi \in \mathcal{R}_r$. These decision variables will be used for resource allocation and signaling protocols in each node. We will discuss this step further in subsection~\ref{ssec:path_computation} below. Finding these decision variables corresponds to similar tasks in classical networking, namely path computation and path installation. 

    \item A swapping policy $\mathcal{Q}_{\pi}$ for any path $\pi \in \mathcal{P}_r$. This will determine the local signaling needed for swapping at each node in support of multiple paths and multiple requests, which somewhat resembles the forwarding table in classical networking. Given that performing an entanglement swap on two imperfect Bell states results in a long-distance entanglement with reduced fidelity, if such reduction is significant, one might need to take into account entanglement purification or distillation steps to probabilistically convert multiple low-quality entangled pairs into a single high-quality entangled pair. In this case, one needs to design not just a swapping policy but a \textit{forwarding policy} that includes swapping and purification steps, where purification can be done on elementary or distant entanglements to improve fidelity. 
    Further discussions on swapping and purification will be given in subsection~\ref{ssec:forwarding} below. 
\end{itemize}

Here, let us briefly remark on the challenges in solving the routing problem described above that do not have a counterpart in classical networking. In general, it is difficult to solve the above optimization problem exactly and efficiently even in the offline and centralized setting because of the following challenges: 
\begin{itemize}
    \item The combinatoric nature of the solution space for the decision variables, namely paths in graph $\mathcal{G}$ with integer capacity $C_{uv}^{\pi}$ and swapping policy/order, where the latter has no counterpart in classical routing and forwarding. Specifically, in the case of heralded swapping, the number of possible orders for a path of length $n$ scales as $O(4^n)$.\footnote{This is also related to the $n$-th Catalan number \cite{chang2022order} in combinatorial mathematics.} In addition, we note that different orders will also have different levels of parallelism and signaling mechanisms, which in turn can also affect the practical throughput. For simplicity, we do not consider such effects in the formulation above (which is reasonable when the duration of the internal phase is negligible compared to that of the external phase for generating all elementary entanglements). 

    \item Lack of efficient path metrics: As shown in \eqref{eq_link_ent_prob}--\eqref{eq_path_thruput}, the expected throughput of a path is rather complicated, involving the characteristics of each node and every edge on the path as well as the swapping policy employed for the path. As a result, comparing two paths becomes nontrivial, especially when they have different edge capacities, different lengths, and different swapping policies, unless computation for the whole path is finished. However, the computational complexity of the throughput in heralded swapping scales quadratically in the maximum width of the path. More importantly, under any swapping policy, finding a path with maximum expected throughput in a network does not have the subpath optimality property, causing methods like Dijkstra and Bellman-Ford to fail in finding optimal paths \cite{caleffi2017optimal, chang2022order}.
\end{itemize}
Thus, solving the routing problem formulated above remains a challenging task. Most existing works often fix a swapping policy or order (e.g., \cite{shi2020concurrent, caleffi2017optimal, ghaderibaneh2022efficient}) and consider the routing problem using Dynamic Programming and/or replacing exact path throughput with heuristics that are more efficient and/or Dijsktra friendly. For example, path computation can be done on-line or off-line using a heuristic path metric such as hop count, edge width, or link fidelity \cite{pant2019routing, shi2020concurrent, zhao2021redundant, nguyen2022multiple}.

It is important to note that the problem described above assumes a centralized, offline and synchronous setting. The following extensions can be more challenging but also more practical to consider:
\begin{itemize}
\item Distributed: Routing paths and swapping orders are computed in a (semi-)distributed fashion based on nodes' local view of the network. 

\item Online: Routing can be adaptive to request arrival patterns and available elementary link entanglements, which are possibly unknown in advance. In stead of static swapping policy/order, swapping can also be dynamic to better utilize elementary entanglements. 

\item Asynchronous: Different levels of asynchrony can also be considered. For example, nodes can have different time slot duration or operate based on discrete events. For each node, the internal and external phases for each memory can be asynchronous as well to better improve throughput and fidelity.
\end{itemize}

\paragraph{Physical parameters} 
One can illustrate the relevant time constraints involved in entanglement routing through the lens of several key physical parameters.
The coherence time is an important factor in establishing and maintaining high-fidelity entanglements. Currently, quantum coherence times can be up to milliseconds for photons in optical fibers and up to seconds or more for other systems \cite{abobeih2018one, anderson2022five}. In particular, typical coherence times of spin qubits can be milliseconds up to seconds in silicon \cite{pla2013high} and diamond \cite{awschalom2018quantum, abobeih2018one}.\footnote{Trapped ions can have a lifetime from minutes to hours \cite{wang2021single}, but the efficiency of frequency conversion to telecommunication wavelength is still rather low \cite{schneeloch2019introduction}.} The success probability of a single entanglement attempt is dependent on the length of the optical fiber (besides other factors such as efficiencies of detectors and memories, as well as the success rate of BSM) and is typically measured on the order of $10^{-4}$ in a lab environment over a few-meter links \cite{dahlberg2019link} and $10^{-6}$ over $6-33$ km telecom fiber \cite{van2022entangling} at a frequency of several tens of kHz. 
A swapping success probability $q=0.5$ is also feasible. Although BSMs based on linear optics have a maximum success probability of $0.5$, a more complex measurement pattern using ancillary photons has  been experimentally demonstrated recently to achieve a success probability of approximately $0.579$ \cite{bayerbach2023bell}. Depending on the attenuation coefficient of the optical fiber, repeater stations may be spaced at intervals of $L\approx20\,\mathrm{km}$, and the one-way classical communication time is $L/(2 \times 10^5 \; \mathrm{km/s}) = 0.1 \; \mathrm{ms}$.

\section{Entanglement Routing Approaches}\label{sec:quantum_routing}
Drawing inspiration from classical networking, we decompose the E2E entanglement distribution process into two distinct phases, namely routing and forwarding \cite{farahbakhsh2022opportunistic}. 
The decision to examine routing and forwarding separately stems from the realization that these phases can be implemented independently in quantum networks. For instance, in repeater-less QLANs, route computation approaches may be relevant without necessitating entanglement swapping operations (i.e., forwarding) for servicing entanglement requests. Conversely, long-distance connections between QLANs over linear paths with a chain of repeaters might utilize forwarding techniques such as swapping and purification without engaging in path selection.

The routing phase is concerned with determining (optimal) paths for E2E entanglement requests and ensuring the necessary routing instructions are in place across the network to support these paths. This includes all the background processes needed to select paths such as collecting topology. 
The routing phase involves two key aspects: path computation and route installation.

Path computation consists of choosing the sequence of intermediate links (and nodes) that will generate the E2E entanglements. This process should not happen for each individual E2E entanglement. Instead, it should be launched once for a single or a group of E2E entanglement requests.
Path computation is strongly tied to the topology, which can be the physical and/or logical and/or virtual topology, and the resources (memory qubits) already consumed by other paths.
Path computation uses a routing algorithm to calculate paths for the requested E2E entanglements. These algorithms may model various network parameters and optimizations as discussed in Section \ref{ssec:algorithms}.

Route installation follows the path computation and involves the deployment of specific instructions at each node along the path to establish E2E entanglements. These instructions can be static and installed manually on the nodes at configuration time, or dynamically transferred from path computation results, akin to updating forwarding tables or setting up cross-connects in classical networks.

Path computation and route installation can be implemented in a distributed fashion across the nodes, in a centralized controller, or combine the two approaches (see Section \ref{sec:protocols_entanglement_routing}).

The forwarding phase usually takes place after route installation\footnote{It is important to recognize that \textit{forwarding} in the quantum communication context is somewhat of a misnomer since it involves generating and swapping entanglements across the network, diverging from the hop-by-hop forwarding of packets seen in classical networks. However, we use this classical networking term for a more familiar and abstracted description of the quantum networking processes.}.
It includes the external phase for the production of elementary entanglements (which may not yet exist) and the internal phase where swapping are executed, according to installed instructions, to establish E2E entanglements. The forwarding phase may include purification and mechanisms to ensure path robustness.

Routing algorithms can be designed to operate on a partial or local view of the network topology (distributed), or, alternatively, they may require a global view of the topology (centralized). According to the routing problem formulation given in Section \ref{sec:routing_problem}, the routing algorithm may include forwarding operations, such as optimal swapping strategies, in addition to routing decisions (i.e., path computation). Hence, we discuss the routing algorithms separately without limiting them only to the routing or forwarding phase.

\begin{figure}[!h]
\centering
\includegraphics[width=1\linewidth]{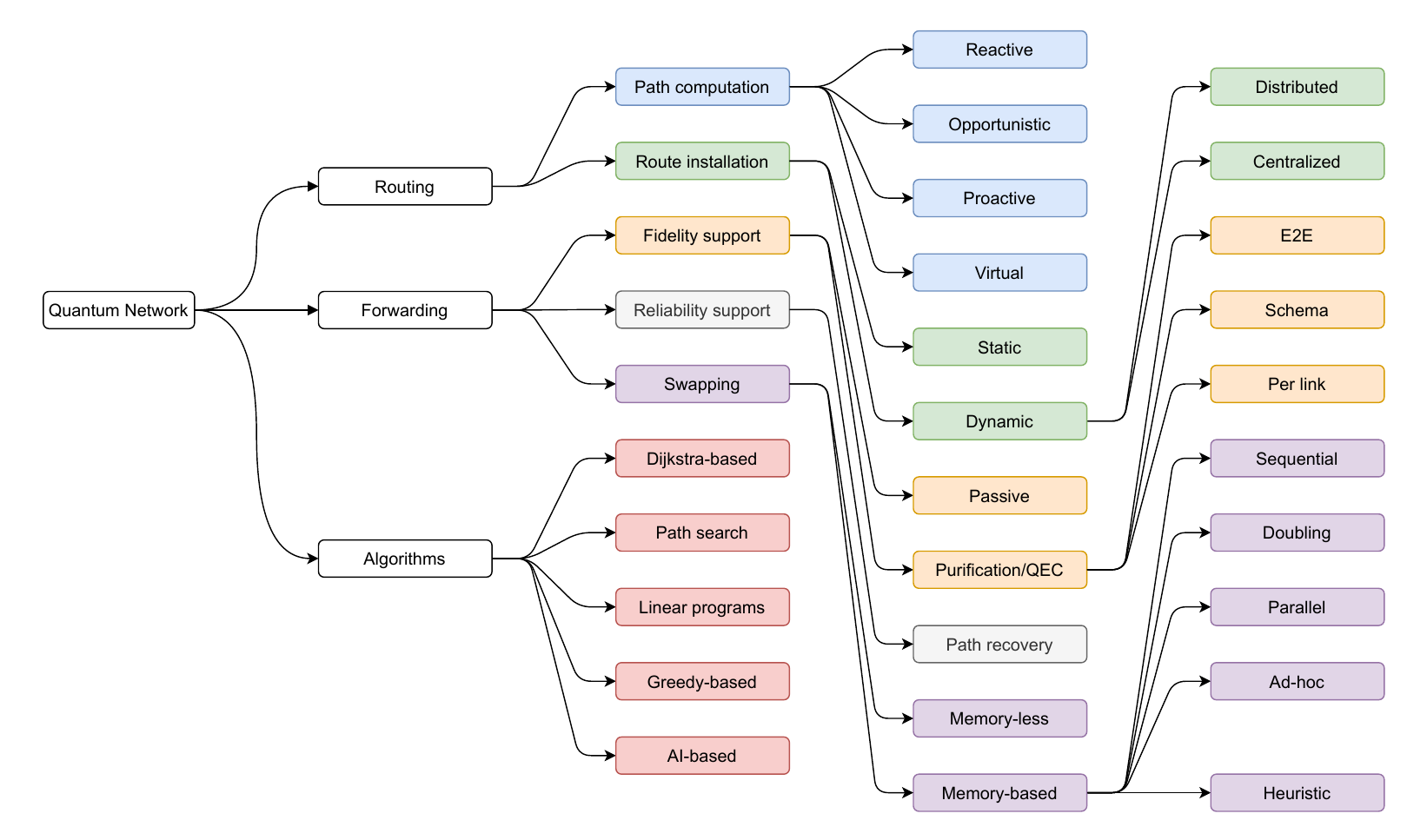}
\caption{\label{fig:classification}Taxonomy of quantum routing concepts discussed in this survey. Partial or complete entanglement distribution protocols can be built by combining approaches from identified classes.}
\end{figure}

The specific concepts and approaches that underpin the routing phase, forwarding phase, and routing algorithms, are organized in the taxonomy outlined in Figure \ref{fig:classification}.
We chose to exclude the notion of a ``routing protocol'' in the taxonomy, viewing it as a result of combining techniques from the categories identified (see Section \ref{sec:protocols_entanglement_routing}). 
To use the taxonomy diagram for routing protocol design, one may start by selecting a path computation scheme for the network's technology and objectives, such as a proactive path computation. Then, choose a route installation that matches the network's capabilities and implementation preferences, like a centralized mode. Next, decide on the swapping strategy for the forwarding and the needed QoS features. Finally, pick or develop a model and algorithm to address the routing problem formulation. 
Note that not all combinations of these elements will be consistent or practical, so careful consideration is needed to ensure compatibility and effectiveness.

\subsection{Routing Phase} \label{ssec:path_computation}
We distinguish four types of routing schemes based on the primary network topology on which paths are computed and the time at which the path computation takes place with respect to the entanglement creation: proactive, reactive, opportunistic, and virtual.

\subsubsection{Proactive}\label{sssec:proactive}
In proactive routing (sometimes designated as \textit{on-demand} entanglement generation \cite{chakraborty2019distributed, dupuy2023survey}) the path computation and route installation take place before elementary entanglements are created\footnote{In this paper, we use ``proactive'' to describe path computation occurring before entanglement generation. However, these terms may be used inversely in other studies such as in \cite{illiano2022quantum, zhang2023segmentrouting}, where what we term ``reactive'' might be referred to as ``proactive'' and vice versa. This inversion stems from the perspective of considering entanglement generation process relative to path computation. Since we are discussing routing schemes, the terminology chosen here considers the path computation process relative to entanglement generation.}.

\begin{figure}[!t]
\centering
\includegraphics[width=0.6\linewidth]{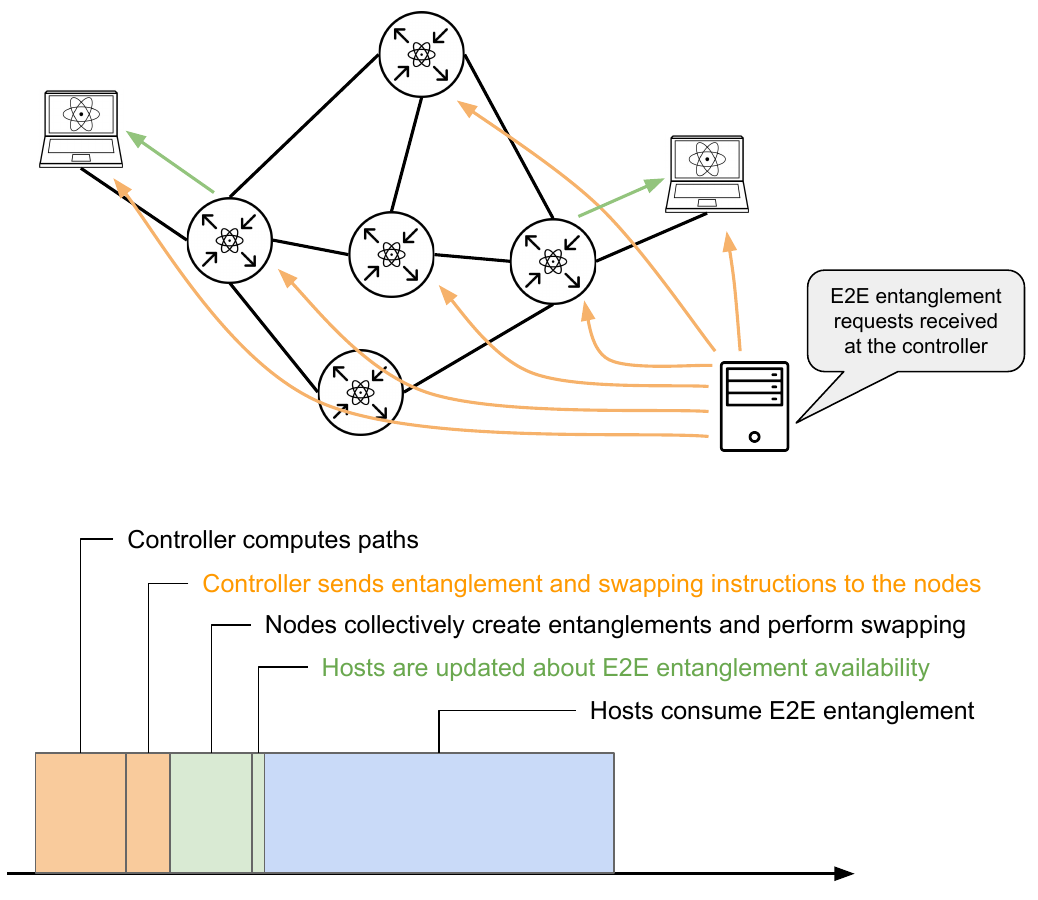}
\caption{\label{fig:routing_proactive}Architecture of a centralized proactive routing. The purpose of the slotted representation is only to illustrate the sequential operations involved in proactive routing.}
\end{figure}

A centralized proactive routing architecture is illustrated in Figure \ref{fig:routing_proactive}. Initially, E2E entanglement requests are received by the controller which computes the paths and provides the nodes (routers and end-nodes) with instructions for entanglement creation and swapping for each path. Subsequently, the nodes coordinate to create E2E entanglements, incorporating purification processes if supported. End-nodes are notified of available E2E entanglements as part of the swapping process.

\begin{figure}[!h]
\centering
\includegraphics[width=0.6\linewidth]{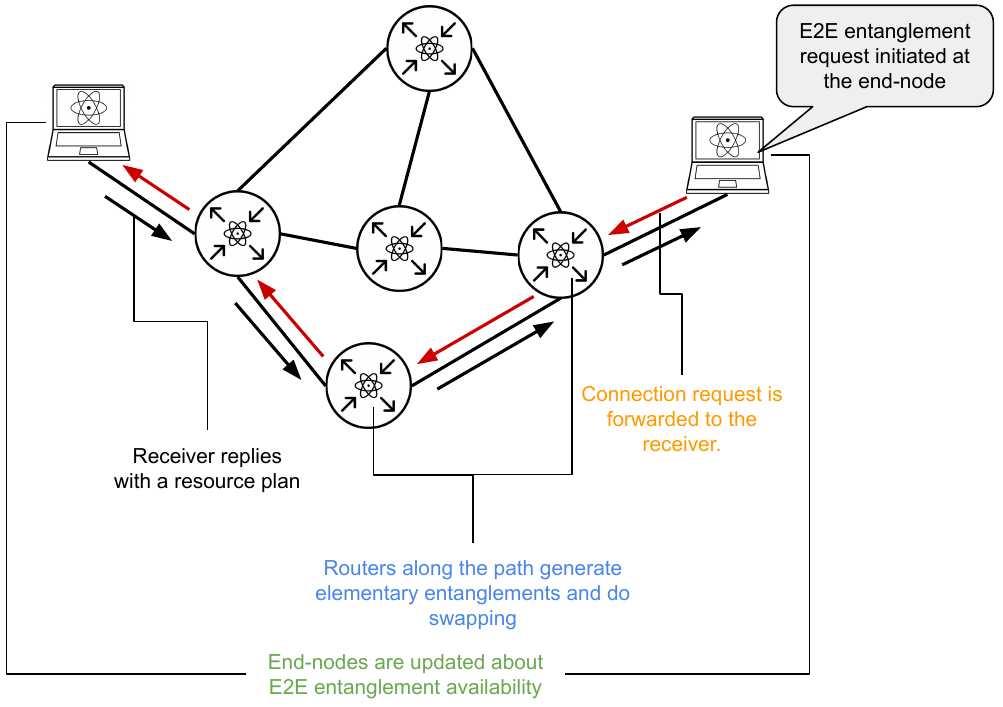}
\caption{\label{fig:routing_proactive_distr}E2E entanglement distribution in distributed proactive routing. Paths can be computed by the requester, the receiver, or hop-by-hop along the path.}
\end{figure}

A distributed proactive routing is also possible and generally uses a connection phase to install paths and negotiate resources among the nodes as illustrated in Figure \ref{fig:routing_proactive_distr}. Path computation can be done at the requester or the receiver, or hop-by-hop along the path as discussed in Section \ref{sec:protocols_entanglement_routing}.

The path computation may take into account parameters for entanglement creation, multiplexing, and swapping, and provides necessary instructions to the nodes. These instructions apply to a large stream of E2E entanglements until different instructions are installed. Once the paths are computed and the relevant instructions installed, E2E entanglements are generated for each path until a stop condition is met (e.g., the request terminates) or new paths and instructions are installed.

Since path computation and route installation are not constrained by entanglement decoherence time, proactive routing allows more flexibility in its design. It can be envisioned in a centralized system where routing instructions are computed and installed from a central entity. Alternatively, proactive routing can be realized as a distributed system, where each node computes and installs its routing instructions. In the latter case, a routing algorithm with a partial knowledge of the physical topology can be used. 
Note that a majority of the proposed proactive routing approaches assume a global knowledge of the physical topology by the path computation algorithm. 
The routing algorithm typically considers the physical topology and can be as simple as pre-computing the paths for all the possible pairs of nodes at initialization time \cite{shi2020concurrent}.

Since the path computation is executed before knowing which elementary entanglements have succeeded, a cost for physical links (paths) must be modeled in the algorithm. 
In its simplest form, link cost reflects the expected entanglement throughput (i.e., generation rate). The throughput is based on the channel loss, which is mainly dependent on the fiber length. For a more realistic throughput estimation, other characteristics may be included such as photon source power, detector efficiency, and quantum memory coherence, frequency, and efficiency \cite{caleffi2017optimal}. 
Based on entanglement generation rates and swapping probabilities, the path throughput can be modeled as in Section~\ref{sec:routing_problem} above, or approximated by using different heuristic metrics such as the sum of node distances or link entanglement generation rates \cite{shi2020concurrent}.
Overall, proactive path selection tends to prefer shorter paths in physical distance, although this does not guarantee selection of paths with the highest throughput in the presence of links with varying capacities. 

Without a-priori knowledge of the existing entanglement links, if the path computation algorithm does not plan for redundant links or paths, the E2E entanglement throughput may quickly degrade due to failures in creating the planned entanglements \cite{zhao2021redundant}.
Moreover, proactive routing may induce a higher latency to satisfy the E2E entanglement requests as multiple attempts may be required before creating entanglement links. This latency may be further increased when path computation is distributed due to additional signaling. These limitations lead to considering the reactive routing scheme discussed next.

\newpage
\subsubsection{Reactive}\label{sssec:reactive}
In reactive routing (sometimes designated as \textit{advance} \cite{dupuy2023survey} or \textit{continuous} \cite{chakraborty2019distributed} entanglement generation), entanglements are constantly produced over each quantum channel across the network and path computation is done on the instant logical topology formed by the created entanglements. 

This scheme frequently assumes a synchronous network where the quantum and routing operations evolve sequentially within discrete time slots of a fixed duration. Such a time-slotted system is used particularly with centralized path computation \cite{cicconetti2021request}, although it can also be used in distributed routing \cite{pant2019routing}.

\begin{figure}[!h]
\centering
\includegraphics[width=0.6\linewidth]{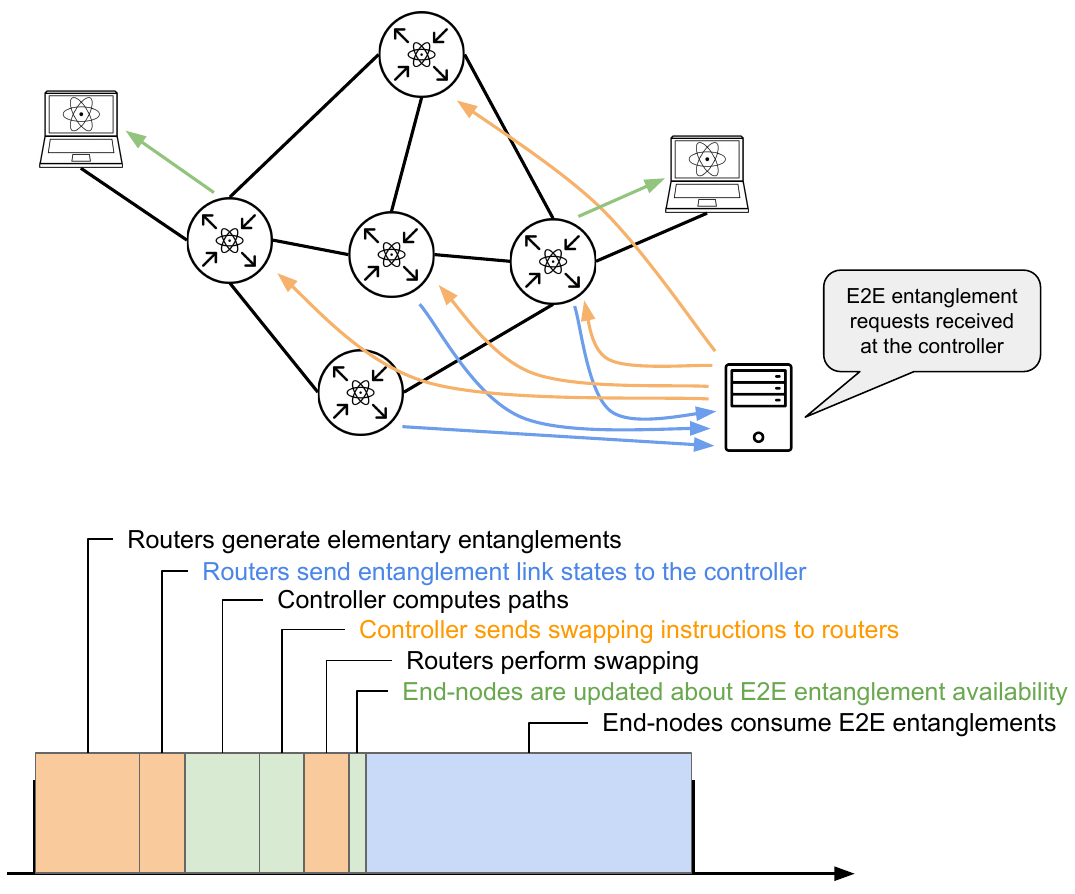}
\caption{\label{fig:routing_reactive}Reactive routing with a central controller and operations executed during a time slot (adapted from \cite{cicconetti2021request}).}
\end{figure}

A centralized slotted reactive routing is illustrated in Figure \ref{fig:routing_reactive}. 
At the beginning of a time slot, each pair of adjacent routers (i.e., sharing a quantum channel) attempts to generate entanglements. Entangled link-states are communicated to the routing element (e.g., controller) which computes paths for the current E2E entanglement requests based on the logical topology formed by the created entanglements. The routing element sends swapping instructions to the routers to establish the E2E entanglements.

As discussed in Section~\ref{sec:routing_problem} (see footnote 1), a time slot typically includes the entanglement attempt/creation, the entanglement outcomes transmission, the path computation, the route installation, and the swapping execution time. 
The remaining time is used to consume the E2E entanglement by the layer above (e.g., application). Hence, the duration of a time slot depends on the hardware characteristics and entanglement protocol. It is upper bounded by the entanglement creation delay, plus the entanglement lifetime (i.e., quantum memories coherence time).
Note that these operations are executed sequentially within the time slot without further synchronization.
Note that the entanglement coherence time must be long enough for the path computation and route installation to take place, and allow enough time for the E2E entanglement to be consumed. However, nodes have a limited time to create elementary entanglements, and must wait until the next time slot to re-attempt failed entanglements. This may lead to non-optimal utilization of network resources.

Reactive routing can also be designed in a distributed system, where routers rely only on the local knowledge of the logical topology to compute paths.
The routing proposed in \cite{pant2019routing} uses a path-finding mechanism in which all elementary entanglements are consumed in each attempt to establish an E2E entanglement.
A more elaborated distributed path selection is designed in \cite{yang2024Asynchronous} where nodes collectively maintain the logical topology as a distributed graph.
As new entanglements arrive and others expire, each node interacts with its neighbors to join the graph by selecting a root node based on link cost (e.g., fidelity). Once the node has joined a graph, it has a route toward the graph root which is used to select entanglements to swap for each E2E entanglement request.

\begin{figure}[!h]
\centering
\includegraphics[width=0.6\linewidth]{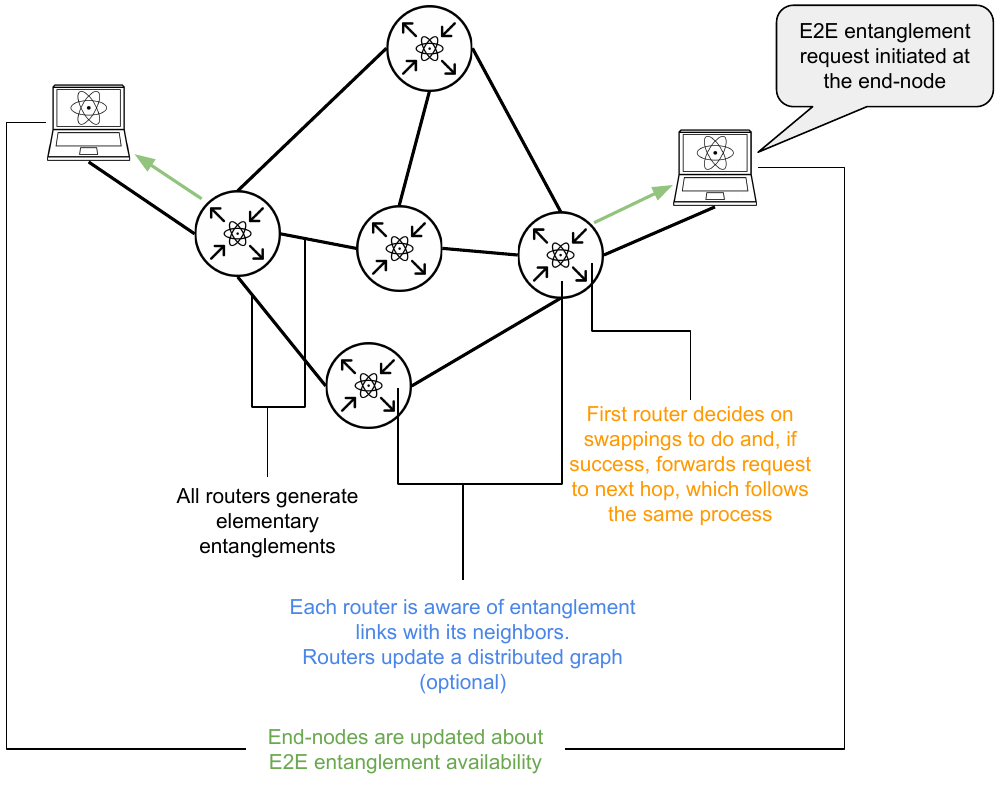}
\caption{\label{fig:routing_reactive_distr}Simplified view of a reactive distributed routing.}
\end{figure}

Figure \ref{fig:routing_reactive_distr} shows a simplified view of a distributed reactive routing scheme where two end nodes want to establish an E2E entanglement.
Since each router is only aware of the status of entanglements with its neighbors, the routers need to perform swapping when possible to increase the probability of establishing an E2E entanglement. By consuming all entanglements in the logical topology, the E2E entanglement is established.
With a more advanced algorithm, a path can then be identified within the distributed graph where each router selects the entanglements to swap.

Note that although reactive routing operates on a partial or global logical topology, it may assume that the global physical topology is also known to the routing algorithm \cite{nguyen2022multiple, li2021effective, zhao2022E2E, cicconetti2021request, zeng2022multi}.

Path computation metrics in reactive routing can be as simple as the hop count \cite{nguyen2022multiple}, although link fidelity may also be used \cite{zhao2022E2E}. To minimize the time elapsed before the remote node applies the swapping corrections and thus maximize fidelity, the routing algorithm can consider the longest physical distance between the intermediate routers and the remote node \cite{cicconetti2021request}. The requisite instructions installed at the routers are essentially swapping instructions for each path.

Selecting paths based on already-created entanglements allows the routing algorithm to support purification schemes \cite{yan2023purification}, swapping strategies, and resource management. However, the logical topology graph formed by the entangled links must be connected enough to satisfy the E2E entanglement requests. Otherwise, most requests must wait until the next period (in the best case) to be satisfied. Therefore, the entanglement technology should provide an efficient entanglement success rate to guarantee sufficiently connected logical graphs, or a high enough frequency to make multiple attempts within a short time period. 

The impact of entanglement generation rate and success rate is still to be investigated in order to determine the technology characteristics required for the reactive routing to be viable. Authors in \cite{semenenko2022entanglement} study the entanglement generation rate in a quantum network and demonstrate that the performance of E2E entanglement distribution may be affected by the scheduling of entanglement generation and swapping. 

Since entanglements are already created at path computation time, the delay to satisfy E2E entanglement requests may be reduced in reactive routing. However, this approach faces significant concerns regarding its feasibility and scalability. Communicating with a controller or neighbor routers for every E2E entanglement creation introduces latency and overhead, challenging the network's efficiency and practicality. 

\subsubsection{Virtual}\label{sssec:virtual}
Entanglements can be created between any pair of non-adjacent routers with swapping at intermediate nodes. This allows the network to form arbitrary virtual topologies with entanglements spanning beyond one physical link. Such entanglements are referred to as virtual links or $l$-level entanglements \cite{illiano2022quantum, gyongyosi2018decentralized}. For a $l$-level entanglement, the hop distance between its two end nodes $x$ and $y$ is $ 2^{l-1} $ \cite{gyongyosi2019adaptive}. Elementary entanglements are considered 1-level entanglements. 

\begin{figure}[!h]
\centering
\includegraphics[width=0.7\linewidth]{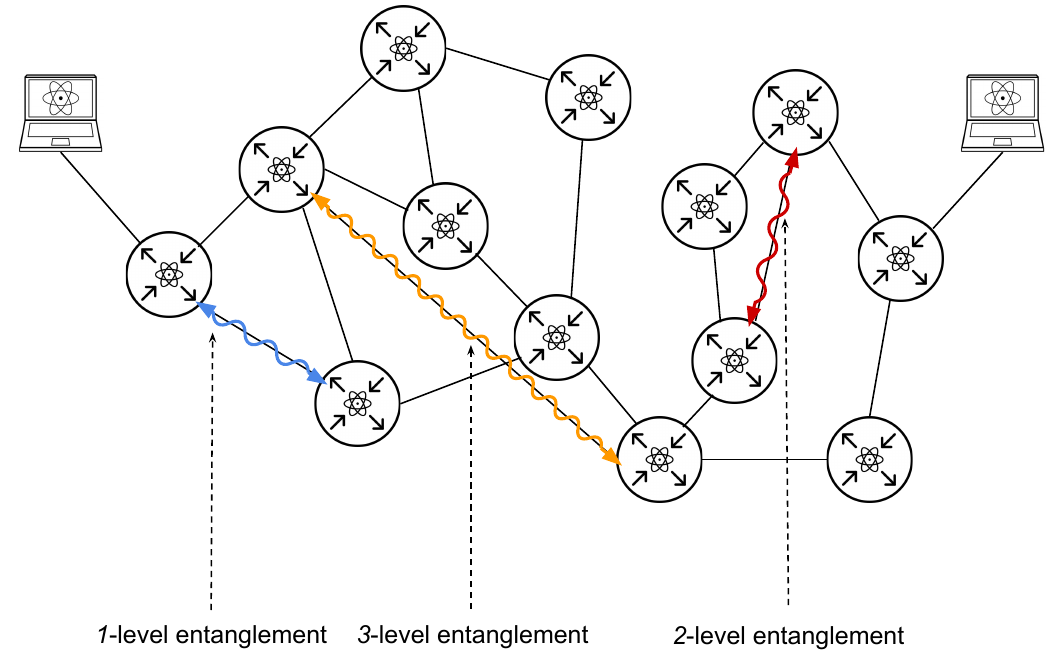}
\caption{\label{fig:routing_virtual}Virtual routing on a quantum network with various levels of entanglement links.}
\end{figure}

Virtual links can be combined, or with elementary entanglements, to create E2E entanglements as illustrated in Figure \ref{fig:routing_virtual}. 
The virtual links can be chosen deterministically or randomly \cite{chakraborty2019distributed}, allowing the routing problem to be divided into two subproblems. On one side, the virtual links are (pre-)calculated and created to form the virtual topology \cite{ghaderibaneh2022predistributed}. On the other side, the routing algorithm uses the virtual topology to compute paths for requested E2E entanglements. 

Entanglements (elementary and virtual) can be continuously generated or created after paths are computed. 
However, virtual entanglements expire and need to be regenerated just like elementary entanglements. This can be a time-consuming and challenging process, involving entanglement generation and swapping.

Using virtual links can increase the connectivity of the network. It also reduces the diameter of the network topology, which may reduce the routing algorithm complexity and execution time. Study \cite{schoute2016shortcuts} showed that an efficient construction of the virtual topology can reduce the latency of E2E entanglement creation. 

Virtual routing approaches are proposed in \cite{gyongyosi2019adaptive,gyongyosi2018decentralized, gyongyosi2017entanglement}. 
The work in \cite{gyongyosi2018decentralized} starts by establishing a physical topology and categorizing nodes based on their capabilities for generating entanglements. Following this, a virtual topology is computed in a decentralized manner, employing the small-world network concept as outlined by Kleinberg \cite{kleinberg2000small}. 
In the virtual topology graph, routing is performed using a distributed greedy algorithm with the entanglement success probability as the metric. The same approach is adopted in \cite{gyongyosi2017entanglement} using a metric derived from entanglement throughput statistics.

Researchers in \cite{chakraborty2019distributed} adopt a similar approach but introduce a maximum limit on the distance of a pre-shared virtual link relative to the physical links. Additionally, they set a cap on the storage duration for the entanglement link.
Considering a network that continuously generates entanglements, they designed hybrid routing algorithms that select a next hop even if no virtual link is established yet (similar to proactive routing). In this case, the entanglement generation is attempted over the selected virtual link. 

This approach helps mitigate the downside of reactive routing resulting from its sub-optimal resource utilization. According to the reported results, when there is only a single request in the network, relying on the continuously generated entanglements yields a lower latency than creating new ones. In the case of multiple concurrent requests, however, relying only on continuously generated entanglements can exhaust pre-shared entanglements before creating new ones to replace them, thus requesting new entanglements can improve performance. 

Another virtual routing approach that addresses the lack of entanglements in reactive routing is proposed in \cite{pouryousef2022quantumoverlay}, where the network accumulates l-level entanglements between nodes with storage capacity (either randomly or based on node degree) and makes them available when needed.

Overall, virtual routing can be seen as an enhancement of the reactive scheme, which can combine the advantages of proactive and reactive routing while limiting their shortcomings. On the one hand, the latency to establish E2E entanglements due to the entanglement creation in proactive routing can be reduced by the continuous generation of virtual links pre-configured on routers. On the other hand, virtual entanglements can artificially improve entanglement availability (e.g., \cite{chakraborty2019distributed, pouryousef2022quantumoverlay}) and accommodate path computation since the virtual topology to process can be simplified.

\subsubsection{Opportunistic}
In opportunistic routing, nodes try to generate the E2E entanglement without a preliminary path computation phase. At each router, the next entanglement (i.e., next hop) is selected based on the results of the elementary entanglements attempted with neighbor routers. Referring to proactive and reactive routing, one may think of quantum opportunistic routing as a scheme where path computation and entanglement creation are executed in parallel, at each hop along the path.

\begin{figure}[!h]
\centering
\includegraphics[width=0.7\linewidth]{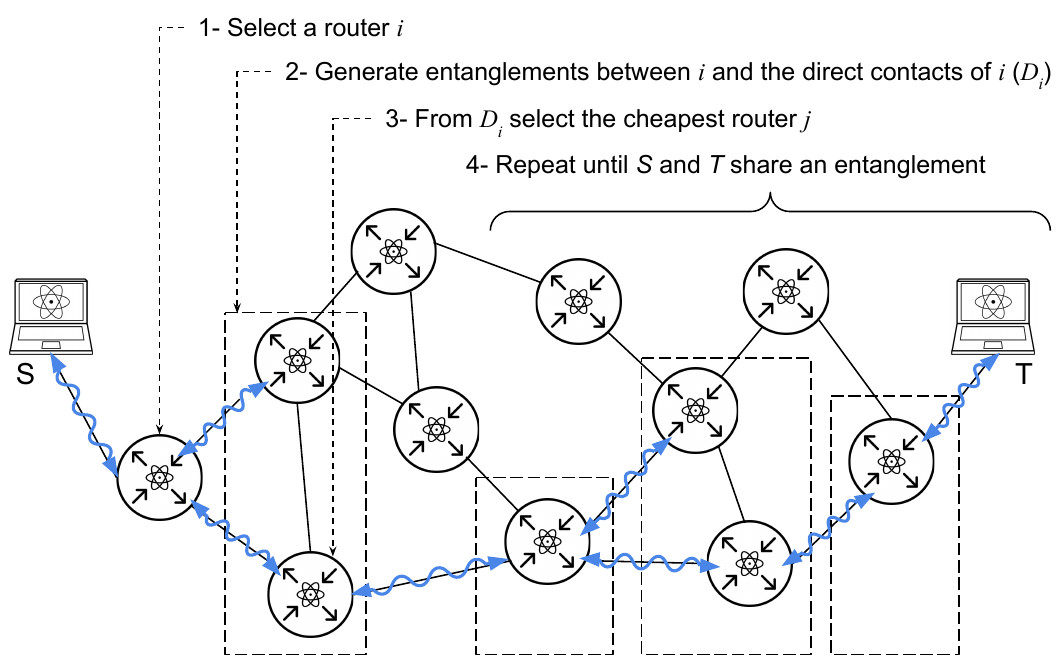}
\caption{\label{fig:routing_opportunistic}Opportunistic routing proposed in \cite{gyongyosi2019opportunistic}.}
\end{figure}

Figure \ref{fig:routing_opportunistic} illustrates an opportunistic routing scheme as proposed in \cite{gyongyosi2019opportunistic}. Starting with the first end-node of the requested E2E entanglement, each node selects a distribution set of nodes among its neighbors. The distribution set of node $i$ is formed by nodes with which it shares entanglements in the direction of the other end node of the request. Node $i$ then selects one neighbor to forward the request based on criteria such as entanglement fidelity. The process is repeated at each hop until the entanglement with the end node is created. 
Note that opportunistic routing handles only the selection of entanglement links to create along the path, and does not assume any specific swapping approach. Swapping can use a certain order or policy after all entanglement links are created, or the swapping can be executed opportunistically while forwarding the E2E entanglement request (see Section \ref{ssec:forwarding}).
Quantum opportunistic routing typically considers partial physical topology or the global physical topology if shared beforehand. However, the logical topology knowledge is always local; within the next neighbor or \textit{k} neighbors.

\subsection{Forwarding Phase} \label{ssec:forwarding}
Forwarding encompasses two main steps: the \textit{external phase} for generating elementary entanglements and the \textit{internal phase} for establishing E2E entanglements via swapping. The execution of these steps may differ depending on the routing scheme and supported services.
In the proactive (and often virtual) routing, the external phase occurs after path computation. In contrast, in reactive and opportunistic routing, it happens before and concurrently to path selection, respectively. 

\begin{figure}[h]
\centering
\includegraphics[width=0.9\linewidth]{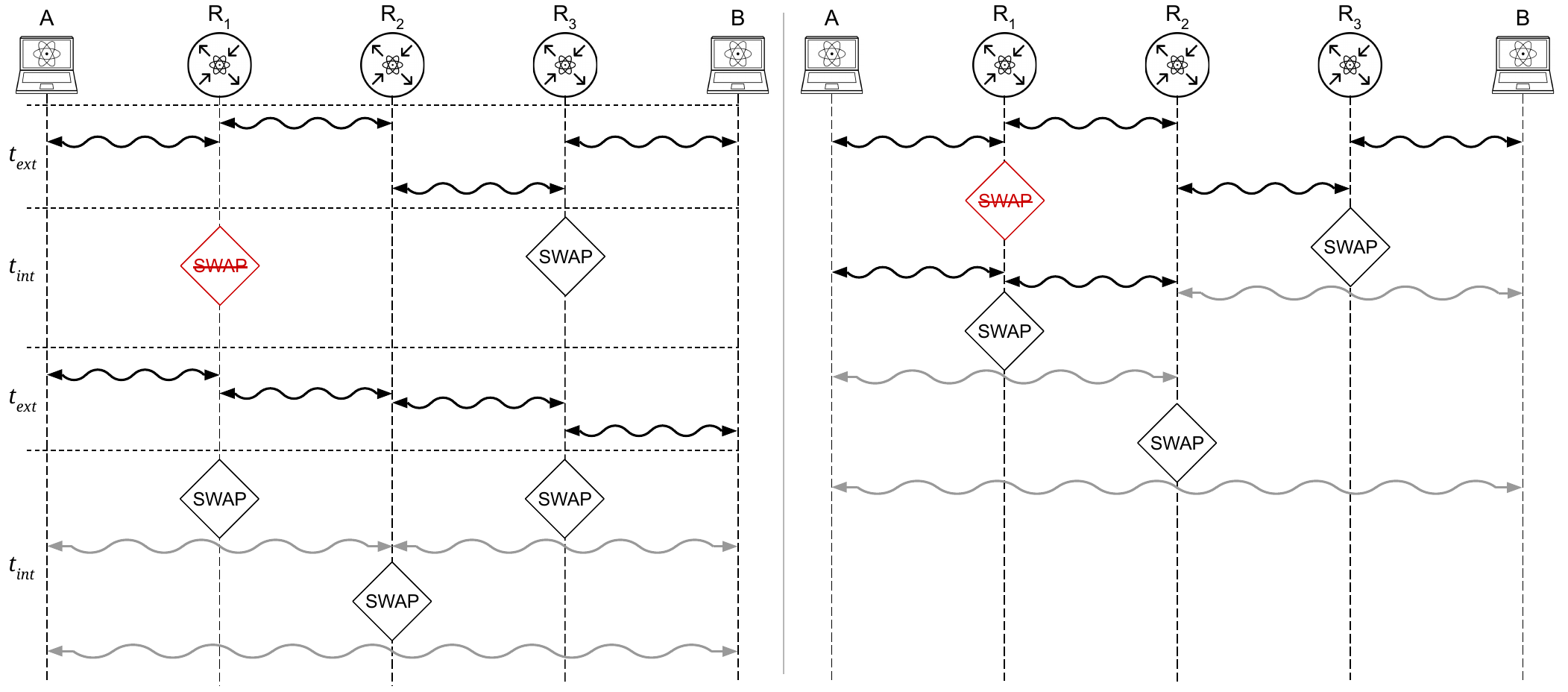}
\caption{\label{fig:sync_async}Entanglement generation and swapping. Synchronous mode (left) and asynchronous mode (right). Crossed SWAP indicates a failed swapping.}
\end{figure}

The internal and external phases can be executed synchronously or asynchronously, as illustrated in Figure \ref{fig:sync_async}. 
In the synchronous mode, nodes perform the external phase within a fixed duration, followed by the internal phase. If a swapping attempt fails, all nodes involved in the path re-execute the external phase again. Note that elementary entanglements do not need to succeed at the same time, but within the duration of the internal phase.
In the asynchronous mode, each pair of nodes creates elementary entanglements independently as soon as quantum memories are available, and swapping is performed immediately when conditions are met. If a swapping attempt fails, only the involved entanglements are regenerated while the other swapping attempts continue.
Consequently, the asynchronous mode provides better resource utilization and achieves higher throughput compared to synchronized entanglement generation and swapping.

The primary operation in forwarding is swapping. Purification and other operations may be included depending on the QoS and reliability features as discussed below. 

\subsubsection{Entanglement Swapping} \label{sssec_swapping}
The fidelity of E2E entanglements drops with each swapping operation \cite{chakraborty2020entanglement} and the time elapsed before applying the swapping correction at the end-node \cite{van2014quantum}. Additionally, the E2E entanglement throughput may be affected by the order in which swappings are performed \cite{chang2022order}. 
In the following, we will discuss different swapping approaches studied in the literature. 
In general, swapping can be either memoryless or based on quantum memory. 

\paragraph{Memory-less swapping:} This is often referred to as \textit{synchronous} swapping or \textit{waitless} protocol \cite{chakraborty2020entanglement}. Here, strict synchronization among routers is required so that all entanglements along a path must be successfully created at the same time and all swapping operations can be carried out simultaneously. Thus, memoryless swapping is unheralded. 
Clearly, an E2E entanglement can only be generated if all of the underlying processes are synchronized and succeed; otherwise, any failure will cause the whole process to restart from the generation of elementary entanglements. As a result, although this approach can provide long-distance entanglements with high fidelity, its generation rate is very low, hindering its practicality. 

\begin{figure}[!h]
\centering
\includegraphics[width=0.67\linewidth]{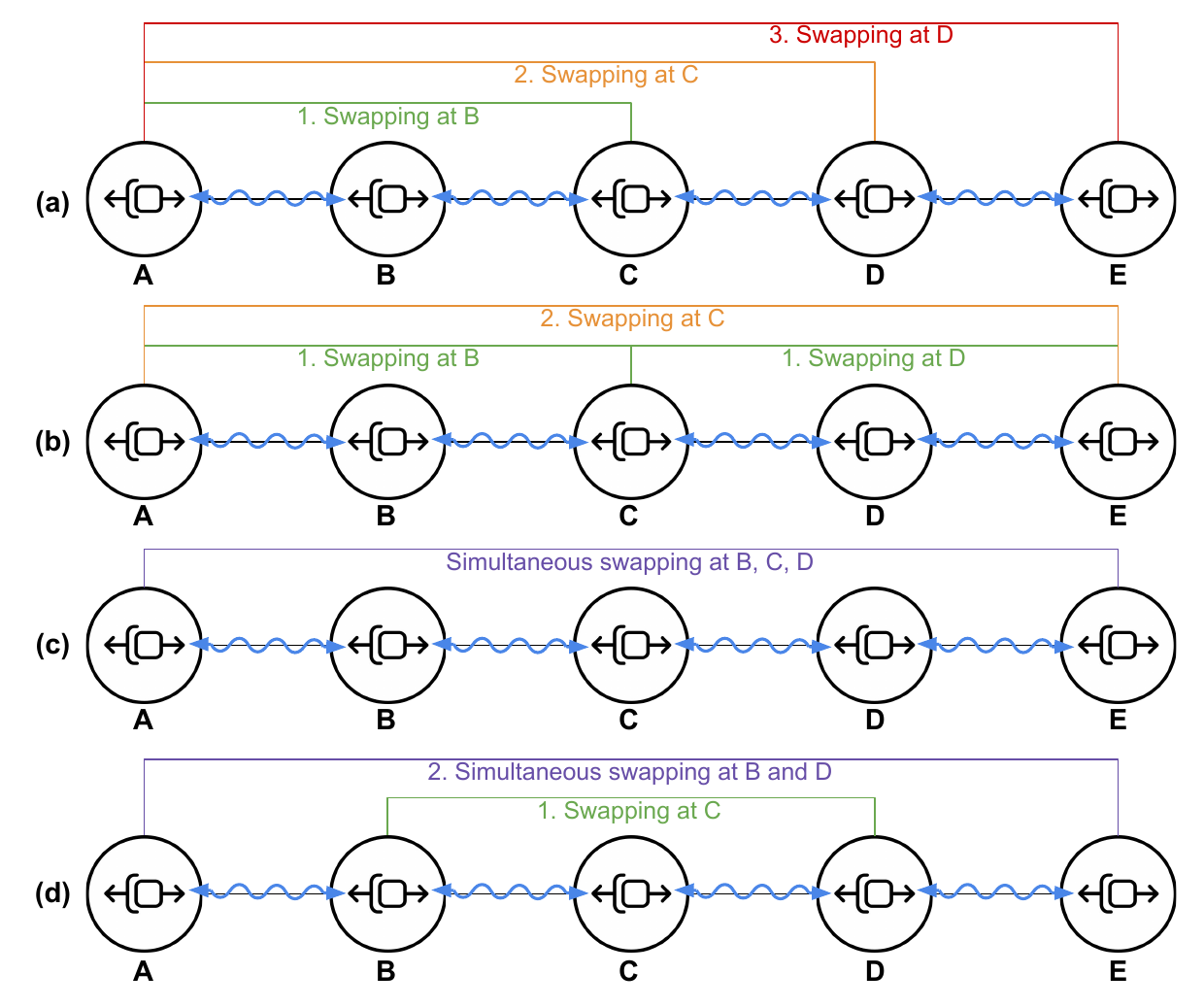}
\caption{\label{fig:swapping_strategies}Examples of swapping orders in a single path without multi-hop purification (adapted from \cite{matsuo2019quantum}).}
\end{figure}

\paragraph{Memory-based swapping:} This is also known as \textit{asynchronous} swapping or \textit{waiting} protocol \cite{chakraborty2020entanglement}, where a qubit of an entangled Bell pair may wait in memory for its swapping counterpart to become available and certain conditions to be met so that a swapping operation can be carried out. This is possible because coherence times of several seconds to minutes (and even hours, depending on the technology) have been demonstrated \cite{bartling2022entanglementMinute}. As a result, swapping can happen at different times at intermediate nodes along a path following certain rules and orders (including random ones).\footnote{Multiple paths going through the same sequence of routers might employ different swapping policies as well.} Thus, different swapping policies can lead to different wait times, fidelity, and throughput in generating E2E entanglements.

Asynchronous mode is expected to perform better than synchronous mode because it does not require all entanglements to succeed at the same time and allows the regeneration of only failed entanglements instead of all entanglements. Additionally, it may be useful for the routers to know if a swapping operation has failed as soon as possible to avoid unnecessary waiting (at the cost of additional classical signaling). 
In the following, we will focus on asynchronous swapping and discuss several common swapping schemes using an example of a path with $n=4$ hops shown in Figure~\ref{fig:swapping_strategies} above.

    \begin{itemize}
    \item \textbf{Sequential:} Nodes perform swapping sequentially from left to right (or right to left) \cite{li2022connection}; see Figure~\ref{fig:swapping_strategies}(a). This scheme requires $n-1$ steps with one node performing swapping operations in each step after its immediate neighbor. As a result, this scheme yields the highest wait time due to swapping steps. However, it is often the most useful scheme in a path-finding algorithm based on simple metrics (such as hop count, node distance, link entanglement creation rate, or link bottleneck capacity as described in Section~\ref{sssec_path_search} above) because the contribution of a new hop to the path length when extending the path is often much easier to find compared to other schemes below. 

    \item \textbf{Doubling:} This is one of the most studied schemes, where the order of swapping corresponds to a balanced binary tree of height $\lceil \log_2 n\rceil$ and nodes on the same level can perform swapping in parallel; see e.g., \cite{briegel1998quantum, van2013path, caleffi2017optimal, ghaderibaneh2022efficient}.\footnote{This scheme is also referred to as \textit{parallel order} in some references such as \cite{chang2022order}; we refer to it as  \textit{doubling} instead so as to be consistent with the literature.} Thus, this scheme takes only $\lceil \log_2 n\rceil$ steps. Figure~\ref{fig:swapping_strategies}(b) shows the case for $n=4$ where Node B and Node D perform swapping first, and Node C connects the end nodes afterward. Note that when $n$ is a power of $2$, the binary tree is called a perfect binary tree and, assuming sufficiently long decoherence time, the doubling scheme gives the optimal generation rate for homogeneous chains \cite{dai2020optimal,ghaderibaneh2022efficient}. In general, however, this scheme is not necessarily optimal.
    
    \item \textbf{Parallel:} When all elementary entanglements have been successfully established on a path, all repeater nodes can perform entanglement swapping simultaneously and independently in the sense that one node's swapping does not depend on the swapping results of others, hence unheralded swapping; see Figure~\ref{fig:swapping_strategies}(c). This scheme is more suitable for reactive routing as it takes only one step to finish, but its generation rate may be low because, taking Figure~\ref{fig:swapping_strategies}(c) for example, if one swapping fails, the whole process needs to restart from scratch (from generation of all elementary entanglements just like in synchronous swapping). 

    \item \textbf{Ad-hoc:} Entanglement swapping is performed based mainly on the availability of local resources (locally adjacent elementary entanglements) instead of following any predefined order. A popular policy in this scheme is often referred to as \textit{opportunistic} \cite{farahbakhsh2022opportunistic} or \textit{swap-as-soon-as-possible} \cite{kamin2023exact}. Here, the order and the number of swapping steps can vary from execution to execution; as a result, any order in Figure~\ref{fig:swapping_strategies} can happen. This policy is the default in opportunistic routing but can be adopted in the other routing approaches. It allows part of the waiting time for the entanglement generation to be used to perform some of the swapping operations. Hence, resources are consumed and freed more quickly, resulting in less waiting time for requests that are waiting for the completion of prior requests. The scheme can be applied after the whole path is computed by the routing phase (in proactive routing) or it can be combined with the opportunistic routing approach; see also \cite{farahbakhsh2022opportunistic}. 

    \item \textbf{Heuristic:} For a general non-homogeneous path, entanglement swapping can be performed in a particular order based on certain rules in each time slot; see Figure~\ref{fig:swapping_strategies}-(d) for example. Oftentimes, this is done to optimize certain performance objectives such as throughput or generation latency \cite{li2022connection, chang2022order,ghaderibaneh2022efficient}. However, since optimizing swapping orders is combinatoric, this scheme is more computationally expensive than the other schemes described above, even when approximation/relaxation is in place. As a result, this scheme is more suitable for proactive routing where swapping orders are usually fixed and  determined along with path computations. Recently, \cite{haldar2024fast} has used Reinforcement Learning (RL) to design swapping policies for optimizing either wait time or fidelity of the end-to-end entanglement in a path with single channels (i.e., unit capacity) and cutoff (coherence) time of quantum memories. In this case, the swapping order varies between time slots as each swapping action depends not only on the policy but also on the state of the whole path. 
\end{itemize}

Except for ad-hoc and certain heuristic (e.g., RL-based) swapping, all the other schemes need the routers to agree on a particular order to perform swapping along each path. Since these orders are predefined, they can be considered \textit{static-swapping} policies. In this case, swapping instructions (excluding entanglement readiness messages between routers) for parallel schemes contain only the qubits to swap along the path, while those for other schemes must include some schedule or rules for the steps in addition to the qubits to swap. 

Ad-hoc schemes and certain heuristic policies such as RL-based policies can be seen as \textit{dynamic swapping}; or more precisely, the swapping sequence is a stochastic process driven by the randomness of link entanglement generation and node entanglement swapping operations, as well as the swapping policy. 
Note however that swapping instructions for ad-hoc schemes can be as simple as satisfying certain local conditions \cite{haldar2024reducing}, while RL-based swapping policies as in \cite{haldar2024fast} must rely on non-local instructions because swapping actions in each step depend on the current state of the whole path.

\subsubsection{Fidelity Support}
Many routing algorithms operate under the assumption that entanglement links are successfully established and maintained in their ideal state. However, more realistic models take into account the quality of these links, acknowledging the possibility of creating lower-quality entanglements. To offer a more accurate assessment of path quality, certain research incorporates entanglement fidelity in the routing algorithm and forwarding process. 

Estimating fidelity in a quantum network is generally a challenging task due to the presence of various sources of noise and errors. Network benchmarking protocols \cite{helsen2023benchmarking} estimate the average fidelity of quantum channels by sending and measuring quantum states through the network links. Machine learning techniques \cite{zhang2021directFidelity, qbgp} have shown promise in estimating the fidelity of quantum states using fewer measurements than traditional methods like full-state tomography.

Entanglements suffer from inherent imperfections due to hardware limitations. These imperfections are magnified during the entanglement swapping process, leading to a reduction in the quality of the resulting E2E entanglement \cite{kozlowski2020designing}. To address this issue, it is essential to input high-fidelity entanglements into the swapping process to ensure the E2E entanglement meets the desired fidelity. Moreover, the noise introduced by quantum operations can be reduced with better hardware, though such improvements fall outside the scope of network protocols. Ensuring high-fidelity entanglements requires a strategy that includes both hardware advancements and several optimizations at the protocol level. 

An effective purification scheme is hard to determine before knowing the entanglement quality. 
During the routing phase, the length of computed paths can be limited to meet the requested fidelity \cite{chakraborty2020entanglement}. 
Additionally, the routing algorithm can decide on the required fidelity of elementary entanglements to ensure a sufficiently high E2E entanglement fidelity \cite{kozlowski2020designing}.
Decoherence significantly impacts the fidelity of qubits while in memory. This can be mitigated during forwarding by minimizing the duration that qubits remain unused in memory \cite{kozlowski2020designing}.
These approaches improve the chances of satisfying the E2E fidelity requirements without using purification on the created entanglements. Hence, they can be considered as a \textit{passive} fidelity support.

A more advanced approach to satisfy the requested fidelity is to use purification on elementary entanglements to individually increase their fidelity (purify-then-swap). Alternatively, purification can be used after generating multiple E2E entanglements between the two end-nodes (swap-then-purify). Since the approaches use purification, they can be classified as \textit{active} fidelity support techniques. However, active fidelity support may require several rounds of purification, which in turn requires more entanglements (rate) and impacts the E2E entanglement throughput. 

An active fidelity support in reactive routing is proposed in \cite{li2021effective}. A purification is systematically applied on all entanglement links with fidelities below the threshold required by the E2E entanglement request. Then, only the entanglement links with sufficient fidelity are included in the path computation. However, systematically purifying all elementary entanglements up to a certain threshold may not guarantee the satisfaction of the requested fidelity in the E2E entanglement, which would result in resource wastage \cite{zhao2022E2E}. 
Alternatively, the routing algorithm can compute a purification scheme for each path, indicating the entanglement links to purify and the number of rounds \cite{victora2020purification}. The scheme is executed by the routers at the forwarding phase. In \cite{zhao2022E2E}, a purification scheme considering the E2E path fidelity is determined during the path computation to ensure the request fidelity is satisfied. 

Generally speaking, active fidelity support is easier to provide in reactive routing where paths are computed based on created entanglements. In this scheme, the authors in \cite{li2022fidelity} use a routing metric that reflects the resources needed to produce an entanglement satisfying the fidelity in each E2E entanglement request as a metric to determine the optimal path.

\subsubsection{Reliability Support}\label{ssec:reliability}
It is evident that mitigating entanglement generation and swapping failures are critical in ensuring the robustness and efficiency of quantum communications. To this extent, various path recovery procedures are envisioned.

Due to time constraints imposed by entanglement decoherence, a path recovery procedure may be more realistic in a distributed way where each router cooperates with its neighboring nodes to find alternative links. However, the effectiveness of distributed/local path recovery is limited because nodes may not be able to optimize the usage of existing entanglements with only neighboring knowledge. This may lead to more frequently fragmented paths and unused entanglements \cite{shi2020concurrent, zhao2021redundant}.

\begin{figure}[!t]
\centering
\includegraphics[width=\linewidth]{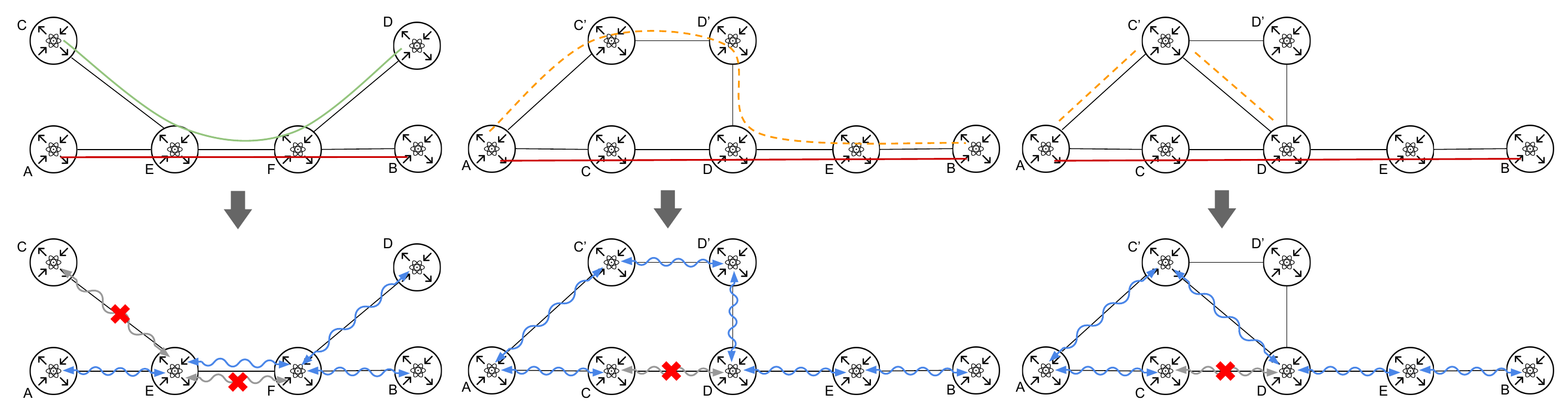}
\caption{\label{fig:path_recovery}Path recovery strategies. Defragmentation (left), Secondary path (center), Redundant links (right).}
\end{figure}

Path recovery is more critical in proactive routing where entangled links are not known at path computation time, which increases uncertainty in forwarding. During the forwarding phase, routers can use various information (e.g., the physical topology, entanglement links outcomes, neighbors, etc.) to create alternative/redundant entanglements or re-affect unused entanglements to complete the creation of E2E entanglements (see Figure \ref{fig:path_recovery}).

Path recovery can be enhanced through its integration with the path computation \cite{shi2020concurrent, zhao2021redundant}. In this case, the routing algorithm can be designed to provide routers with alternative links or entire paths to use for path recovery.

The research \cite{shi2020concurrent} proposes path recovery for proactive routing based on local entanglement link-state outcome following entanglement generation. In the routing phase, nodes use the global topology knowledge to choose a consistent set of paths and address link failures with entanglement generation information during the forwarding phase. For each E2E entanglement request, they identify multiple major paths that can be fully reserved according to available resources, and partial paths which share resources with major paths and thus lack complete reservation. The entanglements are created for both path types, with the partial paths serving as a fallback during the forwarding phase to compensate for any failed entanglement in a major path.
As illustrated in Figure \ref{fig:path_recovery} (center), a routing algorithm computes two concurrent paths for request A-B. Path ACDEB is the main one while path AC’D’DEB does not have enough resources but some of its entanglements can be created and used as a recovery path. In the forwarding phase, node D finds that the main path is disconnected. It chooses to route through AC’D’D, and swaps link DE with link DD’, instead of CD-DE.

Authors in \cite{zhao2021redundant} assume that when sufficient network resources are available, the path computation should involve provisioning extra, potentially redundant resources for establishing entanglement links. This strategy ensures that in the event of failures in creating some entanglements, alternative ones can be utilized to establish E2E entanglements subsequently. To avoid creating extra entanglements on every quantum channel, an algorithm determines an optimal set of backup entanglement links to be created, considering resource and path constraints. 
This approach is depicted in Figure \ref{fig:path_recovery} (right) where an algorithm computes path ACDEB and provisions extra links entanglements (AC’ and C’D) for redundancy. As entanglement CD failed, the backup links are used to connect A to B.

Some path recovery aspects can be supported by the forwarding phase without requiring special features from the path computation algorithm. Typically, entanglements that cannot be used to establish paths can be re-affected to complete other paths. 
These entangled but unused qubit pairs, called fragments \cite{zhang2021fragmentation}, negatively impact resource utilization efficiency and reduce overall network throughput. Defragmentation in quantum networks presents significant challenges due to the unpredictable nature of entanglements and their limited lifetime, preventing link states from being propagated throughout the network. Consequently, nodes must make local decisions to connect entanglement links based on k-hop entanglement states as proposed in \cite{zhang2021fragmentation}. As depicted in Figure \ref{fig:path_recovery} (left), nodes E and F possess the information of selected paths and the states of entanglements within a few hops. This enables them to assign entanglement between E and F to path AB. This approach requires that path computation does not assign specific qubits to specific paths.
A similar strategy is adopted in \cite{farahbakhsh2022opportunistic} in which available entanglements that are not used to establish a path are affected to create other paths. For that, the outcome of all entanglement links is shared by each node to all other nodes within a few hops.

Note that path recovery is also useful in reactive and virtual routing to overcome unexpected failures that may occur due to memory issues or errors incurred during swapping and purification routines. To overcome link failures in the virtual routing scheme, the strategy proposed in \cite{gyongyosi2019adaptive} selects alternative replacement paths for primary paths. These replacement paths provide provisional routes to be used when the main path fails until all disrupted entanglement links are fully restored.

\subsection{Routing Algorithms} \label{ssec:algorithms}
A typical objective of routing is to maximize the overall Entanglement Generation Rate (EGR); i.e., the number of E2E entangled pairs generated per unit of time. To this end, the routing algorithm includes the definition of a metric that is easy to calculate, general enough to be used independently of the physical technology, and reliably leads to choosing reasonable, if not optimal, paths. 

The first-generation quantum repeater networks impose specific constraints that a routing algorithm must consider to effectively solve the entanglement routing problem. As discussed above, these constraints include the limited time available to utilize entanglement before it begins to decohere, the probabilistic nature of entanglement creation and swapping, the variable fidelity of entanglement which may necessitate one or multiple rounds of purification to meet path fidelity requirements, and the exclusive reservation of qubits for a single entanglement path, preventing their concurrent consideration for multiple paths. 
In the following, we provide a classification of the routing algorithms and models proposed in the reviewed studies.

\subsubsection{Dijkstra Algorithms}
The applicability of the Dijkstra algorithm in quantum routing is certainly the oldest approach \cite{di2012optimal, van2013path}. The Dijkstra algorithm operates on the principle that the total cost of a path is derived from the sum of the costs associated with each edge along that path. This does not always match the formulation of the entanglement routing problem regarding quantum network characteristics. However, by defining an appropriate cost for paths, Dijkstra’s algorithm offers a straightforward method for selecting paths. 
Using Dijkstra, one simple link cost could be the inverse of the throughput of the link, measured in seconds per Bell pair of a particular fidelity (e.g., in proactive routing) \cite{van2013path}. 
While lower-level metrics like the quantum measurements and hardware operations serve as useful indicators for gauging the actual work expended to establish a path, they fall short as criteria for link prioritization \cite{van2014quantum}. This is because they mainly mirror the physical attributes of a link rather than the overall E2E entanglement distribution.
Overall, utilizing a variant of Dijkstra's algorithm, where the link cost is represented by the inverse of each hop's throughput, creates a relatively accurate correlation between the simplified path cost and the actual throughput, as well as between the cost and the quantum physical operations involved. Furthermore, this approach can be implemented with a manageable level of computational complexity.

A variant of Dijkstra’s algorithm is examined in \cite{di2012optimal} aiming at maximizing the E2E fidelity of entanglements. The study shows that the final E2E fidelity cannot be systematically inferred from the fidelity of each hop. 
Ideally, entanglement routing should be a function of both EGR and fidelity. Hence, more recent studies \cite{shi2020concurrent,li2022fidelity} argue that reducing path selection to a simple shortest path problem may not always achieve optimal routing decisions. Recent studies frequently use Dijkstra in combination with other methods for path computation.

\subsubsection{Path Search on Graph} \label{sssec_path_search}
Path search and enumeration in graphs are common approaches used to compute paths in both proactive and reactive routing with multiple concurrent requests. Path search algorithms such as Dijkstra, Yen’s algorithm, and Bellman-Ford are extended with mechanisms to satisfy multiple requests without resource contention, provide fairness among requests, and optimize entanglement resource utilization. 
Yen's algorithm is used to find the $k$ shortest paths between nodes, providing multiple alternative routes for route diversity and fault tolerance. Bellman-Ford handles graphs with negative weight edges, ensuring the shortest paths are found while also detecting any negative weight cycles.
However, these algorithms require knowledge of the global topology, whether the routing is implemented in a central controller or distributed across routers. 

In the reactive routing scheme, the research \cite{nguyen2022multiple,pant2019routing} uses the Node-Disjoint-Path (NDP) problem in the logical topology graph to find a set of paths linking a specified pair of nodes such that no two paths share a node. The hop count is used as a metric. 
In \cite{nguyen2022multiple}, several algorithms are proposed for the NDP problem: Sequential Multi-Path Scheduling Algorithm (SMPSA), Min-Cut-based Multi-Path Scheduling Algorithm (MCSA), and Random and Distance (physical) Scheduling Algorithm. 
In \cite{li2022fidelity}, path search with $k$-shortest path and an extended Dijkstra algorithm are used in the logical topology graph to find paths for a single request. The algorithm is extended with a resource allocation mechanism to support multiple E2E entanglement requests.

In proactive routing, the most commonly used metric in path search algorithms is the expected EGR \cite{zhao2021redundant, chakraborty2020entanglement} estimated according to the network model similar to the one presented in Section \ref{sec:routing_problem}.

The research in \cite{caleffi2017optimal} relies on path enumeration within a physical topology graph with the expected E2E entanglement throughput as a metric. The model incorporates low-level components involved in entanglement distribution, including optical fiber attenuation length, efficiency of Bell state measurements, and duration of atom pulses. Despite its advantages, the proposed routing scheme faces challenges in scaling across arbitrary topologies due to its computational time complexity.

Study \cite{shi2020concurrent} applies Yen's algorithm for identifying paths within the physical topology graph, alongside an augmented version of Dijkstra's algorithm to mitigate resource contention. It investigates multiple metrics for path evaluation. The sum of node distances, which sums up the lengths of channels at each hop, reflects path difficulty due to the exponential decrease in channel success rate with physical distance. The creation rate is calculated as the inverse of the product of success rates for channels along the path, offering insight into path width, while the bottleneck capacity is defined to favor paths with greater width. The creation rate is used to resolve ties among paths of equal width. However, the algorithms do not address the scheduling of demands for multi-path routing to optimize the use of resources.

\subsubsection{Linear Programs}
Linear programs are also used to formulate the routing problem whether or not fidelity requirements are included. The problem is usually formulated as mixed-integer linear programs (MILP) that capture the traffic pattern for all demands in the network. Solving exactly these problems for large-scale networks is impractical as they are usually NP-hard and thus cannot be solved in polynomial time. As a result, a relaxation step is usually employed in conjunction with some rounding techniques to obtain approximated solutions possibly with certain provable approximation ratios for the objective functions. Note however that as the time complexity of most LP solvers is polynomial in the problem size, this approach might not be scalable, especially under fidelity and decoherence time conditions.

Sometimes the routing problem is formulated as a multicommodity flow optimization, in both proactive and reactive routing. Since the formulated problems are complex to resolve in a reasonable routing algorithm, a mix of linear programming optimization and path search in a graph is used in the routing algorithm. 
For example, \cite{chakraborty2020entanglement} considers a multicommodity flow-based approach for maximizing the flow rate of entanglement distribution for all demands subject to certain fidelity requirements. The paper proposed to replace the fidelity requirements with hop limits and then formulate the problem as an edge-based LP with $O(|\mathcal V||\mathcal E||\mathcal R|)$ variables and $O(|\mathcal V|^2|\mathcal E||\mathcal R|)$ constraints, where $\mathcal{R}$ is the set of E2E entanglement requests. The solution to this LP can then be converted to path selection and rate allocation using an algorithm with a time complexity of $O(|\mathcal V|^4|\mathcal E||\mathcal R|)$. The overall complexity of this approach appears to be high; in particular, for a network with 70 nodes, solving the edge-based LP alone already takes roughly 45 seconds. 

In reactive routing, \cite{zhao2022E2E} combines the $k$-shortest paths found with Yen’s algorithm and linear program resolution with link fidelity as a metric while including a purification scheme to satisfy the requested fidelity. In particular, this paper considers maximizing the network throughput defined as the number of entanglement connections among multiple E2E pairs subject to fidelity constraints. This is done by first preparing multiple candidate entanglement paths, determining optimal purification schemes, and then selecting a final set of entanglement paths that can maximize network throughput. The entanglement path selection problem therein is solved through iterative LP relaxation and incremental rounding, which results in high time complexity. 

In proactive routing, \cite{zeng2022multi} uses Yen's algorithm and linear programming resolution with the expected rate as a metric. Specifically, this paper aims to simultaneously maximize the number of user pairs and their combined expected throughput. This problem is formulated as two sequential MILP steps, the first of which maximizes the number of user pairs that can be served with a main routing path selected from a limited subset of paths computed from Yen’s algorithm, and the second step is for maximizing the expected throughput of user pairs selected from the first one. Here, both steps involve solving LP relaxations in $O(|\mathcal R|^3 |\mathcal V|^3 + |\mathcal R|^{4.7})$ and using a Branch-and-Bound technique for rounding. For a network with $|\mathcal V|=200$ nodes, the runtime of the proposed approach is in order of hundreds of seconds.

Study \cite{zhao2021redundant} also formulates the routing problem as a linear program, and searches the shortest path using the expected EGR as a metric, where the routing protocol includes request scheduling and selects redundant paths to overcome link failures (see Section \ref{ssec:reliability}). Specifically, this study proposed to provision extra entanglement links for redundancy and then select paths and links for maximizing the expected throughput of multiple E2E entanglement pairs. The link provisioning and entanglement path selection problems are formulated as MILP with moderate sizes and then solved using LP relaxation and randomized rounding techniques while the entanglement link selection is heuristic simply based on the number of entanglement paths selected and probabilities of success. In simulations with networks up to 500 nodes, their algorithm runs in tens of milliseconds due mainly to relatively small problem sizes (in terms of the number of variables and constraints).

\subsubsection{Greedy Algorithms}
Greedy routing algorithms operate by choosing the neighboring node that is closest to the destination according to a specific metric, ensuring that no node is selected more than once. Greedy algorithm variants are used to find near-optimal paths using local link-state knowledge in the reactive routing in \cite{chakraborty2019distributed, gyongyosi2018decentralized, gyongyosi2019adaptive, pant2019routing}. 
Greedy algorithms do not guarantee finding optimal paths although they represent a solution to implement decentralized reactive routing. 

As with path search on a graph, greedy algorithms commonly use the hop count or the link’s physical distance as a metric. 
In \cite{chakraborty2019distributed}, a greedy algorithm is applied by nodes with a global physical topology knowledge and local entanglement link-state knowledge, using the hop count as a metric. 
In \cite{gyongyosi2018decentralized}, the entangled network topology (virtual routing) is represented with a base graph where the Manhattan distance between nodes corresponds to the entanglement existence probability. Using the Manhattan distance function on the base graph, a greedy routing finds the shortest path. 
Similarly to \cite{gyongyosi2018decentralized}, the routing algorithm in \cite{gyongyosi2019adaptive} finds the shortest NDP for temporary replacement path in case of memory failure. In \cite{pant2019routing}, to find multiple paths for a request, a greedy routing considers the subgraph induced by the successfully generated entanglements and the repeater nodes, and finds in it the shortest path connecting the end-nodes using hop counts. Then, all the links of the path are pruned from the subgraph, and another shortest path is computed in the pruned subgraph.

\subsubsection{AI-based Routing}
In proactive routing, \cite{le2022dqra} formulates the routing problem as a reinforcement learning problem. The path selection is implemented with a deep neural network to select a request to fulfill and the shortest path algorithm to find the best path.

Study \cite{gyongyosi2017entanglement} introduces a decentralized approach based on swarm intelligence for path selection in quantum networks. This method employs routing metrics derived from statistical measures of entanglement success. Initially, the relevance of each entangled link is determined by an entanglement utility coefficient, reflecting the link's throughput statistics and its contribution to reaching the current node. 
Additionally, the attractiveness of a quantum node is defined by the link entanglement gradient coefficient, calculated from the deviation in entanglement throughput from statistics and the utility coefficient. 
This entanglement gradient is further applied to entire paths, creating a path entanglement gradient coefficient for routes composed of entangled links. Pathfinding is then achieved through the deployment of multiple threads that navigate local segments of the network topology, leveraging these defined metrics.

\subsection{Discussion}\label{sec:discussion}
Table \ref{tab:routing} summarizes the discussed routing approaches with their models and features, and Table \ref{tab:forwarding} summarizes the forwarding approaches.

Overall, the routing phase of entanglement distribution seems to attract more interest among the reviewed studies, with a more exploration of the forwarding phase in recent works. 
The definition of the network topology appears as an important design aspect. The common assumption across the routing schemes of a knowledge of the global physical topology seems viable for intermediate-scale quantum networks. However, as networks scale, the feasibility and scalability of maintaining global physical topology knowledge (at each node or at a controller) become increasingly complex, hinting at a pivotal challenge for the evolution towards a quantum Internet \cite{Van_Meter_2022}.

Proactive routing's reliance on global physical topology knowledge at the time of routing stands out. It contrasts with reactive routing's focus on the logical topology, which is dynamically formed by successfully established entanglements. This logical post-entanglement routing is less dependent on the underlying, potentially very heterogeneous, hardware technology. Although much less explored, the virtual and opportunistic routing's potential for leveraging both global and local topology knowledge introduces a flexible framework, allowing for trade-offs between scalability, adaptability, and optimality in routing decisions.

\begin{table}[!ht]
\centering
\resizebox{\columnwidth}{!}{%
{\rowcolors{2}{}{lightgray}
\begin{tabular}{|c|c|c|c|c|c|c|c|c|c|}
\hline
Study & Year & Routing & Metric & Phy. topo. & Log. topo. & Request & Path & Fidelity & Model \\
\hline
\cite{shi2020concurrent} & 2020 & Proactive & Link rate & Global & N/A & Multiple & Single & No & PS \\

\cite{caleffi2017optimal} & 2017 & Proactive & E2E rate & Global & N/A & Single & Single & No & PS \\

\cite{zeng2022multi} & 2022 & Proactive & Link rate & Global & N/A & Multiple & Single & No & LP \\

\cite{chakraborty2020entanglement} & 2020 & Proactive & Link rate & Global & N/A & Multiple & Multiple & Passive & MCF \\

\cite{le2022dqra} & 2022 & Proactive & Link capacity & Global & N/A & Multiple & Single & No & ML \\

\cite{zhao2021redundant} & 2021 & Proactive & Link rate & Global & N/A & Multiple & Multiple & No & LP \\

\cite{nguyen2022multiple} & 2022 & Reactive & Hops & Global & Global & Multiple & Multiple & No & PS  \\

\cite{yang2024Asynchronous} & 2024 & Reactive & Hops & N/A & Local & Single & Multiple & No & PS  \\

\cite{pant2019routing} & 2019 & Reactive & Hops & Global & Local & Multiple & Multiple & No & PS \\

\cite{cicconetti2021request} & 2021 & Reactive & Hops/Distance & Global & Global & Multiple & Single & No & MCF \\

\cite{li2021effective} & 2021 & Reactive & Hops & Global & Global & Multiple & Multiple & Link & PS/MCF \\

\cite{zhao2022E2E} & 2022 & Reactive & Fidelity & Global & Global & Single & Multiple & Scheme & PS/LP \\

\cite{li2022fidelity} & 2022 & Reactive & Entang. cost & Global & Global & Multiple & Single & Passive & PS \\

\cite{chakraborty2019distributed} & 2019 & Virtual & Hops & Global & Local & Single & Single & No & Greedy \\

\cite{gyongyosi2018decentralized} & 2018 & Virtual & Entang. prob. & Global & Local & Single & Single & No & Greedy \\

\cite{gyongyosi2019adaptive} & 2019 & Virtual & Entang. prob. & Global & Local & Single & Single & No & Greedy/PS \\

\cite{gyongyosi2017entanglement} & 2017 & Virtual & Entang. prob. & Local & Local & Multiple & Single & No & ML \\

\cite{pouryousef2022quantumoverlay} & 2022 & Virtual & Link rate & Global & Global & Multiple & Multiple & No & LP \\   

\cite{gyongyosi2019opportunistic} & 2019 & Opportunistic & Fidelity/Coherence & Global & Local & Single & Single & No & Greedy \\
\hline
\end{tabular}}%
}\vspace{0.2cm}
\caption{\label{tab:routing}\centering Notable path computation approaches and routing algorithms. PS: Path Search, LP: Linear Program, MCF: Multi-commodity Flow Optimization, ML: Machine Learning}
\end{table}

\begin{table}[!ht]
\centering
{\rowcolors{2}{}{lightgray}
\begin{tabular}{|c|c|c|c|c|c|}
\hline
Study & Year & Log. topo. & Swapping & Path recovery & Fidelity \\
\hline
\cite{shi2020concurrent} & 2020 & Local & Heuristic & Yes & No \\

\cite{li2022connection} & 2022 & Local & Sequential & No & No \\

\cite{kozlowski2020designing} & 2020 & Local & Sequential & No & Passive \\

\cite{li2022connection} & 2020 & Global & Heuristic & No & No \\

\cite{farahbakhsh2022opportunistic} & 2022 & Local & Ad-hoc & No & No \\

\cite{farahbakhsh2022opportunistic} & 2021 & Local & Heuristic & Yes & No \\

\cite{zhao2021redundant} & 2021 & Global & Heuristic & Yes & No \\

\cite{wang2022asynchronous} & 2022 & Local & Heuristic & No & No \\
\hline
\end{tabular}}
\caption{\label{tab:forwarding}Notable forwarding approaches.}
\end{table}

The design strategies also vary. Both proactive and reactive routing can be realized through centralized or distributed systems, each with its advantages and challenges. Proactive routing benefits from not being constrained by the ephemeral nature of entanglement, allowing for flexibility in implementation and optimization. 
Conversely, reactive routing, while more abstracted, must contend with the challenges posed by entanglement's short existence. This does not prevent reactive schemes from being envisioned with both centralized and distributed systems.
Path computation tends to favor the global knowledge of topology, either logical or physical, for feasibility and efficiency. However, the rapid pace at which the logical topology can evolve and the ephemeral nature of entanglements, render approaches that rely solely on the logical topology less practical if not unrealistic. Hence, a hybrid approach that combines global management with localized distributed computation may be the most realistic. 

The study of routing algorithms and models unveils a rich set of strategies covering various quantum networking aspects. Reactive routing strategies, particularly with local topology knowledge, emphasize metrics like hop count for a straightforward approach for path selection, with a service limited to a single path for one or a few requests. Conversely, when global topology knowledge is considered, more complex metrics such as channel capacity, E2E fidelity, and purification costs come into play, allowing for a deeper optimization of routing decisions for multiple paths and multiple concurrent requests. This underscores the importance of detailed topology knowledge (more than in classical) on quantum routing efficiency and features.
Proactive routing, with an emphasis on the expected entanglement rate as a key metric, allows for more optimization of resources. Advanced algorithms, such as k-shortest paths and multi-commodity flow are suitable to support multi-request scheduling and multi-path routing, further highlighting the adaptability of the proactive routing approach. 

Regarding route installation and forwarding, the process of establishing E2E entanglements goes beyond the simplicity of traditional networking tasks, such as populating forwarding tables with route prefixes and matching routes based on the longest prefix. Instead, it involves the orchestration of point-to-point elementary entanglements to create E2E connections, a task that is inherently more complex and stateful. The state information and signaling required to handle this process is, therefore, significantly more challenging than basic forwarding protocols.
The swapping policies and strategies, relatively less explored, present their own sets of challenges and strategic considerations. The degradation of entanglement fidelity with each swapping operation, alongside the varied impact of swapping strategies on E2E entanglement throughput and fidelity, emphasizes the need for a deeper understanding of these mechanisms and their implications for quantum routing tasks. Our intuition is that the degree of network heterogeneity, such as node capabilities and link quality, would reflect in the complexity of the swapping scheme, which can range from arbitrarily pre-defined orders (e.g., left-to-right), to topology-adapted orders, to predefined local (e.g., swap asap) or quasi-local (e.g., wider neighbor first) policies, to dynamic per-path strategies.

\section{Protocols for Entanglement Routing}\label{sec:protocols_entanglement_routing}
The routing and forwarding approaches discussed above need a combination of various protocols to be effectively implemented in a quantum network. As mentioned, the quantum forwarding phase does not have a direct counterpart in classical networks due to the coordination and processing necessary for swapping and purification.
However, route computation and installation and the resource reservation processes share many similarities with classical networking protocols, and quantum routers behave similarly to classical routers in many ways \cite{shi2020concurrent, cicconetti2021request}.
Overall, routing protocols are used to collect network state information and instruct routers on the best downstream and upstream entanglements to create in order to establish E2E entanglements.

In this section, we adopt an engineering perspective by exploring how classical networking protocols and architectures might be adapted to support the entanglement routing methods discussed earlier. Although classical routing protocols cannot be directly applied to entanglement routing, this discussion aims to present a realistic view of the practical strategies that could be employed to create a suite of protocols for routing entanglements in quantum networks.

\subsection{Distributed Packet Switching}
Quantum and classical wired networks share similarities through their physical topology formed by routers and optical links. In quantum networks, however, the connections between routers through entanglements represent a dynamic and probabilistic logical topology, contrasting with the relatively stable link connectivity of classical networks. Notably, the physical topology in quantum networks might also exhibit some dynamism due to variations in fiber quality under environmental effects (e.g., temperature) which in turn affects the entanglement creation success and thus link quality.

Distributed wired packet-switching networks rely on various protocols to learn routes to destinations. Typical routing protocols include Border Gateway Protocol (BGP) where routes are selected based on deterministic criteria such as distance or origin from the received route advertisements. BGP provides path selection without global knowledge of the topology and can scale to large networks with many routes. 
Open Shortest Path First (OSPF), on the other hand, is used for routing in smaller networks, where routers share information about their links with all other routers in the network. This ensures all routers have a consistent view of the global topology, and each router uses Dijkstra’s algorithm to compute paths. These protocols are sufficient to realize a functioning packet-switching classical network providing best-effort forwarding. 

Quantum routing algorithms that rely on simple cost metrics such as link entanglement throughput with a global physical topology knowledge can be supported by adapting the OSPF protocol in proactive scenarios with limited efficiency, such as a small number of concurrent requests with a single path per request without fidelity guarantees \cite{chakraborty2020entanglement}. However, many routing algorithms assume that the network state shared among routers needs to extend beyond simple link cost (e.g., link fidelity \cite{liu2024linkselfie}, E2E capacity \cite{vardoyan2024capacity}) to select high-quality paths, and may require more frequent updates than in classical networks \cite{ghaderibaneh2022efficient, dai2020optimal}. This is the case in QBGP \cite{qbgp, qbgp_code} where the BGP protocol is adapted to entanglement routing, but the path selection rules had to be replaced by a more dynamic decision process. At the request time, QBGP adapts its path preference according to the link fidelity using an online learning algorithm instead of deterministic rules.

Routing solutions can also be borrowed from the wireless world. Similar to quantum networks, routing in wireless multi-hop networks such as wireless sensor networks often works with partial knowledge of the network topology and handles intermittent connectivity. The Routing Protocol for Low-Power and Lossy Networks (RPL) is designed for low-power and lossy networks (LLNs). RPL maintains a distributed Destination-Oriented Directed Acyclic Graph (DODAG) to organize nodes in a tree-like structure where each node has a single path to the root. 
Asynchronous Entanglement Routing \cite{yang2024Asynchronous} adopts a similar approach to RPL for distributed entanglement routing in the reactive scheme. The logical topology is established and maintained as a DODAG and follows a path computation approach similar to RPL by tracing the path through the most current DODAG’s root. When a node receives an E2E entanglement request, it determines the path to the destination and attempts swapping between its previous and next entanglement links. If the swapping succeeds, the node forwards the request to the next hop, which repeats the same process until the destination is reached. Otherwise, the node waits until necessary entanglements are available to continue. As a reactive scheme, this approach shows a promising distributed protocol for entanglement routing but still requires high entanglement coherence times and generation rates to be effective. 

With distributed protocols, the E2E entanglement request is forwarded to the next hop parallel to the entanglement-link creation (similar to opportunistic routing \cite{gyongyosi2019opportunistic}). Swapping measurements are also routed along the path to the end-nodes \cite{qbgp, gyongyosi2017entanglement} (see Figure \ref{fig:swapping_protocol}). Hence, this scheme requires memory-based swapping as entanglement creation along the path is asynchronous. 
Given the various swapping schemes available in memory-based scenarios, the relatively limited metrics supported in distributed routing may be overcome with a more advanced swapping policy.

In another routing approach discussed below, quantum versions of packet-switching distributed routing protocols can serve to pre-share network state information among the routers to compute paths for more robust and resilient protocols, similar to the classical virtual circuit switching architectures.

\subsection{Virtual Circuit Switching}
\begin{figure}[h]
\centering
\includegraphics[width=0.6\linewidth]{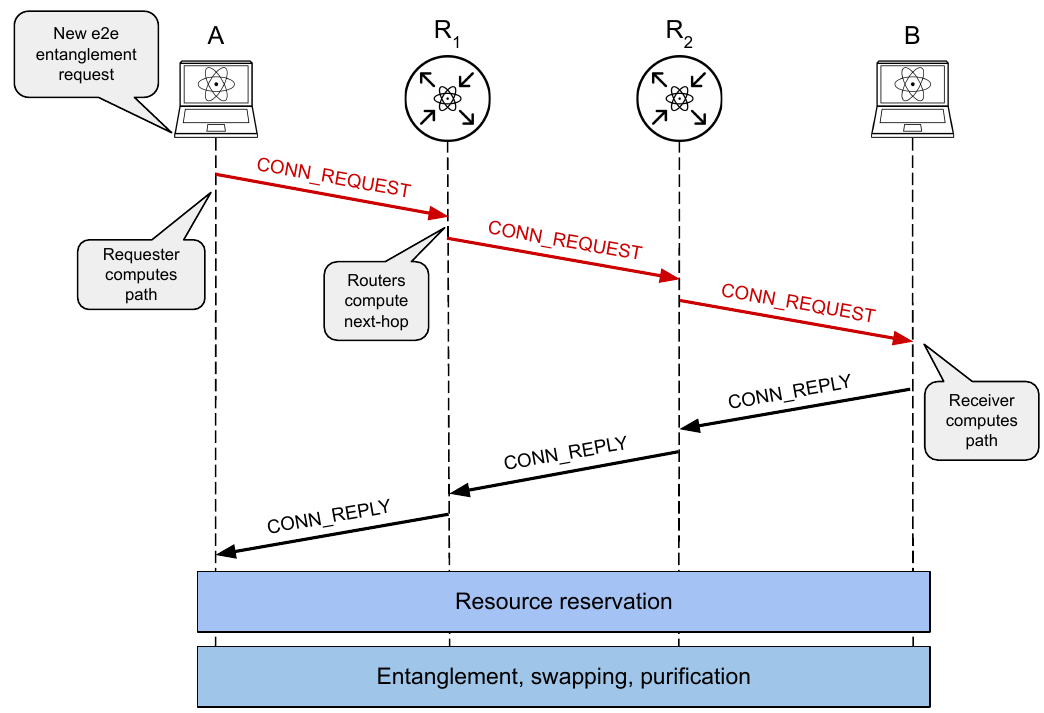}
\caption{\label{fig:routing_proactive_distr_protocol}Distributed E2E entanglement generation protocol (adapted from \cite{bacciottini2024redip}).}
\end{figure}

Virtual Circuit Switching (VCS) is an architecture where a path is established before data transfer begins. This connection-oriented approach is common in classical networks through data plane protocols such as Multiprotocol Label Switching (MPLS), which forwards packets based on their labels and preinstalled forwarding instructions. 
Generalized Multiprotocol Label Switching (GMPLS), extends the notion of labels to represent packet flows, optical fibers or wavelengths, timeslots, etc. The label-switched path scheme represents a natural choice to manage E2E entanglement paths \cite{rfc9340}.
The connection-oriented E2E entanglement generation in a distributed network is illustrated in Figure \ref{fig:routing_proactive_distr_protocol}.

VCS protocols use signaling mechanisms to negotiate paths and install labels. Resource Reservation Protocol Traffic Engineering (RSVP-TE) is a distributed signaling scheme that uses paths computed via distributed routing such as OSPF with traffic engineering extension (OSPF-TE). OSPF-TE distributes information about link attributes such as bandwidth, delay, administrative constraints, and available resources, and can use constrained shortest path first (CSPF) to take these factors into account to compute traffic-engineered paths for RSVP-TE.

With a similar architecture, the RuleSet protocol \cite{Van_Meter_2022} uses a two-pass communication initiated from a requester node to provision E2E entanglements. The outbound request is forwarded to the responder based on a standard next-hop table while collecting information about the nodes and links it traverses. The responder node replies with a set of rules for each intermediate node along the path. A rule has conditions (e.g., entanglement created, timeout) and corresponding actions (e.g., measure qubit). The RuleSet approach can be seen as a reversed SR mechanism that uses an outbound request to avoid the drawbacks of relying on globally pre-shared topology state via quantum versions of distributed routing protocols \cite{van-meter-qirg-quantum-connection-setup-01}.

Building on a similar architecture to RuleSet, the REDiP protocol \cite{bacciottini2024redip} provides a way to specify swapping orders (as node ranks) and purification instructions (i.e., purification rounds to be performed at each rank) during the connection phase. The authors also provide elaborated guidelines and metrics on defining swapping and purification strategies to meet E2E fidelity requirements and maximize throughput.

Segment Routing (SR) is an alternative to RSVP-TE in which routing instructions are carried within packet headers, eliminating the need for a per-flow state in intermediate routers. This design choice reduces the need for additional signaling protocols and simplifies the architecture. SR leverages a stateless model, with the head-end router managing paths from either centralized or distributed routing.

Quantum-SR \cite{zhang2023segmentrouting} adapts the Segment Routing mechanism to provision E2E entanglement paths using centralized routing to select paths based on topology state shared via distributed routing protocols such as OSPF.

Entanglement distribution protocols proposed in this direction are considered more adapted for the current and near-term technologies (e.g., memory-less repeaters, short coherence time, limited qubits) \cite{rfc9340}. The path reservation process allows for both memory-less and memory-based swapping. Moreover, by reserving the entire path beforehand, this approach can support more elaborated path computation algorithms with E2E entanglement throughput \cite{ghaderibaneh2022efficient}, path length \cite{chakraborty2020entanglement}, and multipath \cite{vardoyan2024capacity} while providing routers with the capability to enforce specific paths through the network.

\subsection{Software Defined Networking}
Given the complexity of quantum path selection and the diversity of parameters it considers, many routing approaches are developed assuming a Software Defined Networking (SDN) architecture for various scenarios \cite{nguyen2022multiple, aguado2020qkdsdn, aguado2019engineering, chiti2021towards, picchi2020satellite}. This brings to quantum networks a familiar and easy-to-design framework which positively impacted classical networking. The controller can dynamically select paths in response to changing network conditions such as link quality, demands, congestion, or failures.

SDN is most commonly considered in the reactive entanglement routing scenario, where elementary entanglement outcomes are collected by the controller which uses them to satisfy E2E entanglement requests. Such architecture requires a quantum technology that provides high enough entanglement coherence time combined with a high enough entanglement generation success probability to satisfy requests.
In the proactive scenario, the SDN controller collects the network topology and provides path computation in the same way Segment Routing can rely on centralized routing to install forwarding instructions \cite{zhang2023segmentrouting}. The controller can also be used for time-scheduling the E2E entanglements to provision similar to the Quantum-adapted Time Sensitive Networks architecture envisioned in \cite{bush2021perspective}.

As mentioned, factors other than throughput per fidelity can be relevant in path selection, such as channel capacities and their distribution (especially for swapping strategies), multiplexing, and routing algorithm fairness. Supporting these factors requires more coordination between nodes and a global view of the network with complex routing algorithms such as Linear Programs or Machine Learning models \cite{zhao2021redundant, zeng2022multi}.

Moreover, distributed traffic engineering is usually sub-optimal in classical networks \cite{jain2013b4, rexford2002traffic} and one could expect this to be exacerbated in quantum networks. 
Distributed routing protocols and virtual circuit mechanisms could certainly benefit from the global view of the SDN architecture for centralized traffic engineering. Such a hybrid architecture is adopted in Google’s B4 SD-WAN \cite{hong2018b4andafter}, where an SDN controller is used to improve traffic engineering and resource management while keeping a distributed BGP routing among the devices \cite{khorsandroo2021hybridsdn}.
Similarly in quantum networks, rather than attempting to dynamically control the network’s logical topology and E2E entanglement requests, the SDN controller could be leveraged to understand communication demands, manage network resources, predict failures, select adapted swapping policies \cite{haldar2024fast}, and efficiently pre-distribute entanglements in the virtual routing scheme \cite{gyongyosi2019adaptive, ghaderibaneh2022predistribution}.

\subsection{Swapping}
\begin{figure}[h]
\centering
\includegraphics[width=0.5\linewidth]{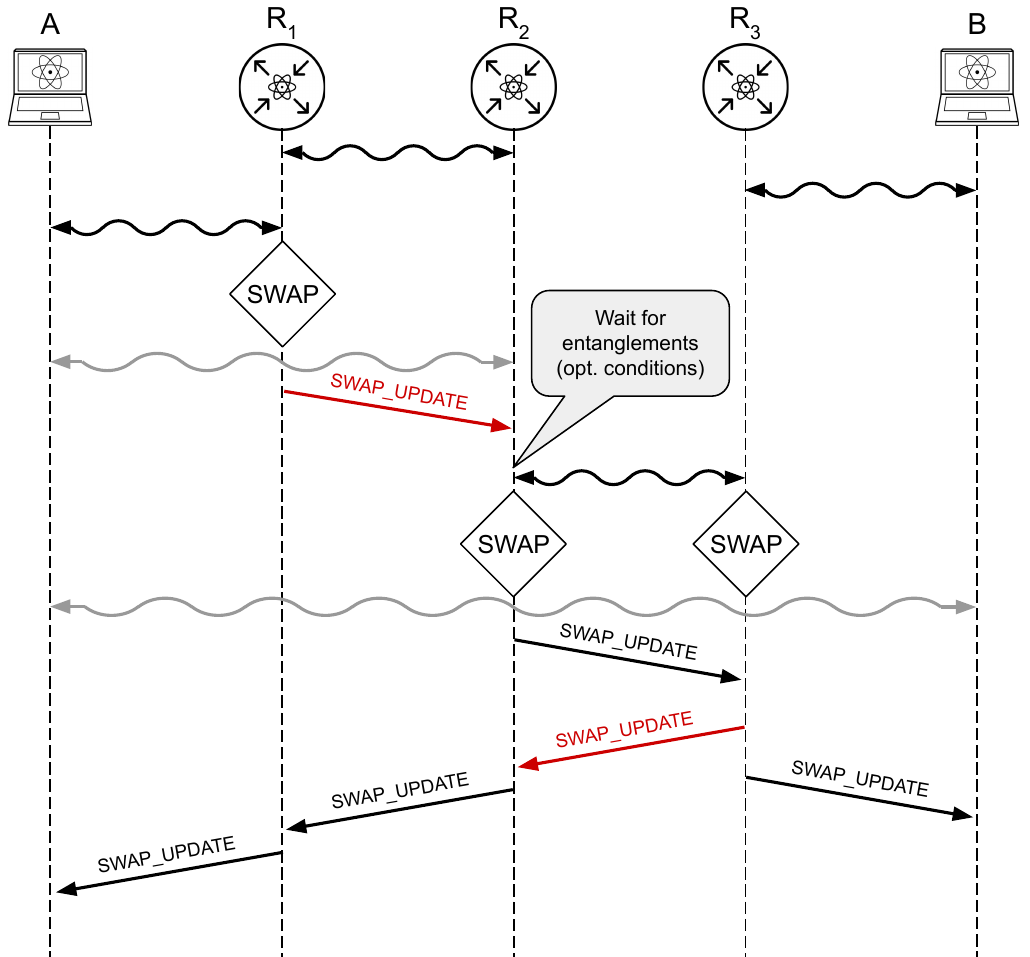}
\caption{\label{fig:swapping_protocol}Swapping process along a path (adapted from \cite{bacciottini2024redip}). Dark wavy lines indicate elementary entanglements and light wavy lines indicate post-swapping entanglements.}
\end{figure}

Entanglement swapping can be viewed as a distributed computation task executed once an E2E path is reserved. Figure \ref{fig:swapping_protocol} illustrates the implementation of a \textit{swap-as-soon-as-possible} policy along a path of three routers between end-nodes A and B as commonly envisioned \cite{kozlowski2020designing, bacciottini2024redip}. 
Router R1 is the first to get two neighboring entanglements available, swaps immediately, and sends a \texttt{SWAP\_UPDATE} to R2 containing the swapping measurement. 
Upon receiving the message, R2 waits until an upstream entanglement is available and swaps it with the entanglement specified in the message received from R1. 
Then, R2 adds the swapping measurement to R1’s \texttt{SWAP\_UPDATE} and forwards it to R3. When R2 receives the SWAP\_UPDATE from R3 (which swapped its local entanglements in the meantime), it can immediately include its swapping measurement and forward it to R1. 
This process continues until both \texttt{SWAP\_UPDATE}s are delivered to end nodes A and B.

Other swapping policies or orders can be implemented in this scheme using specific swapping conditions instead of “\textit{as soon as upstream and downstream entanglements are available}”. 
For example, a swapping condition in R2 specifying the upstream entanglement only with B will prevent R2 from swapping until it receives a SWAP\_UPDATE from R3, thereby achieving a \textit{doubling} swapping policy.

\section{Operating Quantum Networks}\label{sec:classical_control}
After exploring routing and forwarding approaches, path computation algorithms, and practical protocol implementations, we turn our attention to the design and management of a quantum network. 
In the following, We discuss the classical control of quantum networks, propose key metrics for network monitoring, and the introduce expected common types of failures.

\subsection{Classical Control}
Classical communication and data processing are an indispensable part of quantum protocols like HEG, HEP, swapping, and teleportation, as well as for device configuration, route installation, and failure recovery. These classical operations constitute the control plane of quantum networks \cite{kozlowski2020p4, dasari2016openflow, illiano2021impact}. 

To implement the necessary control plane protocols, quantum networks require at least one classical channel alongside each quantum channel in the network. This classical channel can be realized on the same physical media as the quantum channel (e.g. using wavelength or time-division multiplexing) or can live on a physically independent channel.

Although the control planes of quantum and classical networks serve similar purposes, they differ significantly in their abstraction and implementation.
For example, while classical networks use abstractions based on packet headers and interfaces to express matching conditions and forwarding actions, the instructions in quantum networks include receiving/sending measurements and creating/swapping entanglements.

Due to the fundamental way quantum signals are generated, processed, and consumed, there are much tighter time and latency constraints placed on the quantum network control plane that are important to highlight. 
In near- and medium-term quantum networks, the fidelity of generated entanglements decays quickly over time, leading to coherence times on the order of milliseconds to seconds depending on the quality of the quantum memories storing the entanglement. Additionally, quantum networks rely on the indistinguishability of photon pairs requiring tight clock synchronization and timestamp accuracy. These latency constraints imply that not all classical control communications can be passed through an entire TCP/IP stack -- a frequent design in today’s Internet -- nor can all classical data processing be executed on traditional CPUs. As such, portions of the control plane may necessitate leveraging low-latency communications technology and fast data processing on FPGAs or ASICs.
Where possible, implementations should leverage classical communications technologies (e.g., standardized protocols and interfaces) to maximize compatibility with standard telecommunications infrastructure and devices. However, due to the constraints posed by quantum mechanics and quantum communications protocols, implementations of the control plane will employ quantum-network-specific functionality.

\begin{figure}[!t]
\centering
\includegraphics[width=0.85\linewidth]{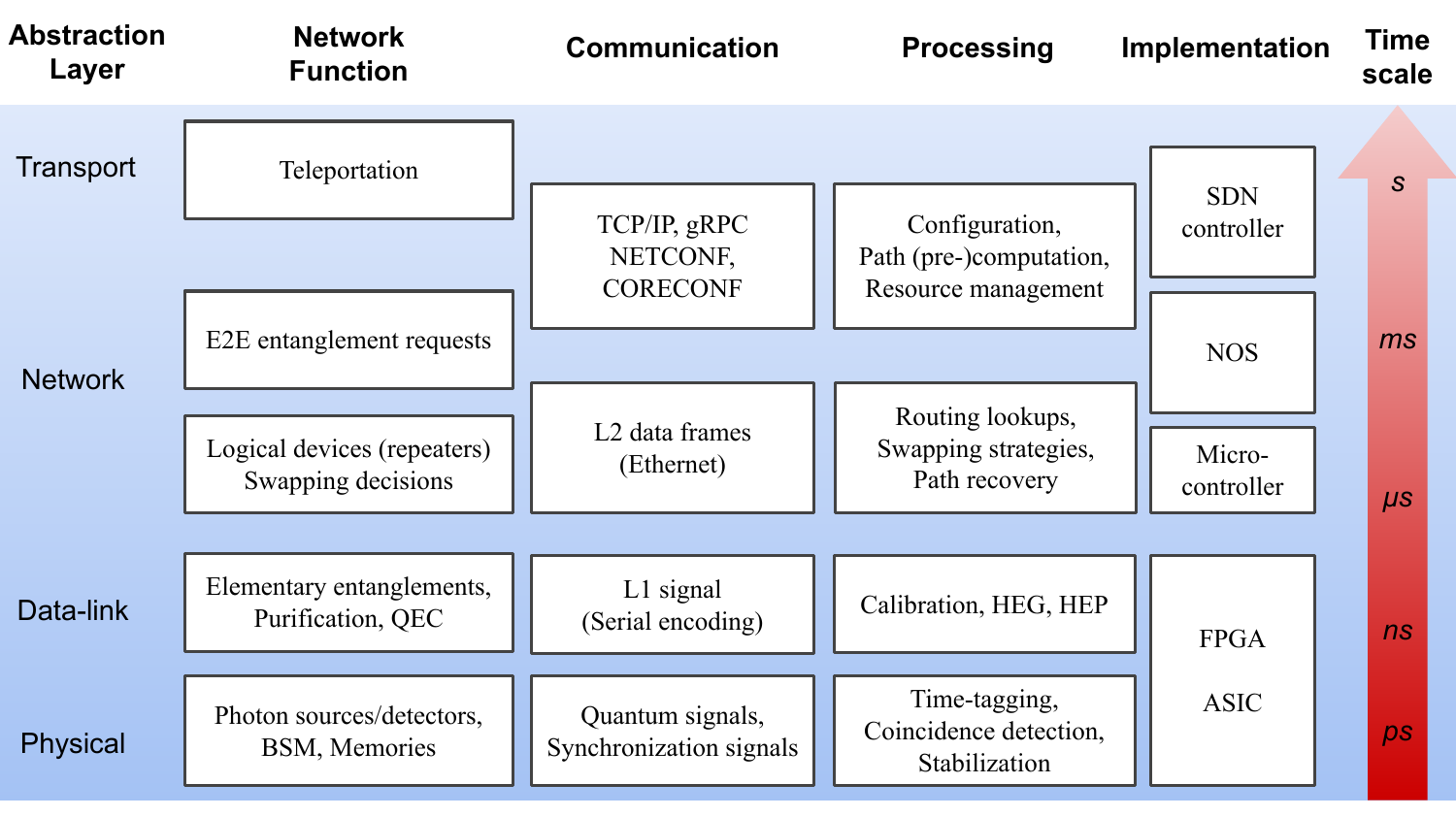}
\caption{\label{fig:quantum_classical}Abstractions, communications, and classical processing for operating a quantum repeater network with SDN principles: implementation and timescale.}
\end{figure}

Figure \ref{fig:quantum_classical} illustrates a layered abstraction of quantum network functions, communications, classical processing, and timescales for operating a quantum repeater network. 

Several types of events in quantum networks need to be communicated in real-time and processed with picosecond-level accuracy such as time-stamping photodetector clicks, coincidence count data, and heralding signals.
Portions of the classical communication and processing comprising the control plane can be implemented in traditional software on CPUs and microcontroller hardware (top layers in Figure \ref{fig:quantum_classical}). 
However, to meet the latency constraints imposed by the more time-sensitive network functions (bottom layers in Figure \ref{fig:quantum_classical}), implementation requires high-performance hardware such as FPGAs, and the time-sensitive communications need low-latency data formatting such as direct L1 encoding or a more-efficient L2 data framing. 

\begin{table}[!h]
    \renewcommand{\arraystretch}{1.3}
    \newcolumntype{?}[1]{!{\vrule width #1}}
    \definecolor{titlerowcolor}{rgb}{0.74, 0.83, 0.9}
    \resizebox{\columnwidth}{!}{%
    \centering
    {\rowcolors{2}{}{lightgray}
    \begin{tabular}{|>{\centering\arraybackslash}m{0.12\linewidth}?{0.5pt}>{\small\centering\arraybackslash}m{0.23\linewidth}|>{\small\centering\arraybackslash}m{0.23\linewidth}|>{\small\centering\arraybackslash}m{0.14\linewidth}|>{\small\centering\arraybackslash}m{0.15\linewidth}|}
    \hline
         \rowcolor{titlerowcolor}\textbf{Metric}& \textbf{Definition}& \textbf{Indication}& \textbf{Classical unit}& \textbf{Quantum unit}\\ \hline
         Throughput&Effective rate at which data is successfully transferred over the network&Informs about the efficiency of congestion control, the reliability of the network, and the capacity utilization.&Bits per second (bps)&E2E entanglements with a given fidelity per second (eps)\\ \hline
         Uptime&Duration or percentage of time that a network is operational and available for use&Indicates the overall reliability and availability of the network& \% & \% \\ \hline
         Bandwidth&Maximum rate at which data can be transferred over a network connection&Suggests the capacity of the network infrastructure and sets the upper limit on the potential throughput&Bits per second (bps)&Link/E2E capacity (eps), Link fidelity \\ \hline
         Error Rate&The frequency of errors in data transmission over the network&Provides insights into the integrity and efficiency of data transmission, and potentially external undesirable effects&Bit error rate (BER) or Frame error rate (FER)&Qubit Error Rate (QBER)\\ \hline
         Data Loss&Percentage of data units that are sent, but fail to reach their intended destination&Highlights problems with traffic congestion, potentially inadequate routing, or hardware issues&Packet Error Rate (PER)&Qubit/Photon Loss Rate\\ \hline
         Latency&Time taken for data units to travel from sources to destinations across the network&Can be indicative of network congestion, insufficient resources, or poor routing&Milliseconds&Milliseconds\\ \hline
         Jitter&Variability in the latency of data unit delivery within the network&Indicates issues with network stability and consistency&Milliseconds&Milliseconds/ Nanoseconds\\ \hline
         Redundancy&The inclusion of extra or duplicate network devices, connections, or pathways&Informs about the network's fault-tolerance, reliability, and overall robustness&Depends on verification tools&Depends on verification tools\\ \hline
    \end{tabular}}}
    \vspace{1ex}
    \caption{\centering Comparison of analogous metrics for classical networks and quantum networks}
    \label{tab:metrics}
\end{table}

\subsection{Performance Metrics}
The physical processes involved in quantum communication protocols exhibit properties and sensitivity that bring a set of performance metrics that differ from those used in classical networks. The main differences between the two network technologies are rooted in the fact that quantum network operations are probabilistic, and, as such, quantum networking protocols are riddled with randomness and uncertainty. Thus, when defining metrics for quantum networks, one must take into account these unique properties and expected failures.

Classical packets can be buffered for theoretically infinite time on any router for future transmission. However, quantum signals cannot be amplified or copied, nor can they be stored for indefinite amounts of time due to their susceptibility to decoherence -- buffering entanglements is limited by the quantum memory coherence time. Thus, all entanglements along a path must be successfully established within a relatively short time period if they are to be used to create the E2E entanglement. 

Table \ref{tab:metrics} outlines the parallels and divergences in how critical network metrics are conceptualized and quantified across classical and quantum domains.
Throughput in classical networks is measured in bits per second to evaluate the efficiency of congestion control, network reliability, and capacity utilization. In contrast, throughput in quantum networks corresponds to the number of entangled pairs with a certain fidelity delivered per second to the end nodes.
Bandwidth in classical networks refers to the maximum rate of data transfer, usually in bits per second, setting the upper limit on throughput and indicating the physical infrastructure’s capacity.
The bandwidth in quantum networks is a more complex metric, measured as link or E2E capacity \cite{gyongyosi2018survey_capacity, vardoyan2024capacity}, which may involve the link fidelity and the underlying architecture of quantum routers (e.g., memory size and management, multiplexing).
Error rate, another critical metric, accounts for the frequency of transmission errors. Classical networks often report this as a bit or frame error rate, while quantum networks use the qubit error rate. Similarly, data loss, which indicates the percentage of data failing to reach its destination, is an important metric in assessing network robustness against congestion, routing inefficiencies, or hardware issues. Quantum networks might measure this as qubit or photon loss rate. To account for quantum-specific phenomena like decoherence, one can evaluate novel metrics such as the rate of entanglement links not used by the routing or forwarding protocol.
Latency and jitter address the time efficiency and consistency of data transmission. Latency is the delay before data reaches its destination; it is crucial for performance analysis and is usually measured in milliseconds. Jitter, or the variability in latency, highlights network stability and is also measured in milliseconds, although quantum networks may use nanoseconds to picoseconds due to the high precision required in quantum operations.
Lastly, redundancy emphasizes the network's fault tolerance by incorporating extra or duplicate devices and links. Classical networks rely on control plane simulation tools \cite{matt2023batfish} and formal data plane verification tools to verify routing properties such as multi-path inconsistency, loop-free routing, and traffic isolation \cite{zhang2022dna}. 
Simulating even modestly sized quantum networks demands computational resources that exceed conventional capabilities due to the need for controlling individual qubits instead of packets/flows, and the need for detailed modeling of rapidly changing qubit states \cite{chen2023simqn}.

\subsection{Network Failures}
Network design and operation must consider operational and infrastructural failures as the two major types of network issues. 
Operational failures are inherent to the network technology and are expected to occur frequently due to issues such as photon loss, unsuccessful entanglement swapping operations, or entanglement expiry due to decoherence. These failures should be factored into the network design, for example when computing the paths and establishing communications. 
Infrastructural failures are unexpected and may happen when parts of the physical infrastructure are damaged or tampered with. For example, links may become unusable due to fiber breaks, requiring an alternate path for recovery. The support of these failures depends on the desired network reliability and robustness.

\begin{table}[!t]
    \renewcommand{\arraystretch}{1.3}
    \newcolumntype{?}[1]{!{\vrule width #1}}
    \definecolor{titlerowcolor}{rgb}{0.74, 0.83, 0.9}
    \definecolor{highcolor}{rgb}{0.97, 0.51, 0.47}
    \definecolor{moderatecolor}{rgb}{1.0, 0.65, 0.0}
    \definecolor{lowcolor}{rgb}{0.98, 0.93, 0.36}
    \definecolor{notafailurecolor}{rgb}{0.56, 0.93, 0.56}
    \definecolor{lightgray}{gray}{0.9}
    \resizebox{\columnwidth}{!}{
    \centering
    \begin{tabular}{?{1.2pt}>{\centering\arraybackslash}m{0.13\linewidth}?{1.15pt}>{\centering\arraybackslash}m{0.18\linewidth}|>{\centering\arraybackslash}m{0.18\linewidth}?{1.15pt}>{\centering\arraybackslash}m{0.1\linewidth}|>{\centering\arraybackslash}m{0.1\linewidth}?{1.15pt}>{\centering\arraybackslash}m{0.14\linewidth}?{1.15pt}} \hlinewd{1.15pt}  
     \rowcolor{titlerowcolor}\textbf{Failure} &\textbf{Classical Network Causes} &\textbf{Quantum Network Causes} &\textbf{Classical Sensitivity} &\textbf{Quantum Sensitivity} &\textbf{Quantum Mitigations}\\ \hlinewd{1.15pt}        
     \cellcolor{lightgray}Calibration Error & \multicolumn{2}{>{\centering\arraybackslash}m{0.36\linewidth}?{1.15pt}}{Misaligned system settings, improper device configuration, link calibration errors}&\cellcolor{moderatecolor}Moderate&\cellcolor{highcolor}High&Photon interference maximization (recalibration)\\ \hline  
    \cellcolor{lightgray}Device Desynchronization& \multicolumn{2}{>{\centering\arraybackslash}m{0.36\linewidth}?{1.15pt}}{Clock drift, synchronization protocol failures}&\cellcolor{moderatecolor}Moderate&\cellcolor{highcolor}High&Periodic or event-driven resynchronization\\ \hline
     \cellcolor{lightgray}Routing Errors&{\small Misconfigurations, protocol incompatibilities, software bugs}&{\small Misconfigurations, protocol incompatibilities, software bugs, inaccurate fidelity estimation}&\cellcolor{moderatecolor}Moderate&\cellcolor{moderatecolor}Moderate&Alternate path installation, routing verification\\ \hline  
     \cellcolor{lightgray}Forwarding Errors& \small Software failures in switches and routers& \small Entanglement and swapping failures, software failures in switches and routers&\cellcolor{lowcolor}Low&\cellcolor{highcolor}High&Telemetry, robust control, path recovery\\ \hline 
     \cellcolor{lightgray}Security Breach&\small Unauthorized packet interception and modification&\small Photon interception, modification, and measurement&\cellcolor{highcolor}High&\cellcolor{notafailurecolor}High (feature)&QBER estimation, authentication\\ \hlinewd{0.8pt}
     Signal Interference& \multicolumn{2}{>{\centering\arraybackslash}m{0.36\linewidth}?{1.15pt}}{Noise, cross-talk}&\cellcolor{notafailurecolor}Not a Failure&\cellcolor{highcolor}High&Wavelength allocation, time-division muxing\\ \hline
 Signal Attenuation& \multicolumn{2}{>{\centering\arraybackslash}m{0.36\linewidth}?{1.15pt}}{Fiber length, quality, environmental factors, fiber coupling inefficiency}&\cellcolor{notafailurecolor}Not a Failure&\cellcolor{highcolor}High&High-rate sources, muxing, repeaters\\ \hline  
 Photon Depolarization or Dephasing& \multicolumn{2}{>{\centering\arraybackslash}m{0.36\linewidth}?{1.15pt}}{Environmental interactions,\newline thermal fluctuations}&\cellcolor{notafailurecolor}Not a Failure&\cellcolor{highcolor}High&Calibration \& compensation, QEC\\ \hline  
 Loss of Connectivity& \multicolumn{2}{>{\centering\arraybackslash}m{0.36\linewidth}?{1.15pt}}{Fiber cuts, hardware failures, Denial-of-Service attacks}&\cellcolor{highcolor}High&\cellcolor{highcolor}High&Multiple path installation, disjoint paths\\ \hline  
 Hardware Failure& \multicolumn{2}{>{\centering\arraybackslash}m{0.36\linewidth}?{1.15pt}}{Physical damage, device malfunction, wear and tear, manufacturing defects}&\cellcolor{highcolor}High&\cellcolor{highcolor}High&Device \& link redundancy\\ \hlinewd{1.15pt}
    \end{tabular}}
    \vspace{1ex}
    \caption{\centering Comparison of analogous operational (light gray) and infrastructural (white) failures, causes, and sensitivity for classical networks and quantum networks}
    \label{tab:failures}
\end{table}

Table \ref{tab:failures} summarizes the key failures in quantum networks with their causes, sensitivity, and mitigation measures. The failures counterpart in classical networks is provided for reference and comparison. 
The sensitivity is labeled as \textit{low}, \textit{moderate}, or \textit{high} considering the likelihood of failure, its impact on the overall functioning, and the complexity of the impacted underlying system.
Calibration errors and desynchronization, caused by misaligned system settings or clock drifts, have a high sensitivity in quantum networks due to the need for precise timing in operations. 
Routing failures, stemming from misconfigurations or protocol incompatibilities, have a moderate sensitivity in both network types, mitigated by alternate path installation and routing verification. 
However, forwarding failures are more critical in quantum networks due to the coordination, signaling, and complex decision-making process of swapping. In contrast, forwarding in classical networks is a much simpler approach that benefits from caching and retransmission features.
Security breaches, although serious in both networks, are uniquely leveraged in quantum networks for securing communications by detecting photon interceptions.
Lastly, some infrastructural issues like signal attenuation and photon depolarization usually do not affect classical communications, whereas, in quantum networks, they represent big challenges necessitating high-quality equipment, error correction, and strict signal isolation.
Naturally, loss of connectivity and hardware failures have a high impact on both network technologies, which can be addressed with measures such as multiple disjoint paths and hardware redundancy.

\section{Open Questions}\label{sec:open_questions}
\subsection{Network Design and Topology}
Identifying suitable topologies for quantum networks remains unresolved, as their effectiveness varies based on specific application needs and technology limitations \cite{gyongyosi2018topology, choi2023scalable}.
Moreover, the challenge of ensuring compatibility across the diverse components of inherently heterogeneous intermediate-scale quantum networks is considerable \cite{kumar2023routing}. This diversity impacts the abstraction level of routing models considering the hardware, operating wavelengths, physical link architectures, and multiplexing schemes to ensure efficient interconnection of quantum devices and networks, facilitate maintenance and upgrades, and minimize communication disruptions and inefficiencies.

Routing protocols and architectures could be designed to ensure the continuity and interoperability of first-generation quantum networks with upcoming (different) generations. This may concern the appearance of second and third-generation repeaters (i.e., shifting from circuit-switched to packet-switched quantum paradigm) \cite{munro2022designing} or the integration of all-photonic repeaters \cite{azuma2015allphotonic, benchasattabuse2024architecture}. 
Quantum network design may benefit from the Recursive InterNetwork Architecture (RINA), an approach that treats the network as a single, recursive programmable layer, to manage heterogeneity and scalability and envisioned in \cite{Van_Meter_2022}.

\subsection{Routing and Forwarding Interdependence}
The complexity of routing and forwarding, each composed of multiple functions and considering various network parameters, raises the question of their interdependence. As discussed, many routing (resp. forwarding) approaches must include forwarding (resp. routing) considerations in their models to achieve higher performance. Alternatively, other studies assume independent routing (resp. forwarding) in the proposed design. 
Therefore, quantum network designs might either integrate routing and forwarding closely, akin to traditional networking, or distinguish them as separate functions linked by a standard interface, as seen in SDN-based quantum networks \cite{aguado2019engineering, chiti2021towards}. Additionally, the underlying multiplexing scheme for managing E2E paths remains relatively unexplored, as well as its consideration in path computation and swapping.

\subsection{Evaluation and Prototype Implementation}
The absence of a reference architecture for quantum repeaters introduces difficulties in standardizing control and configuration interfaces, crucial for ensuring interoperability and scalability across quantum networks. 
Moreover, the lack of specialized hardware platforms further impedes the rapid prototyping and testing of repeater technologies. This makes it difficult to define clear assumptions on underlying mechanisms for repeaters (e.g., memory management, multiplexing, heterogeneity) which in turn leads to uncertainties in designing routing models, signaling mechanisms, and interoperable protocols.

Evaluating proposed routing schemes through simulations or analytical methods \cite{wu2021sequence, satoh2022quisp}, while useful, does not fully capture the complexities of a real-world implementation. Moreover, current tools are not designed for quick evaluation of dynamic routing protocols and swapping strategies over various abstractions of entanglement distribution technologies. Consequently, many entanglement routing approaches are considered in isolation, and not evaluated under realistic network scenarios such as random topologies with heterogeneous networks.

A more comprehensive approach, possibly involving an emulation environment for quantum repeater networks, could facilitate the development and evaluation of effective routing protocols, device APIs, network controllers, and management systems \cite{mininet, optical-mininet}.

Traditional network simulation and emulation methods may be inefficient for accurately modeling and running a virtual quantum network. New evaluation paradigms such as digital twins \cite{zhuge2023dt, wang2024dt} and generative artificial intelligence \cite{huynh2024genai_phy} may be considered to address these challenges.

\subsection{Reliability and Availability}
The fragility of the processes involved in entanglement routing underscores the critical need for robust mechanisms to ensure reliability and availability under the probabilistic nature of quantum communications. Proposed routing and forwarding mechanisms often neglect E2E entanglement distribution failures, and mostly ignore related operational and infrastructural aspects such as classical communication and computation issues and overhead.
Investigating methods to enhance the robustness of quantum networks is crucial. This includes developing mechanisms for efficient path recovery, access control, and other strategies to ensure uninterrupted service and secure communications.

\subsection{Metrology and Hardware Characterization}
Given the unprecedented precision required in synchronization and the susceptibility of quantum states to noise, identifying and prioritizing the relevant metrology metrics is crucial for enhancing the design and efficacy of quantum networks.
For example, relevant metrics related to fibers in quantum networks are already emerging, such as polarization stability, photon time-of-travel, and noise. These are equivalent to, for example, the detection efficiency, timing jitter, or dark counts of a single photon detector. 

The advancement of measurement tools and protocols will not only impact the standardization of quantum network measurements, but also drive innovative solutions to address practical challenges in real-world quantum deployments. 
This raises the critical question of how best to leverage these metrics and tools to refine entanglement routing schemes, ultimately optimizing quantum network performance and resilience.

\section{Conclusions and Perspectives}
\label{sec:conclusion}
In concluding this survey, it is apparent that entanglement routing is rich with theoretical innovations and practical challenges. Our exploration ties quantum networking concepts to classical terminology to provide a grounding framework that makes these complex concepts more accessible and aligned with familiar networking paradigms. The separation of routing and forwarding, and the modular approach to network design reflected in our proposed taxonomy, could help understanding and organizing the domain of quantum networks.

The formal definition of the quantum routing problem presented is abstract but captures the major aspects necessary for developing entanglement routing models. This is complemented by our discussion on practical deployment and implementation strategies, which brings the theoretical concepts closer to real-world application. The integration of topology knowledge with routing and forwarding mechanisms, alongside the consideration of routing algorithms and metrics, provides a holistic view of current and possible directions in quantum network design.

From our analysis, while it is clear that there is a rich body of work on routing, the area of forwarding within quantum networks remains less explored. The importance of disaggregating control and data planes, a principle that significantly advanced classical networks, remains attractive in quantum networking as it simplifies network operations and enhances flexibility. However, the boundaries of such disaggregation will certainly be different from the ones known in classical networks, and a clear distinction is still to emerge.

Our observations suggest that a software-defined and centralized approach to routing, similar to SDN, could streamline operations and optimize performance by including multiple aspects of the network, such as classical communication delays, physical topology, and unused resources, rather than focusing on a single aspect such as logical topologies. However, the real-time requirements of entanglement distribution and decoherence pose significant challenges, suggesting that a hybrid model incorporating both centralized control and localized decision-making at each node could be more effective. Such approach would mitigate the rapid decoherence of quantum states and the unpredictability of entanglement generation and swapping, while providing efficient communication and resource management. 

The management and monitoring of entanglement-based quantum networks will necessitate more complex strategies that are distinct from classical approaches \cite{cobo2022management}, such as the use of simulation-aided management systems, for example through the digital twins technology \cite{bush2021perspective}, given the impossibility of duplicating qubits, the inadequacy of traditional traffic monitoring techniques, and the huge amount of log data generate by quantum processes.

Furthermore, the necessity for the synchronization operations across the network and the application of multiplexing strategies are essential for deploying viable, reliable, and scalable quantum network \cite{spiess2023sync, bathaee2023multiplexing, vanmilligen2024entanglement}. Additionally, robust security protocols cannot be overstated, as quantum networks introduce new vulnerabilities and attack vectors \cite{satoh2021attacking, zhou2022security}. The development of robust protocols and secure authentication schemes is crucial for protecting these emerging networks against sophisticated attacks.

As we look to the longer-term future, the promise of third-generation quantum repeaters and the exploration of multipartite entanglements are expected to profoundly impact quantum network design and routing protocols. The ongoing research and development in this field promise to lead to significant breakthroughs and innovative solutions that will (re-)shape the evolution of quantum routing.

Finally, it becomes apparent that the traditional concepts and terminologies rooted in classical networking—such as the distinctions between control and data planes, as well as the concepts of forwarding, route metrics and costs may not fully encapsulate the quantum networking characteristics. Quantum networks, particularly the forwarding phase, introduce important shifts in communication paradigms, necessitating a reevaluation of these foundational definitions to better align with all aspects of quantum communications. 
While in this survey we adhere to classical networking terms for familiarity and reference points for classical networking engineers, this also highlights the transformative impact of quantum technologies on communication, urging a reimagined understanding of network operations that bridges the gap between classical precedents and quantum innovations.

\section*{Disclaimer}
Any mention of commercial products or reference to commercial organizations is for information only; it does not imply recommendation or endorsement by NIST, nor does it imply that the products mentioned are necessarily the best available for the purpose.

\bibliography{ref}

\begin{thebibliography}{100}

\bibitem{IBM50qubit}
Ibm raises the bar with a 50-qubit quantum computer.
\newblock \url{https://www.technologyreview.com/2017/11/10/147728/ibm-raises-the-bar-with-a-50-qubit-quantum-computer/}.
\newblock Accessed: 2024-01-11.

\bibitem{Google72qubit}
Google's 72-qubit chip is the largest yet.
\newblock \url{https://www.newscientist.com/article/2162894-googles-72-qubit-chip-is-the-largest-yet/}.
\newblock Accessed: 2024-01-11.

\bibitem{castelvecchi2017china}
Davide Castelvecchi.
\newblock China’s quantum satellite clears major hurdle on way to ultrasecure communications.
\newblock {\em Nature}, 15, 2017.

\bibitem{bennett1993first}
Charles~H. Bennett, Gilles Brassard, Claude Cr\'epeau, Richard Jozsa, Asher Peres, and William~K. Wootters.
\newblock Teleporting an unknown quantum state via dual classical and einstein-podolsky-rosen channels.
\newblock {\em Phys. Rev. Lett.}, 70:1895--1899, Mar 1993.

\bibitem{zukowski1993swapping}
M.~\ifmmode~\dot{Z}\else \.{Z}\fi{}ukowski, A.~Zeilinger, M.~A. Horne, and A.~K. Ekert.
\newblock ``event-ready-detectors'' bell experiment via entanglement swapping.
\newblock {\em Phys. Rev. Lett.}, 71:4287--4290, Dec 1993.

\bibitem{riedmatten2005swapping}
H.~de~Riedmatten, I.~Marcikic, J.~A.~W. van Houwelingen, W.~Tittel, H.~Zbinden, and N.~Gisin.
\newblock Long-distance entanglement swapping with photons from separated sources.
\newblock {\em Phys. Rev. A}, 71:050302, May 2005.

\bibitem{illiano2022quantum}
Jessica Illiano, Marcello Caleffi, Antonio Manzalini, and Angela~Sara Cacciapuoti.
\newblock Quantum internet protocol stack: A comprehensive survey.
\newblock {\em Computer Networks}, 213:109092, 2022.

\bibitem{craddock2024automated}
Alexander~N. Craddock, Anne Lazenby, Gabriel~Bello Portmann, Rourke Sekelsky, Mael Flament, and Mehdi Namazi.
\newblock Automated distribution of high-rate, high-fidelity polarization entangled photons using deployed metropolitan fibers, 2024.

\bibitem{bathaee2023multiplexing}
Marzieh Bathaee and Jawad~A. Salehi.
\newblock Entangled-based quantum wavelength-division-multiplexing and multiple-access networks.
\newblock {\em Entropy}, 25(12), 2023.

\bibitem{vanmilligen2024entanglement}
Emily A~Van Milligen, Eliana Jacobson, Ashlesha Patil, Gayane Vardoyan, Don Towsley, and Saikat Guha.
\newblock Entanglement routing over networks with time multiplexed repeaters, 2024.

\bibitem{sichen2022firstRequest}
Si-Chen Li, Bang-Ying Tang, Han Zhou, Hui-Cun Yu, Bo~Liu, Wan-Rong Yu, and Bo~Liu.
\newblock First request first service entanglement routing scheme for quantum networks.
\newblock {\em Entropy}, 24(10), 2022.

\bibitem{van2013path}
Rodney Van~Meter, Takahiko Satoh, Thaddeus~D Ladd, William~J Munro, and Kae Nemoto.
\newblock Path selection for quantum repeater networks.
\newblock {\em Networking Science}, 3:82--95, 2013.

\bibitem{di2012optimal}
Carlo Di~Franco and D~Ballester.
\newblock Optimal path for a quantum teleportation protocol in entangled networks.
\newblock {\em Physical Review A}, 85(1):010303, 2012.

\bibitem{van2008system}
Rodney Van~Meter, Thaddeus~D Ladd, William~J Munro, and Kae Nemoto.
\newblock System design for a long-line quantum repeater.
\newblock {\em IEEE/ACM Transactions On Networking}, 17(3):1002--1013, 2008.

\bibitem{kozlowski2019towards}
Wojciech Kozlowski and Stephanie Wehner.
\newblock Towards large-scale quantum networks.
\newblock In {\em Proceedings of the sixth annual ACM international conference on nanoscale computing and communication}, pages 1--7, 2019.

\bibitem{dupuy2023survey}
Fabrice Dupuy, Claire Goursaud, and Fabrice Guillemin.
\newblock A survey of quantum entanglement routing protocols—challenges for wide-area networks.
\newblock {\em Advanced Quantum Technologies}, page 2200180, 2023.

\bibitem{kar2023routing}
Binayak Kar and Pankaj Kumar.
\newblock Routing protocols for quantum networks: Overview and challenges.
\newblock {\em arXiv preprint arXiv:2305.00708}, 2023.

\bibitem{yan2023purification}
Pei-Shun Yan, Lan Zhou, Wei Zhong, and Yu-Bo Sheng.
\newblock Advances in quantum entanglement purification.
\newblock {\em Science China Physics, Mechanics \& Astronomy}, 66(5):250301, 2023.

\bibitem{singh2021quantum}
Amoldeep Singh, Kapal Dev, Harun Siljak, Hem~Dutt Joshi, and Maurizio Magarini.
\newblock Quantum internet—applications, functionalities, enabling technologies, challenges, and research directions.
\newblock {\em IEEE Communications Surveys \& Tutorials}, 23(4):2218--2247, 2021.

\bibitem{muralidharan2016optimal}
Sreraman Muralidharan, Linshu Li, Jungsang Kim, Norbert L{\"u}tkenhaus, Mikhail~D Lukin, and Liang Jiang.
\newblock Optimal architectures for long distance quantum communication.
\newblock {\em Scientific reports}, 6(1):20463, 2016.

\bibitem{pant2019routing}
Mihir Pant, Hari Krovi, Don Towsley, Leandros Tassiulas, Liang Jiang, Prithwish Basu, Dirk Englund, and Saikat Guha.
\newblock Routing entanglement in the quantum internet.
\newblock {\em npj Quantum Information}, 5(1):25, 2019.

\bibitem{shi2020concurrent}
Shouqian Shi and Chen Qian.
\newblock Concurrent entanglement routing for quantum networks: Model and designs.
\newblock In {\em Proceedings of the Annual conference of the ACM Special Interest Group on Data Communication on the applications, technologies, architectures, and protocols for computer communication}, pages 62--75, 2020.

\bibitem{chakraborty2020entanglement}
Kaushik Chakraborty, David Elkouss, Bruno Rijsman, and Stephanie Wehner.
\newblock Entanglement distribution in a quantum network: A multicommodity flow-based approach.
\newblock {\em IEEE Transactions on Quantum Engineering}, 1:1--21, 2020.

\bibitem{chang2022order}
Alena Chang and Guoliang Xue.
\newblock Order matters: On the impact of swapping order on an entanglement path in a quantum network.
\newblock In {\em IEEE INFOCOM 2022-IEEE Conference on Computer Communications Workshops (INFOCOM WKSHPS)}, pages 1--6. IEEE, 2022.

\bibitem{bayerbach2023bell}
Matthias~J Bayerbach, Simone~E D’Aurelio, Peter van Loock, and Stefanie Barz.
\newblock Bell-state measurement exceeding 50\% success probability with linear optics.
\newblock {\em Science Advances}, 9(32):eadf4080, 2023.

\bibitem{van2014quantum}
Rodney Van~Meter.
\newblock {\em Quantum networking}.
\newblock John Wiley \& Sons, 2014.

\bibitem{caleffi2017optimal}
Marcello Caleffi.
\newblock Optimal routing for quantum networks.
\newblock {\em Ieee Access}, 5:22299--22312, 2017.

\bibitem{ghaderibaneh2022efficient}
Mohammad Ghaderibaneh, Caitao Zhan, Himanshu Gupta, and CR~Ramakrishnan.
\newblock Efficient quantum network communication using optimized entanglement swapping trees.
\newblock {\em IEEE Transactions on Quantum Engineering}, 3:1--20, 2022.

\bibitem{zhao2021redundant}
Yangming Zhao and Chunming Qiao.
\newblock Redundant entanglement provisioning and selection for throughput maximization in quantum networks.
\newblock In {\em IEEE INFOCOM 2021-IEEE Conference on Computer Communications}, pages 1--10. IEEE, 2021.

\bibitem{nguyen2022multiple}
Tu~N Nguyen, Kashyab~J Ambarani, Linh Le, Ivan Djordjevic, and Zhi-Li Zhang.
\newblock A multiple-entanglement routing framework for quantum networks.
\newblock {\em arXiv preprint arXiv:2207.11817}, 2022.

\bibitem{li2021effective}
Changhao Li, Tianyi Li, Yi-Xiang Liu, and Paola Cappellaro.
\newblock Effective routing design for remote entanglement generation on quantum networks.
\newblock {\em npj Quantum Information}, 7(1):10, 2021.

\bibitem{abobeih2018one}
Mohamed~H Abobeih, Julia Cramer, Michiel~A Bakker, Norbert Kalb, Matthew Markham, Daniel~J Twitchen, and Tim~H Taminiau.
\newblock One-second coherence for a single electron spin coupled to a multi-qubit nuclear-spin environment.
\newblock {\em Nature communications}, 9(1):2552, 2018.

\bibitem{anderson2022five}
Christopher~P Anderson, Elena~O Glen, Cyrus Zeledon, Alexandre Bourassa, Yu~Jin, Yizhi Zhu, Christian Vorwerk, Alexander~L Crook, Hiroshi Abe, Jawad Ul-Hassan, et~al.
\newblock Five-second coherence of a single spin with single-shot readout in silicon carbide.
\newblock {\em Science advances}, 8(5):eabm5912, 2022.

\bibitem{pla2013high}
Jarryd~J Pla, Kuan~Y Tan, Juan~P Dehollain, Wee~H Lim, John~JL Morton, Floris~A Zwanenburg, David~N Jamieson, Andrew~S Dzurak, and Andrea Morello.
\newblock High-fidelity readout and control of a nuclear spin qubit in silicon.
\newblock {\em Nature}, 496(7445):334--338, 2013.

\bibitem{awschalom2018quantum}
David~D Awschalom, Ronald Hanson, J{\"o}rg Wrachtrup, and Brian~B Zhou.
\newblock Quantum technologies with optically interfaced solid-state spins.
\newblock {\em Nature Photonics}, 12(9):516--527, 2018.

\bibitem{wang2021single}
Pengfei Wang, Chun-Yang Luan, Mu~Qiao, Mark Um, Junhua Zhang, Ye~Wang, Xiao Yuan, Mile Gu, Jingning Zhang, and Kihwan Kim.
\newblock Single ion qubit with estimated coherence time exceeding one hour.
\newblock {\em Nature communications}, 12(1):233, 2021.

\bibitem{schneeloch2019introduction}
James Schneeloch, Samuel~H Knarr, Daniela~F Bogorin, Mackenzie~L Levangie, Christopher~C Tison, Rebecca Frank, Gregory~A Howland, Michael~L Fanto, and Paul~M Alsing.
\newblock Introduction to the absolute brightness and number statistics in spontaneous parametric down-conversion.
\newblock {\em Journal of Optics}, 21(4):043501, 2019.

\bibitem{dahlberg2019link}
Axel Dahlberg, Matthew Skrzypczyk, Tim Coopmans, Leon Wubben, Filip Rozpundefineddek, Matteo Pompili, Arian Stolk, Przemys\l{}aw Pawe\l{}czak, Robert Knegjens, Julio de~Oliveira~Filho, Ronald Hanson, and Stephanie Wehner.
\newblock A link layer protocol for quantum networks.
\newblock In {\em Proceedings of the ACM Special Interest Group on Data Communication}, SIGCOMM '19, page 159–173, New York, NY, USA, 2019. Association for Computing Machinery.

\bibitem{van2022entangling}
Tim van Leent, Matthias Bock, Florian Fertig, Robert Garthoff, Sebastian Eppelt, Yiru Zhou, Pooja Malik, Matthias Seubert, Tobias Bauer, Wenjamin Rosenfeld, et~al.
\newblock Entangling single atoms over 33 km telecom fibre.
\newblock {\em Nature}, 607(7917):69--73, 2022.

\bibitem{farahbakhsh2022opportunistic}
Ali Farahbakhsh and Chen Feng.
\newblock Opportunistic routing in quantum networks.
\newblock In {\em IEEE INFOCOM 2022-IEEE Conference on Computer Communications}, pages 490--499. IEEE, 2022.

\bibitem{chakraborty2019distributed}
Kaushik Chakraborty, Filip Rozpedek, Axel Dahlberg, and Stephanie Wehner.
\newblock Distributed routing in a quantum internet.
\newblock {\em arXiv preprint arXiv:1907.11630}, 2019.

\bibitem{zhang2023segmentrouting}
Ling Zhang and Qin Liu.
\newblock Concurrent multipath quantum entanglement routing based on segment routing in quantum hybrid networks.
\newblock {\em Quantum Information Processing}, 22, 03 2023.

\bibitem{cicconetti2021request}
Claudio Cicconetti, Marco Conti, and Andrea Passarella.
\newblock Request scheduling in quantum networks.
\newblock {\em IEEE Transactions on Quantum Engineering}, 2:2--17, 2021.

\bibitem{yang2024Asynchronous}
Zebo Yang, Ali Ghubaish, Raj Jain, Hassan Shapourian, and Alireza Shabani.
\newblock Asynchronous entanglement routing for the quantum internet.
\newblock {\em AVS Quantum Science}, 6(1), January 2024.

\bibitem{zhao2022E2E}
Yangming Zhao, Gongming Zhao, and Chunming Qiao.
\newblock {E2E} fidelity aware routing and purification for throughput maximization in quantum networks.
\newblock In {\em IEEE INFOCOM 2022-IEEE Conference on Computer Communications}, pages 480--489. IEEE, 2022.

\bibitem{zeng2022multi}
Yiming Zeng, Jiarui Zhang, Ji~Liu, Zhenhua Liu, and Yuanyuan Yang.
\newblock Multi-entanglement routing design over quantum networks.
\newblock In {\em IEEE INFOCOM 2022-IEEE Conference on Computer Communications}, pages 510--519. IEEE, 2022.

\bibitem{semenenko2022entanglement}
Vyacheslav Semenenko, Xuedong Hu, Eden Figueroa, and Vasili Perebeinos.
\newblock Entanglement generation in a quantum network with finite quantum memory lifetime.
\newblock {\em AVS Quantum Science}, 4(1), 2022.

\bibitem{gyongyosi2018decentralized}
Laszlo Gyongyosi and Sandor Imre.
\newblock Decentralized base-graph routing for the quantum internet.
\newblock {\em Physical Review A}, 98(2):022310, 2018.

\bibitem{gyongyosi2019adaptive}
Laszlo Gyongyosi and Sandor Imre.
\newblock Adaptive routing for quantum memory failures in the quantum internet.
\newblock {\em Quantum Information Processing}, 18:1--21, 2019.

\bibitem{ghaderibaneh2022predistributed}
Mohammad Ghaderibaneh, Himanshu Gupta, C.R. Ramakrishnan, and Ertai Luo.
\newblock Pre-distribution of entanglements in quantum networks.
\newblock In {\em 2022 IEEE International Conference on Quantum Computing and Engineering (QCE)}, pages 426--436, 2022.

\bibitem{schoute2016shortcuts}
Eddie Schoute, Laura Mancinska, Tanvirul Islam, Iordanis Kerenidis, and Stephanie Wehner.
\newblock Shortcuts to quantum network routing.
\newblock {\em arXiv preprint arXiv:1610.05238}, 2016.

\bibitem{gyongyosi2017entanglement}
Laszlo Gyongyosi and Sandor Imre.
\newblock Entanglement-gradient routing for quantum networks.
\newblock {\em Scientific reports}, 7(1):14255, 2017.

\bibitem{kleinberg2000small}
Jon Kleinberg.
\newblock The small-world phenomenon: An algorithmic perspective.
\newblock In {\em Proceedings of the thirty-second annual ACM symposium on Theory of computing}, pages 163--170, 2000.

\bibitem{pouryousef2022quantumoverlay}
Shahrooz Pouryousef, Nitish~K. Panigrahy, and Don Towsley.
\newblock A quantum overlay network for efficient entanglement distribution, 2022.

\bibitem{gyongyosi2019opportunistic}
Laszlo Gyongyosi and Sandor Imre.
\newblock Opportunistic entanglement distribution for the quantum internet.
\newblock {\em Scientific Reports}, 9(1):2219, 2019.

\bibitem{matsuo2019quantum}
Takaaki Matsuo, Cl{\'e}ment Durand, and Rodney Van~Meter.
\newblock Quantum link bootstrapping using a ruleset-based communication protocol.
\newblock {\em Physical Review A}, 100(5):052320, 2019.

\bibitem{bartling2022entanglementMinute}
H.~P. Bartling, M.~H. Abobeih, B.~Pingault, M.~J. Degen, S.~J.~H. Loenen, C.~E. Bradley, J.~Randall, M.~Markham, D.~J. Twitchen, and T.~H. Taminiau.
\newblock Entanglement of spin-pair qubits with intrinsic dephasing times exceeding a minute.
\newblock {\em Phys. Rev. X}, 12:011048, Mar 2022.

\bibitem{li2022connection}
Jian Li, Qidong Jia, Kaiping Xue, David~SL Wei, and Nenghai Yu.
\newblock A connection-oriented entanglement distribution design in quantum networks.
\newblock {\em IEEE Transactions on Quantum Engineering}, 3:1--13, 2022.

\bibitem{briegel1998quantum}
H-J Briegel, Wolfgang D{\"u}r, Juan~I Cirac, and Peter Zoller.
\newblock Quantum repeaters: the role of imperfect local operations in quantum communication.
\newblock {\em Physical Review Letters}, 81(26):5932, 1998.

\bibitem{dai2020optimal}
Wenhan Dai, Tianyi Peng, and Moe~Z Win.
\newblock Optimal remote entanglement distribution.
\newblock {\em IEEE Journal on Selected Areas in Communications}, 38(3):540--556, 2020.

\bibitem{kamin2023exact}
Lars Kamin, Evgeny Shchukin, Frank Schmidt, and Peter van Loock.
\newblock Exact rate analysis for quantum repeaters with imperfect memories and entanglement swapping as soon as possible.
\newblock {\em Physical Review Research}, 5(2):023086, 2023.

\bibitem{haldar2024fast}
Stav Haldar, Pratik~J Barge, Sumeet Khatri, and Hwang Lee.
\newblock Fast and reliable entanglement distribution with quantum repeaters: principles for improving protocols using reinforcement learning.
\newblock {\em Physical Review Applied}, 21(2):024041, 2024.

\bibitem{haldar2024reducing}
Stav Haldar, Pratik~J. Barge, Xiang Cheng, Kai-Chi Chang, Brian~T. Kirby, Sumeet Khatri, Chee~Wei Wong, and Hwang Lee.
\newblock Reducing classical communication costs in multiplexed quantum repeaters using hardware-aware quasi-local policies, 2024.

\bibitem{helsen2023benchmarking}
Jonas Helsen and Stephanie Wehner.
\newblock A benchmarking procedure for quantum networks.
\newblock {\em npj Quantum Information}, 9(1):17, 2023.

\bibitem{zhang2021directFidelity}
Xiaoqian Zhang, Maolin Luo, Zhaodi Wen, Qin Feng, Shengshi Pang, Weiqi Luo, and Xiaoqi Zhou.
\newblock Direct fidelity estimation of quantum states using machine learning.
\newblock {\em Phys. Rev. Lett.}, 127:130503, Sep 2021.

\bibitem{qbgp}
Maoli Liu, Zhuohua Li, Kechao Cai, Jonathan Allcock, Shengyu Zhang, and John~C.S. Lui.
\newblock Quantum bgp with online path selection via network benchmarking.
\newblock In {\em Proceedings of the IEEE Conference on Computer Communications (INFOCOM)}, Vancouver, Canada, May 2024.
\newblock (AR: 256/1307=19.6\%).

\bibitem{kozlowski2020designing}
Wojciech Kozlowski, Axel Dahlberg, and Stephanie Wehner.
\newblock Designing a quantum network protocol.
\newblock In {\em Proceedings of the 16th international conference on emerging networking experiments and technologies}, pages 1--16, 2020.

\bibitem{victora2020purification}
Michelle Victora, Stefan Krastanov, Alexander~Sanchez de~la Cerda, Steven Willis, and Prineha Narang.
\newblock Purification and entanglement routing on quantum networks.
\newblock {\em arXiv preprint arXiv:2011.11644}, 2020.

\bibitem{li2022fidelity}
Jian Li, Mingjun Wang, Kaiping Xue, Ruidong Li, Nenghai Yu, Qibin Sun, and Jun Lu.
\newblock Fidelity-guaranteed entanglement routing in quantum networks.
\newblock {\em IEEE Transactions on Communications}, 70(10):6748--6763, 2022.

\bibitem{zhang2021fragmentation}
Shengyu Zhang, Shouqian Shi, Chen Qian, and Kwan~L Yeung.
\newblock Fragmentation-aware entanglement routing for quantum networks.
\newblock {\em Journal of Lightwave Technology}, 39(14):4584--4591, 2021.

\bibitem{le2022dqra}
Linh Le and Tu~N Nguyen.
\newblock Dqra: Deep quantum routing agent for entanglement routing in quantum networks.
\newblock {\em IEEE Transactions on Quantum Engineering}, 3:1--12, 2022.

\bibitem{Van_Meter_2022}
Rodney Van~Meter, Ryosuke Satoh, Naphan Benchasattabuse, Kentaro Teramoto, Takaaki Matsuo, Michal Hajdusek, Takahiko Satoh, Shota Nagayama, and Shigeya Suzuki.
\newblock A quantum internet architecture.
\newblock In {\em 2022 IEEE International Conference on Quantum Computing and Engineering (QCE)}. IEEE, September 2022.

\bibitem{wang2022asynchronous}
Zhaoying Wang, Jian Li, Kaiping Xue, Shaoyin Cheng, Nenghai Yu, Qibin Sun, and Jun Lu.
\newblock An asynchronous entanglement distribution protocol for quantum networks.
\newblock {\em IEEE Network}, 36(5):40--47, 2022.

\bibitem{liu2024linkselfie}
Maoli Liu, Zhuohua Li, Xuchuang Wang, and John~C.S. Lui.
\newblock {LinkSelFiE: Link Selection and Fidelity Estimation in Quantum Networks}.
\newblock In {\em Proceedings of the IEEE Conference on Computer Communications (INFOCOM)}, Vancouver, Canada, May 2024.
\newblock (AR: 256/1307=19.6\%).

\bibitem{vardoyan2024capacity}
G.~Vardoyan, E.~van Milligen, S.~Guha, S.~Wehner, and D.~Towsley.
\newblock On the bipartite entanglement capacity of quantum networks.
\newblock {\em IEEE Transactions on Quantum Engineering}, 5(01):1--14, jan 2024.

\bibitem{qbgp_code}
Zhuohua Li.
\newblock {lizhuohua/quantum-bgp-online-path-selection: Release of QBGP paper codebase}, December 2023.

\bibitem{rfc9340}
Wojciech Kozlowski, Stephanie Wehner, Rodney~Van Meter, Bruno Rijsman, Angela~Sara Cacciapuoti, Marcello Caleffi, and Shota Nagayama.
\newblock {Architectural Principles for a Quantum Internet}.
\newblock RFC 9340, March 2023.

\bibitem{bacciottini2024redip}
Leonardo Bacciottini, Luciano Lenzini, Enzo Mingozzi, and Giuseppe Anastasi.
\newblock Redip: Ranked entanglement distribution protocol for the quantum internet.
\newblock {\em IEEE Open Journal of the Communications Society}, 5:397--411, 2024.

\bibitem{van-meter-qirg-quantum-connection-setup-01}
Rodney~Van Meter and Takaaki Matsuo.
\newblock {Connection Setup in a Quantum Network}.
\newblock Internet-Draft draft-van-meter-qirg-quantum-connection-setup-01, Internet Engineering Task Force, September 2019.
\newblock Work in Progress.

\bibitem{aguado2020qkdsdn}
Alejandro Aguado, Victor López, Juan~Pedro Brito, Antonio Pastor, Diego~R. López, and Vicente Martin.
\newblock Enabling quantum key distribution networks via software-defined networking.
\newblock In {\em 2020 International Conference on Optical Network Design and Modeling (ONDM)}, pages 1--5, 2020.

\bibitem{aguado2019engineering}
Alejandro Aguado, Victor Lopez, Diego Lopez, Momtchil Peev, Andreas Poppe, Antonio Pastor, Jesus Folgueira, and Vicente Martin.
\newblock The engineering of software-defined quantum key distribution networks.
\newblock {\em IEEE Communications Magazine}, 57(7):20--26, 2019.

\bibitem{chiti2021towards}
Francesco Chiti, Romano Fantacci, Roberto Picchi, and Laura Pierucci.
\newblock Towards the quantum internet: Satellite control plane architectures and protocol design.
\newblock {\em Future Internet}, 13(8):196, 2021.

\bibitem{picchi2020satellite}
Roberto Picchi, Francesco Chiti, Romano Fantacci, and Laura Pierucci.
\newblock Towards quantum satellite internetworking: A software-defined networking perspective.
\newblock {\em IEEE Access}, 8:210370--210381, 2020.

\bibitem{bush2021perspective}
Stephen~F Bush, William~A Challener, and Guillaume Mantelet.
\newblock A perspective on industrial quantum networks.
\newblock {\em AVS Quantum Science}, 3(3), 2021.

\bibitem{jain2013b4}
Sushant Jain, Alok Kumar, Subhasree Mandal, Joon Ong, Leon Poutievski, Arjun Singh, Subbaiah Venkata, Jim Wanderer, Junlan Zhou, Min Zhu, Jon Zolla, Urs H\"{o}lzle, Stephen Stuart, and Amin Vahdat.
\newblock B4: experience with a globally-deployed software defined wan.
\newblock In {\em Proceedings of the ACM SIGCOMM 2013 Conference on SIGCOMM}, SIGCOMM '13, page 3–14, New York, NY, USA, 2013. Association for Computing Machinery.

\bibitem{rexford2002traffic}
B.~Fortz, J.~Rexford, and M.~Thorup.
\newblock Traffic engineering with traditional ip routing protocols.
\newblock {\em Comm. Mag.}, 40(10):118–124, oct 2002.

\bibitem{hong2018b4andafter}
Chi-Yao Hong, Subhasree Mandal, Mohammad Al-Fares, Min Zhu, Richard Alimi, Kondapa~Naidu B., Chandan Bhagat, Sourabh Jain, Jay Kaimal, Shiyu Liang, Kirill Mendelev, Steve Padgett, Faro Rabe, Saikat Ray, Malveeka Tewari, Matt Tierney, Monika Zahn, Jonathan Zolla, Joon Ong, and Amin Vahdat.
\newblock B4 and after: managing hierarchy, partitioning, and asymmetry for availability and scale in google's software-defined wan.
\newblock In {\em Proceedings of the 2018 Conference of the ACM Special Interest Group on Data Communication}, SIGCOMM '18, page 74–87, New York, NY, USA, 2018. Association for Computing Machinery.

\bibitem{khorsandroo2021hybridsdn}
Sajad Khorsandroo, Adrián~Gallego Sánchez, Ali~Saman Tosun, JM~Arco, and Roberto Doriguzzi-Corin.
\newblock Hybrid sdn evolution: A comprehensive survey of the state-of-the-art.
\newblock {\em Computer Networks}, 192:107981, 2021.

\bibitem{ghaderibaneh2022predistribution}
Mohammad Ghaderibaneh, Himanshu Gupta, C.R. Ramakrishnan, and Ertai Luo.
\newblock Pre-distribution of entanglements in quantum networks.
\newblock In {\em 2022 IEEE International Conference on Quantum Computing and Engineering (QCE)}, pages 426--436, 2022.

\bibitem{kozlowski2020p4}
Wojciech Kozlowski, Fernando Kuipers, and Stephanie Wehner.
\newblock A p4 data plane for the quantum internet.
\newblock In {\em Proceedings of the 3rd P4 Workshop in Europe}, CoNEXT ’20. ACM, December 2020.

\bibitem{dasari2016openflow}
Venkat~R Dasari and Travis~S Humble.
\newblock Openflow arbitrated programmable network channels for managing quantum metadata.
\newblock {\em The Journal of Defense Modeling and Simulation: Applications, Methodology, Technology}, 16(1):67–77, October 2016.

\bibitem{illiano2021impact}
Jessica Illiano, Angela~Sara Cacciapuoti, Antonio Manzalini, and Marcello Caleffi.
\newblock The impact of the quantum data plane overhead on the throughput.
\newblock In {\em Proceedings of the Eight Annual ACM International Conference on Nanoscale Computing and Communication}, pages 1--6, 2021.

\bibitem{gyongyosi2018survey_capacity}
Laszlo Gyongyosi, Sandor Imre, and Hung~Viet Nguyen.
\newblock A survey on quantum channel capacities.
\newblock {\em IEEE Communications Surveys \& Tutorials}, 20(2):1149--1205, 2018.

\bibitem{matt2023batfish}
Matt Brown, Ari Fogel, Daniel Halperin, Victor Heorhiadi, Ratul Mahajan, and Todd Millstein.
\newblock Lessons from the evolution of the batfish configuration analysis tool.
\newblock In {\em Proceedings of the ACM SIGCOMM 2023 Conference}, ACM SIGCOMM '23, page 122–135, New York, NY, USA, 2023. Association for Computing Machinery.

\bibitem{zhang2022dna}
Peng Zhang, Aaron Gember-Jacobson, Yueshang Zuo, Yuhao Huang, Xu~Liu, and Hao Li.
\newblock Differential network analysis.
\newblock In {\em 19th USENIX Symposium on Networked Systems Design and Implementation (NSDI 22)}, pages 601--615, Renton, WA, April 2022. USENIX Association.

\bibitem{chen2023simqn}
Lutong Chen, Kaiping Xue, Jian Li, Nenghai Yu, Ruidong Li, Qibin Sun, and Jun Lu.
\newblock Simqn: A network-layer simulator for the quantum network investigation.
\newblock {\em IEEE Network}, 37(5):182--189, 2023.

\bibitem{gyongyosi2018topology}
Laszlo Gyongyosi and Sandor Imre.
\newblock Topology adaption for the quantum internet.
\newblock {\em Quantum Information Processing}, 17:1--12, 2018.

\bibitem{choi2023scalable}
Hyeongrak Choi, Marc~G. Davis, Álvaro G.~Iñesta, and Dirk~R. Englund.
\newblock Scalable quantum networks: Congestion-free hierarchical entanglement routing with error correction, 2023.

\bibitem{kumar2023routing}
Vinay Kumar, Claudio Cicconetti, Marco Conti, and Andrea Passarella.
\newblock Routing in quantum repeater networks with mixed noise figures, 2023.

\bibitem{munro2022designing}
W.~J. Munro, Nicolo'~Lo Piparo, Josephine Dias, Michael Hanks, and Kae Nemoto.
\newblock {Designing tomorrow's quantum internet}.
\newblock {\em AVS Quantum Science}, 4(2):020503, 06 2022.

\bibitem{azuma2015allphotonic}
Koji Azuma, Kiyoshi Tamaki, and Hoi-Kwong Lo.
\newblock All-photonic quantum repeaters.
\newblock {\em Nature Communications}, 6(1):6787, 2015.

\bibitem{benchasattabuse2024architecture}
Naphan Benchasattabuse, Michal Hajdušek, and Rodney~Van Meter.
\newblock Architecture and protocols for all-photonic quantum repeaters, 2024.

\bibitem{wu2021sequence}
Xiaoliang Wu, Alexander Kolar, Joaquin Chung, Dong Jin, Tian Zhong, Rajkumar Kettimuthu, and Martin Suchara.
\newblock Se{QU}e{NC}e: a customizable discrete-event simulator of quantum networks.
\newblock {\em Quantum Science and Technology}, 6(4):045027, 2021.

\bibitem{satoh2022quisp}
Ryosuke Satoh, Michal Hajdu{\v{s}}ek, Naphan Benchasattabuse, Shota Nagayama, Kentaro Teramoto, Takaaki Matsuo, Sara~Ayman Metwalli, Poramet Pathumsoot, Takahiko Satoh, Shigeya Suzuki, et~al.
\newblock Qu{ISP}: a quantum internet simulation package.
\newblock In {\em 2022 IEEE International Conference on Quantum Computing and Engineering (QCE)}, pages 353--364. IEEE, 2022.

\bibitem{mininet}
Mininet.
\newblock \url{http://mininet.org/}.
\newblock Accessed: 2024-01-11.

\bibitem{optical-mininet}
Welcome to mininet-optical.
\newblock \url{https://mininet-optical.org/README.html}.
\newblock Accessed: 2024-01-30.

\bibitem{zhuge2023dt}
Qunbi Zhuge, Xiaomin Liu, Yihao Zhang, Meng Cai, Yichen Liu, Qizhi Qiu, Xueying Zhong, Jiaping Wu, Ruoxuan Gao, Lilin Yi, and Weisheng Hu.
\newblock Building a digital twin for intelligent optical networks -- {I}nvited {T}utorial.
\newblock {\em J. Opt. Commun. Netw.}, 15(8):C242--C262, Aug 2023.

\bibitem{wang2024dt}
Danshi Wang, Yuchen Song, Yao Zhang, Xiaotian Jiang, Jiawei Dong, Faisal~Nadeem Khan, Takeo Sasai, Shanguo Huang, Alan Pak~Tao Lau, Massimo Tornatore, and Min Zhang.
\newblock Digital twin of optical networks: A review of recent advances and future trends.
\newblock {\em Journal of Lightwave Technology}, 42(12):4233--4259, 2024.

\bibitem{huynh2024genai_phy}
Nguyen Van~Huynh, Jiacheng Wang, Hongyang Du, Dinh~Thai Hoang, Dusit Niyato, Diep~N. Nguyen, Dong~In Kim, and Khaled~B. Letaief.
\newblock Generative ai for physical layer communications: A survey.
\newblock {\em IEEE Transactions on Cognitive Communications and Networking}, 10(3):706--728, 2024.

\bibitem{cobo2022management}
Iván García-Cobo.
\newblock Quantum network intelligent management system.
\newblock {\em Optics}, 3(4):430--437, 2022.

\bibitem{spiess2023sync}
Christopher Spiess, Sebastian T\"opfer, Sakshi Sharma, Andrej Kr\ifmmode \check{z}\else \v{z}\fi{}i\ifmmode~\check{c}\else \v{c}\fi{}, Meritxell Cabrejo-Ponce, Uday Chandrashekara, Nico~Lennart D\"oll, Daniel Riel\"ander, and Fabian Steinlechner.
\newblock Clock synchronization with correlated photons.
\newblock {\em Phys. Rev. Appl.}, 19:054082, May 2023.

\bibitem{satoh2021attacking}
Takahiko Satoh, Shota Nagayama, Shigeya Suzuki, Takaaki Matsuo, Michal Hajdusek, and Rodney~Van Meter.
\newblock Attacking the quantum internet.
\newblock {\em IEEE Transactions on Quantum Engineering}, 2:1–17, 2021.

\bibitem{zhou2022security}
Hongyi Zhou, Kefan Lv, Longbo Huang, and Xiongfeng Ma.
\newblock Quantum network: Security assessment and key management.
\newblock {\em IEEE/ACM Transactions on Networking}, 30(3):1328–1339, June 2022.

\bibitem{nielsen2010quantum}
M.A. Nielsen and I.L. Chuang.
\newblock {\em Quantum Computation and Quantum Information: 10th Anniversary Edition}.
\newblock Cambridge University Press, 2010.

\bibitem{wilce2007no-broadcasting}
Howard Barnum, Jonathan Barrett, Matthew Leifer, and Alexander Wilce.
\newblock Generalized no-broadcasting theorem.
\newblock {\em Phys. Rev. Lett.}, 99:240501, Dec 2007.

\bibitem{van2017multiplexed}
Suzanne~B van Dam, Peter~C Humphreys, Filip Rozpedek, Stephanie Wehner, and Ronald Hanson.
\newblock Multiplexed entanglement generation over quantum networks using multi-qubit nodes.
\newblock {\em Quantum Science and Technology}, 2(3):034002, 2017.

\bibitem{Barz_2010}
Stefanie Barz, Gunther Cronenberg, Anton Zeilinger, and Philip Walther.
\newblock Heralded generation of entangled photon pairs.
\newblock {\em Nature Photonics}, 4(8):553–556, June 2010.

\bibitem{heshami2016quantum}
Khabat Heshami, Duncan~G England, Peter~C Humphreys, Philip~J Bustard, Victor~M Acosta, Joshua Nunn, and Benjamin~J Sussman.
\newblock Quantum memories: emerging applications and recent advances.
\newblock {\em Journal of modern optics}, 63(20):2005--2028, 2016.

\bibitem{muralidharan2014ultrafast}
Sreraman Muralidharan, Jungsang Kim, Norbert L{\"u}tkenhaus, Mikhail~D Lukin, and Liang Jiang.
\newblock Ultrafast and fault-tolerant quantum communication across long distances.
\newblock {\em Physical review letters}, 112(25):250501, 2014.

\bibitem{bauml2017fundamental}
Stefan B{\"a}uml and Koji Azuma.
\newblock Fundamental limitation on quantum broadcast networks.
\newblock {\em Quantum Science and Technology}, 2(2):024004, 2017.

\bibitem{santra2019quantum}
Siddhartha Santra, Liang Jiang, and Vladimir~S Malinovsky.
\newblock Quantum repeater architecture with hierarchically optimized memory buffer times.
\newblock {\em Quantum Science and Technology}, 4(2):025010, 2019.

\bibitem{cacciapuoti2019quantum}
Angela~Sara Cacciapuoti, Marcello Caleffi, Francesco Tafuri, Francesco~Saverio Cataliotti, Stefano Gherardini, and Giuseppe Bianchi.
\newblock Quantum internet: Networking challenges in distributed quantum computing.
\newblock {\em IEEE Network}, 34(1):137--143, 2019.

\bibitem{multiplexing2011}
Luciano Aparicio and Rodney~Van Meter.
\newblock {Multiplexing schemes for quantum repeater networks}.
\newblock In Ronald~E. Meyers, Yanhua Shih, and Keith~S. Deacon, editors, {\em Quantum Communications and Quantum Imaging IX}, volume 8163, page 816308. International Society for Optics and Photonics, SPIE, 2011.

\bibitem{BBM92}
Charles~H. Bennett, Gilles Brassard, and N.~David Mermin.
\newblock Quantum cryptography without bell's theorem.
\newblock {\em Phys. Rev. Lett.}, 68:557--559, Feb 1992.

\bibitem{coopmans2021netsquid}
Tim Coopmans, Robert Knegjens, Axel Dahlberg, David Maier, Loek Nijsten, Julio de~Oliveira~Filho, Martijn Papendrecht, Julian Rabbie, Filip Rozpedek, Matthew Skrzypczyk, et~al.
\newblock Net{S}quid, a network simulator for quantum information using discrete events.
\newblock {\em Communications Physics}, 4(1):164, 2021.

\bibitem{dahlberg2018simulaqron}
Axel Dahlberg and Stephanie Wehner.
\newblock Simula{Q}ron—a simulator for developing quantum internet software.
\newblock {\em Quantum Science and Technology}, 4(1):015001, 2018.

\bibitem{diadamo2021qunetsim}
Stephen DiAdamo, Janis N{\"o}tzel, Benjamin Zanger, and Mehmet~Mert Be{\c{s}}e.
\newblock Qu{N}et{S}im: A software framework for quantum networks.
\newblock {\em IEEE Transactions on Quantum Engineering}, 2:1--12, 2021.

\end{thebibliography}
\bibliographystyle{unsrt}

\appendix

\section{Quantum Communication Background}\label{appendix:q_comm}

This section introduces the fundamental quantum physics operations and properties essential in quantum communication. Instead of detailing each principle in isolation, as is common in the literature, we will illustrate these principles through the step-by-step design of a hypothetical quantum network, intended to transfer quantum states two end-nodes over a long distance. This allows us to present these operations as they functionally coexist in a quantum network. Some technological specifics and hardware components are intentionally understated to maintain an abstracted perspective on quantum networks and entanglement routing.

\subsection{From Quantum Bits to Quantum Networks}\label{ssec:qbit_to_qnet}

\paragraph{Qubits and Quantum States}
Quantum computing operates on quantum bits (qubits) encoded in the quantum state of particles such as photons, electrons, and atoms. A quantum state has the particularity of being in a superposition; i.e., 0 and 1 at the same time \cite{singh2021quantum}. Various technologies can be used to represent a qubit such as encoding a quantum state in photon polarization or electron spin. 
The principles of quantum mechanics hold regardless of the underlying qubit technology. 

Scaling the computational power of quantum computers and expanding the potential of quantum computing to more applications, requires the transfer of qubits between distant nodes without destroying their superposition states. 
For example, Alice begins by encoding her message into the quantum states of photons, which will be sent to Bob. This initial step introduces the challenge of maintaining the integrity of the qubits during transmission.
The no-cloning principle \cite{nielsen2010quantum} prohibits the exact replication of quantum states, ensuring that quantum information remains fundamentally secure but also challenging to manipulate. Additionally, any measurement of a quantum system inevitably alters its state. 
Consequently, qubits cannot be measured and then replicated to repeat a quantum signal or retry a transmission. These principles do not hold in with classical signals; which is why classical networks can use amplifiers, feed-forward error correction, and retransmission to reliably send units of data across the network. 

\paragraph{Elementary Entanglement}
Bipartite entanglement is a special connection between two quantum particles, like photons, where their properties are linked together regardless of the distance between them. 
This means if one of the entangled qubits is observed or measured, it instantly determines the state of the other, breaking the entangled state, no matter how far apart they are. 
It is as if the two qubits are acting as one unit, even when they are separated by an arbitrary distance. 
Unlike a mix of separate states, this entangled state cannot be expressed by simpler, individual states for each particle \cite{illiano2022quantum}.
Since a qubit cannot be replicated, a third qubit cannot be entangled with either of the two entangled qubits. This is referred to as the no-broadcasting theorem \cite{wilce2007no-broadcasting}. 

Alice and Bob are considered \textit{entangled nodes} if they share one or more pairs of entangled qubits. 
These pairs of entangled qubits are known as \textit{Bell pairs} or Einstein-Podolsky-Rosen (EPR) pairs (terms that we will use interchangeably throughout this paper) symbolizing the four orthogonal states possible for two maximally entangled qubits located with Alice and Bob, respectively. 
As a result, when Alice independently measures her qubit from a given Bell pair, she receives a random result, with the probabilities of zero and one outcomes being equally likely. Bob experiences the same phenomenon with his measurement.
If Alice and Bob measure their respective qubits in the same basis, the results will be perfectly correlated. This is regardless of the distance between Alice and Bob and without any further interaction between the two parties.

Some prominent schemes to generate Bell pairs include spontaneous parametric down-conversion (SPDC), single-atom excitation by laser beam, and two-atom simultaneous excitation by laser beam \cite{singh2021quantum}. Most entangled photon pair sources (EPPS) based on these schemes are nondeterministic as to when Bell pairs are generated. Based on these schemes, so-called elementary entanglement generation (EEG) protocols are designed to attempt to generate Bell pairs and distribute each entangled qubit to one end-node of a direct link \cite{van2017multiplexed}. Elementary entanglement is defined as entanglement between two neighboring nodes (i.e., directly connected through an optical channel).

\paragraph{Entanglement Generation and Heralding}
One way to overcome the limitations of nondeterministic sources is by heralding. Heralding refers to the process of mutually acknowledging the confirmed presence or absence of an entangled pair between the nodes attempting an entanglement creation. For example, one can implement a heralded entangled photon source by generating bipartite entangled states conditioned on the detection of additional auxiliary photons \cite{Barz_2010}. If these auxiliary photons are found to meet the requisite conditional measurement patterns, they herald the successful creation of the entangled pair. Another example scheme would be to integrate heralded quantum memories at each end-node which can confirm the successful storage of a photon upon its arrival. As such, a Heralded Entanglement Generation (HEG) protocol may require a two-way classical signaling message to each qubit recipient node \cite{muralidharan2016optimal} to confirm the presence of an entangled state between the two end-nodes. 
It is important to note that this additional classical communication overhead of HEG protocols has critical implications for higher-layer network protocols such as routing.

\begin{figure}[!h]
\centering
\includegraphics[width=0.5\linewidth]{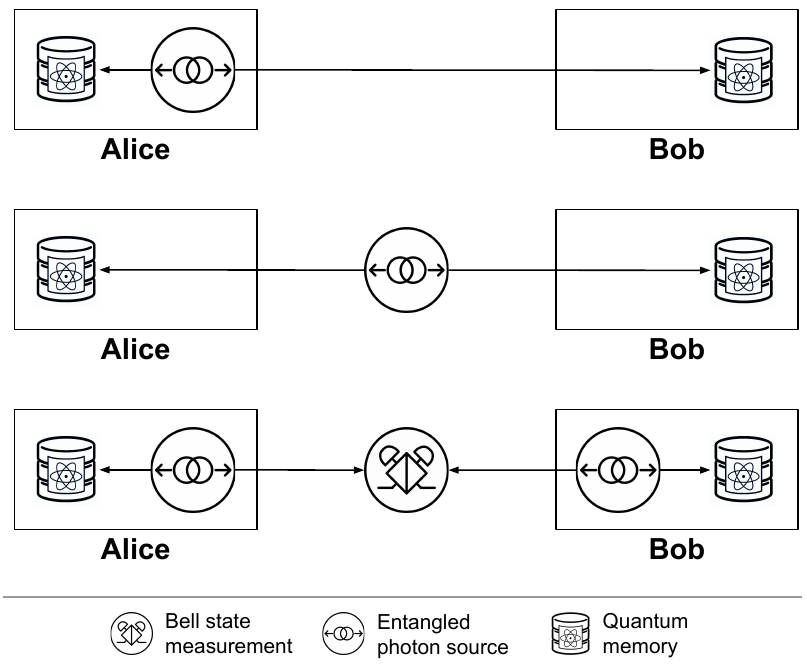}
\caption{\label{fig:heg}Varying physical architectures for entanglement generation between quantum memories on Alice and Bob: node-source (top), midpoint-source (middle), and meet-in-the-middle (bottom).}
\end{figure}

There are several architectures one can consider for HEG between two neighboring nodes, as illustrated in Figure \ref{fig:heg}. We summarize these architectures as node-source, midpoint-source, and meet-in-the-middle. 

Let the physical distance between Alice and Bob be $L$. In a node-source architecture, the sender (Alice) has a local EPPS and a quantum memory that receives one entangled qubit from the EPPS, and the receiver (Bob) has a local quantum memory that receives the other entangled qubit from Alice’s EPPS after it traverses over distance $L$. 

To make an HEG protocol from a node-source architecture, only one-way classical communication from Alice to Bob over distance $L$ is required after each attempt. 
In a midpoint-source architecture, Alice and Bob each receive an entangled qubit into their local quantum memory from an EPPS that lies somewhere on the channel between them. To make a HEG protocol from a midpoint-source architecture, the EPPS must send a heralding signal to both Alice and Bob after each attempt. 

In a meet-in-the-middle architecture, Alice and Bob both have a local EPPS and quantum memory, from which they each emit one entangled qubit from their EPPS to their local memory and send the other entangled qubit to a Bell-state measurement (BSM) station that lies somewhere on the channel between them. If the qubits arriving at the BSM are indistinguishable (i.e., have identical or perfectly correlated properties such as their polarization states, phase, wavelength, and timing), a successful BSM will result in entanglement between the quantum memories at Alice and Bob. 
Depending on the capabilities of the BSM station, the emitted qubits from each pair generation attempt at Alice and Bob may need to arrive at the BSM station simultaneously, requiring a time-synchronized system. 
To make an HEG protocol from a meet-in-the-middle architecture, the BSM station must send a heralding signal to both Alice and Bob after each attempt. In addition to the heralding signal itself, these classical messages may also include data indicating which Bell state was prepared in the process, which allows the end nodes to execute any relevant state transformations, if necessary.

While the HEG protocols described herein assume that the classical communication of heralding signals is reliable, the HEG protocol implementation must be robust in the face of lost or garbled messages. It is straightforward to see that each of these architectures and HEG protocols have different tradeoffs with classical communication overhead, hardware and infrastructure requirements, and time synchronization.

\paragraph{Quantum Memory}
Entanglement generation schemes frequently use photons (flying qubits) to encode quantum information and deliver the generated entangled states to end-nodes over optical fiber, and atoms, or matter qubits, to store the entangled state at each node in a quantum memory. Photons are a promising medium for encoding and transmitting quantum information due to their degrees of freedom, and their fast traversal speed over fiber and free-space optical links. 

Capturing and storing a photon in a quantum memory can be achieved with various technologies \cite{heshami2016quantum}, either with solid-state platforms or free-space platforms. Some platforms support on-demand photon emission from the memory, and others may only emit at set time intervals. 
Memories can be characterized by key parameters such as their storage time representing the time interval beyond which the stored quantum state is irreversibly degraded and can no longer be used. This results from entanglement decoherence where the entangled pair of particles degrades over time due to interactions with their surrounding environment.
Other memory parameters can be storage efficiency, retrieval efficiency, and acceptable wavelength range.

\paragraph{Fidelity and Purification}
Decoherence, fiber loss, or environmental disturbances can prevent a pair of qubits from achieving or maintaining optimal-quality entanglement, thus compromising the utility of the entangled link for its intended application.
The probability that a pair of entangled qubits is in a specific, desired state is quantified by a value known as the fidelity of an entanglement.

\begin{figure}[!h]
\centering
\includegraphics[width=0.5\linewidth]{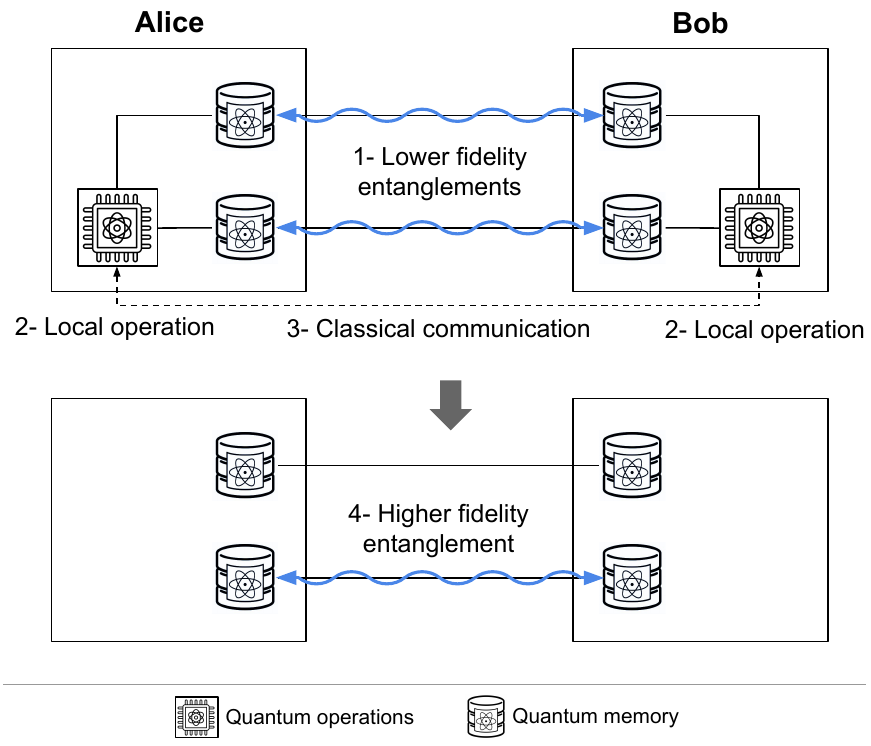}
\caption{\label{fig:purification}Diagram of entanglement purification. With local physical operations and communication through classical channels, an entangled pair with high fidelity can be distilled from two pairs with lower fidelity.}
\end{figure}

Purification (or distillation) techniques have been developed to increase entanglement fidelity \cite{yan2023purification}. One such technique, Heralded Entanglement Purification (HEP), allows for the conversion of two or more low-fidelity Bell pairs into a single pair with higher fidelity, as illustrated in Figure \ref{fig:purification}. The HEP process can sometimes fail and necessitates bidirectional classical communication to notify both end nodes about the outcome of the purification effort \cite{singh2021quantum}.

\paragraph{Quantum Error Correction}
In addition to purification, Quantum Error Correction (QEC) is essential for managing errors in quantum states during transmission \cite{muralidharan2016optimal, singh2021quantum}. 
QEC techniques encode a logical qubit into a block of several physical qubits, which helps in preserving the integrity of the quantum state despite potential errors that may arise from decoherence or other quantum operations. By encoding and then decoding the information, QEC allows for the correction of errors without needing to directly observe the quantum state, thereby not violating the no-cloning principle. 
However, the effectiveness of QEC is constrained by the no-cloning theorem to less than a 50\% threshold \cite{muralidharan2014ultrafast}. This error threshold represents the maximum rate of errors that can be corrected by a QEC protocol. Beyond this threshold, the accumulation of errors outpaces the capability of the error correction process, failing to recover the original quantum information.

\paragraph{Entanglement Swapping}
Given the inherently lossy nature of optical channels, the probability of successfully distributing an entangled pair decreases exponentially as the physical distance of the channel increases.
There are several approaches for mitigating photon loss to extend the distance over which high-quality entanglements are distributed. For example, using high-rate sources or multiplexing techniques (increase the number of attempts per unit time), high-efficiency photon detectors (more accurate measurements), low-loss optical fiber, or free-space channels can all help improve HEG success probability. 
However, these approaches can only partially mitigate the exponential nature of photon loss and rapidly become insufficient \cite{bauml2017fundamental}. 
To distribute entanglement across even longer distances, quantum repeaters are used to chain a sequence of elementary entanglements into a longer one using swapping \cite{santra2019quantum}.

\begin{figure}[!h]
\centering
\includegraphics[width=0.65\linewidth]{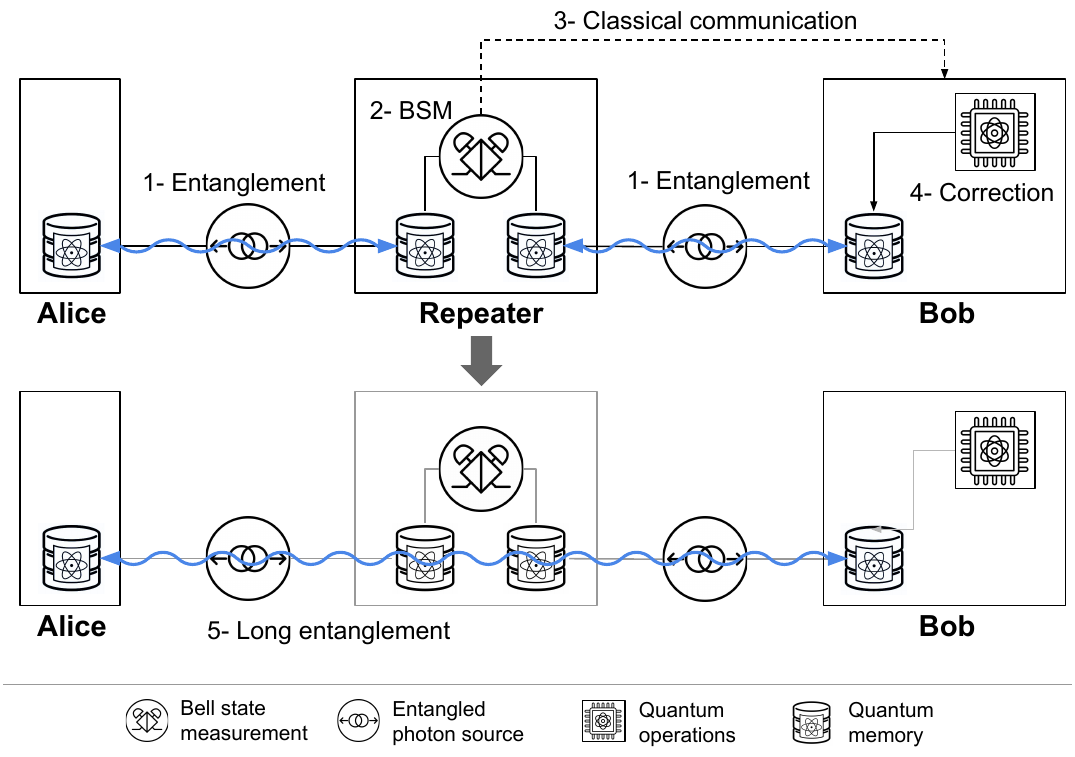}
\caption{\label{fig:swapping}Diagram of entanglement swapping where two distant nodes share an entanglement with the assistance of an intermediate repeater.}
\end{figure}

The swapping process is illustrated in Figure \ref{fig:swapping}.
One repeater is placed between Alice and Bob to split the end-to-end distance into two smaller distances. Two elementary entanglements are generated; one between the Alice and the repeater and one between the repeater and the Bob. 
The repeater then measures its two entangled local qubits and sends the measurement outcome to Bob. Based on the received measurement, Bob applies a correction to its local qubit creating an E2E entanglement between Alice and Bob.
It is important to note that the measurement process at the repeater destroys the two initial entanglements, thus requiring the generation of new elementary entanglements whenever swapping is used. The success of the swapping operation is also probabilistic.

To support longer distances between Alice and Bob, more repeaters can be placed between the two end nodes. Elementary entanglements are created from Alice to the first repeater, between repeaters, and from the last repeater to Bob. These elementary entanglements are consumed as the swapping operations are executed at each repeater creating the E2E entanglement. 

If all swapping operations are successful, the E2E entanglement can be obtained by applying the swapping corrections in any order, given that they are performed within the coherence time of the entangled pairs \cite{illiano2022quantum, cacciapuoti2019quantum}. This allows using different swapping orders for creating the E2E entanglement. However, it is worth noting that the fidelity of the resulting entanglement depends on the time elapsed before applying corrections. That is, the resulting entanglement fidelity is also affected by the time taken by the classical measurements to reach the end node.

\paragraph{Quantum Repeaters}
The main function of a repeater is to implement the swapping process described above. 
Entanglement distribution over quantum repeaters is subject to two main types of errors \cite{muralidharan2016optimal}: loss errors arising from the attenuation of fibers, and operational errors stemming from inaccuracies in manipulating and measuring quantum states. 
 
By repeating the heralded process until the neighboring repeaters confirm the successful creation of an entanglement, the HEG protocol serves as a mechanism for natively mitigating loss errors. During this period, the entangled qubits are maintained by the repeaters until a clear indication of success or failure is communicated.

On the other hand, QEC offers a more efficient approach, though it demands more resources. At each repeater, QEC can be implemented to restore the original logical qubit. In this scenario, the quantum and classical signals flow unidirectionally. That is, the end nodes do not store any quantum states once the logical qubit is restored and transmitted.

To handle operational errors, the choice falls between QEC \cite{nielsen2010quantum} and Heralded Entanglement Purification (HEP) \cite{yan2023purification}. HEP utilizes a set of low-fidelity Bell pairs to probabilistically generate a reduced number of high-fidelity pairs, necessitating a bidirectional classical communication for the verification of success. 
In contrast, QEC addresses these errors through a unidirectional classical communication system but requires quantum gates with higher fidelity.

The evolution of quantum repeaters can be delineated into three distinct generations based on their error mitigation strategies \cite{muralidharan2016optimal}.

The first generation integrates HEG for loss error management and HEP for the correction of operation errors. The communication is based on the production of high-fidelity entangled pairs between neighboring repeaters, and employing HEP at each stage of entanglement swapping to counteract fidelity degradation due to quantum operations. It is important to note that executing an HEP protocol on multi-hop entanglement requires classical signaling between non-adjacent nodes, adding a potentially costly latency overhead.

The second generation employs HEG for the identification of loss errors and QEC for operation error mitigation. In this architecture, entangled qubits distributed through HEG enable the formation of a Bell pair for each qubit within an encoded physical block. QEC is applied to restore the logical qubit from the block of entangled physical qubits, thus obviating the necessity for bidirectional classical communication by substituting HEP with QEC.

The third generation exclusively relies on QEC to address both loss and operational errors, directly encoding the logical qubit within a series of physical qubits dispatched through channels as photons. Provided the errors are minimal, the incoming physical qubits can be employed to reconstruct and transmit the encoding block to the subsequent station. As a result, the signaling is unidirectional and the communication rate can potentially be maximized.

\begin{figure}[!h]
\centering
\includegraphics[width=0.5\linewidth]{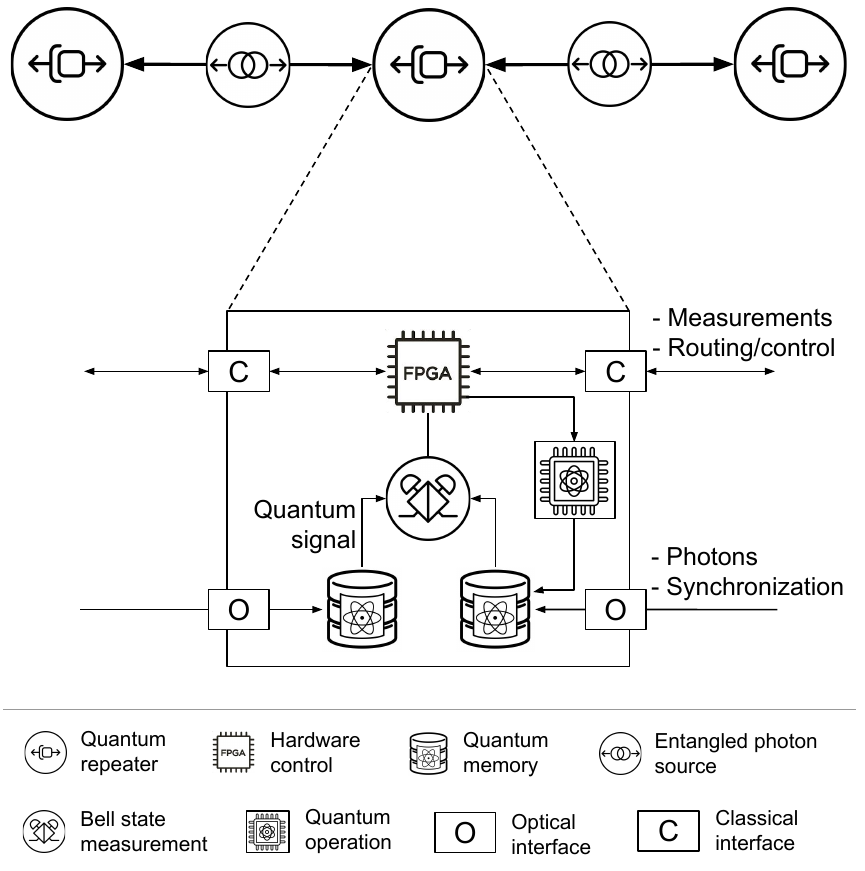}
\caption{\label{fig:repeater}Simplified architecture of a first-generation quantum repeater assuming midpoint-source entanglement architecture.}
\end{figure}

The first-generation repeaters are currently in the development phase. Although not commercially available yet, a high-level view of the architecture of a first-generation quantum repeater is illustrated in Figure \ref{fig:repeater}. 
Such repeater consists of quantum measurement and operation components such as BSM and quantum gates, and quantum memory modules. Depending on the specific features and technology, the architecture may include an entangled photon source and a quantum error correction circuit. 
In addition to optical ports, repeaters also include classical communication ports to coordinate actions and share measurement outcomes.
These components and interactions would be controlled with a pre-programmed hardware component such as an FPGA to meet the real-time requirements of low-level quantum processes. 
At a higher level, software-based control can be used for relatively slower operations such as network instruction processing and forwarding table lookups.

Note that quantum repeaters are limited to only extending entanglement distribution distances over a linear path. Path computation and routing are implemented in quantum routers which, as their classical counterpart, are quantum repeaters with more than two interfaces; i.e., they can service direct quantum connections between more than two neighbors.

\paragraph{Multiplexing} 
First-generation quantum networks utilize circuit switching, which establishes a dedicated path between the two communicating end nodes before transmitting quantum information. To serve E2E entanglements across the network, a policy is needed to allocate the qubits available at repeaters to different paths. Various approaches have been proposed for this purpose.
A straightforward one is to reserve the entire quantum channel along the path for the duration of the communication session. While easy to implement, this pure circuit switching method has limitations in terms of network utilization and scalability.
Alternatively, buffer space multiplexing divides the limited memory or buffer space at each repeater among the different paths passing through it \cite{multiplexing2011}. This method allows multiple communications to be supported concurrently, thereby improving utilization over pure circuit switching.
Statistical multiplexing dynamically shares the available buffer space across different paths based on demand, rather than statically dividing it \cite{multiplexing2011}. 
Time-division multiplexing allocates available time slots at each repeater node among different paths. 

Multiplexing techniques are also used to manage entanglement creation over a single link, for example through time-division \cite{vanmilligen2024entanglement} or wavelength-division \cite{bathaee2023multiplexing}.
Overall, the choice of multiplexing approach involves balancing performance, complexity, and implementation feasibility trade-offs for the specific quantum network design.

\begin{figure}[ht]
\centering
\includegraphics[width=0.65\linewidth]{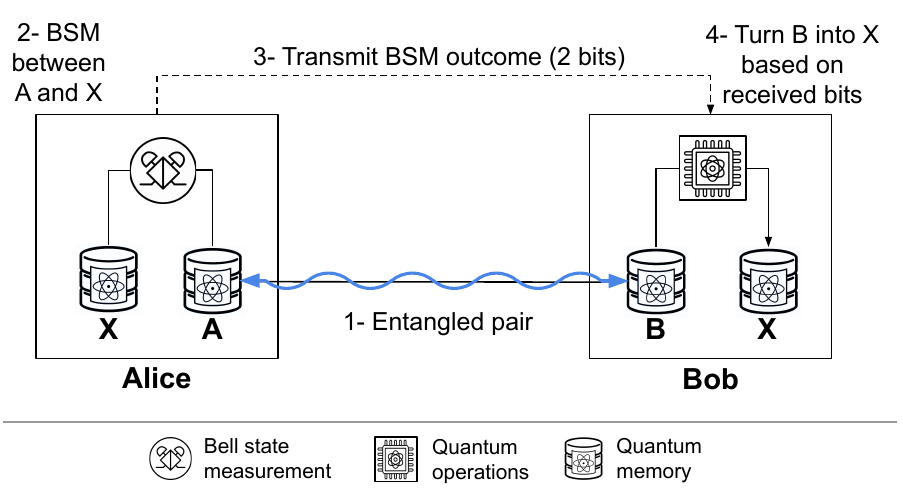}
\caption{\label{fig:teleportation}Quantum teleportation: an unknown qubit is transferred from Alice to Bob by consuming an entangled pair A-B shared between Alice and Bob.}
\end{figure}

\paragraph{Teleportation}
Distributing entanglements between Alice and Bob does not constitute an exchange of quantum data by itself. Instead, it serves as a connection to transmit data qubits through the teleportation process \cite{van2014quantum}. Since qubits cannot be copied, teleportation is used for the transmission of a quantum state without physically transferring the quantum particle that encodes the qubit. 
The teleportation process is depicted in Figure \ref{fig:teleportation}.
The procedure for teleporting an arbitrary qubit $X$ from Alice to Bob involves generating an E2E entanglement between Alice and Bob, where qubit $A$ is located at Alice and qubit $B$ at Bob. Alice then performs a BSM of $X$ and $A$, and the two-bit result of this measurement is communicated to Bob through a classical channel. Upon receiving this information, Bob applies specific operations to qubit $B$ based on the measurement outcome, resulting in the retrieval of qubit $X$ \cite{illiano2022quantum}. 

While the teleportation operation is described to illustrate an end-to-end exchange of quantum information, it is worth noting that there are applications, such as entanglement-based quantum key distribution \cite{BBM92}, that consume E2E entanglements directly without teleportation.

\subsection{Quantum Network Simulators}
Several quantum network simulators are currently available as either open-source packages or commercial products, each supporting different levels of abstraction and catering to various aspects of quantum networking. Some notable examples include:
\begin{itemize}
    \item NetSquid \cite{coopmans2021netsquid}: a Python-based discrete-event simulator for scalable quantum networks featuring the effects of time on quantum devices’ performance. 
    
    \item SeQUeNCe \cite{wu2021sequence}: a full-stack discrete-event simulator specifically designed for studying quantum links, systems, and networks. 
    
    \item QuISP \cite{satoh2022quisp}: a specialized simulation environment focusing on quantum repeater networks. 
    
    \item SimulaQron \cite{dahlberg2018simulaqron}: an application-level simulator, facilitating testing and development of quantum network protocols. 

    \item QuNetSim \cite{diadamo2021qunetsim}: a high-level framework that focuses solely on the network layer and does not implement lower-level quantum device operations, and thus is more suitable for the development and testing of quantum networking protocols.

\end{itemize}
Similar to classical network simulators, quantum network simulators provide significant benefits in modeling and testing, protocol development and optimization, and scalability and security assessments. They play a crucial role in advancing quantum communication research, enabling innovation, and preparing for the future quantum Internet.

\end{document}